\documentclass[aps,prd,onecolumn,preprintnumbers,superscriptaddress,nofootinbib,floatfix,notitlepage,longbibliography]{revtex4-1}
\usepackage[utf8]{inputenc}
\usepackage{amssymb}
\usepackage{amsmath}
\numberwithin{equation}{section}
\usepackage{tensor}
\usepackage{xspace} 
\usepackage[pdftex]{hyperref}
\usepackage{hyperref}
\usepackage{graphicx}
\usepackage[caption=false]{subfig}
\usepackage{breakcites}
\usepackage{ragged2e}
\usepackage{tikz}
\usepackage{orcidlink}
\usepackage{setspace}

\hypersetup{
colorlinks,linkcolor=blue,citecolor=blue,urlcolor=blue
}

\usepackage{natbib}
\usepackage{tablefootnote}
\usepackage{footnote}
\usepackage{amsmath}
\usepackage{amssymb}
\usepackage[T1]{fontenc}

\begin{document}

\title{\boldmath Non-linear Electrodynamics in Blandford-Znajeck Energy Extraction}

\author{A. Carleo }
\email{acarleo@unisa.it}
\affiliation{Dipartimento di Fisica, Universit\`a di Salerno, Via Giovanni Paolo II, 132 I-84084 Fisciano (SA), Italy}
\affiliation{INFN, Sezione di Napoli, Gruppo collegato di Salerno, Italy}
\affiliation{INAF, Osservatorio Astronomico di Cagliari, Via della Scienza 5,
09047 Selargius (CA), Italy}

\author{G. Lambiase  \orcidlink{0000-0001-7574-2330}}
\email{lambiase@sa.infn.it}
\affiliation{Dipartimento di Fisica, Universit\`a di Salerno, Via Giovanni Paolo II, 132 I-84084 Fisciano (SA), Italy}
\affiliation{INFN, Sezione di Napoli, Gruppo collegato di Salerno, Italy}

\author{A. \"Ovg\"un
\orcidlink{0000-0002-9889-342X}
}
\email{ali.ovgun@emu.edu.tr}
\affiliation{Physics Department, Eastern Mediterranean University, Famagusta, 99628 North Cyprus via Mersin 10, Turkey}







\begin{abstract}

Non-linear electrodynamics (NLED) is a  generalization of Maxwell’s electrodynamics for strong fields. It could have significant implications for the study of black holes and cosmology and have been extensively studied in the literature, extending from quantum to cosmological contexts. Recently, its application to black holes, inflation and dark energy has caught on, being able to provide an accelerated Universe and address some current theoretical inconsistencies, such as the Big Bang singularity. In this work, we report two new ways to investigate these non-linear theories. First, we have analyzed the Blandford-Znajeck mechanism in light of this promising theoretical context, providing the general form of the extracted power up to second order in the black hole spin parameter $a$. We have found that, depending on the NLED model, the emitted power can be extremely  increased or  decreased, and that the magnetic field lines around the black hole seems to become vertical quickly. Considering only separated solutions, we have found that no monopole solutions exist and this could have interesting astrophysical consequences (not considered here). 
Last but not least, we attempted to confine the NLED parameters by inducing the amplification of primordial magnetic fields (`seeds'), thus admitting non-linear theories already during the early stages of the Universe. However,  the latter approach  proved to be useful  for NLED research only in certain models. Our (analytical) results emphasize that the existence and behavior of non-linear electromagnetic phenomena strongly depend on the physical context and that only a power-low model seems to have any chance to compete with Maxwell. 

\end{abstract}


\maketitle
\flushbottom


\section{Introduction}
\label{sec1}
\setcounter{equation}{0}
Maxwell's electromagnetic theory (MED) is a widely used fundamental theory in both quantum physics and the context of cosmology. It is a well-known and recognized theory. In 1933 and 1934 Born and Infeld made the first attempts to change equations of MED \cite{Born1933ModifiedFE,Born:1934gh} and tried to eliminate the divergence of the electron's self-energy in classical electrodynamics. The Born-Infeld electrodynamics model does not contain any singularities because its electric field starts at its highest value at the center (which is equal to the nonlinearity parameter $b$), and then gradually decreases until it behaves like the electric field of Maxwell at longer distances. This model also ensures that the energy of a single point charge is limited. The parameter $b$ has a connection to the tension of strings in the theory \cite{Gibbons:2001gy,Fradkin:1985qd}, and there have been studies done to determine potential constraints for the value of $b$ in \cite{Davila:2013wba,Ellis:2017edi,NiauAkmansoy:2017kbw,NiauAkmansoy:2018ilv,Neves:2021tbt,Neves:2021jdy,DeFabritiis:2021qib}. In contrast to the Euler-Heisenberg electrodynamics \cite{Heisenberg:1936nmg}, the Born-Infeld model does not show vacuum birefringence when subjected to an external electric field. The Born-Infeld theory maintains both causality and unitarity principles. The Born-Infeld electrodynamics has served as inspiration for other models that are free of singularities and possess similar properties. For instance, various models presented by Kruglov in \cite{Kruglov:2007bh,KRUGLOV_B,Kruglov:2014iwa,Kruglov:2014iqa,Kruglov:2015yua,Kruglov:2015fbl,Kruglov:2016cdm,Kruglov:2016ymq,Kruglov:2016ezw,Kruglov:2017mpj,Kruglov:2017fck,Kruglov:2017xmb,Kruglov:2017xmb}. Fang and Wang have presented a fruitful method for finding black hole solutions that have either electric or magnetic charges, in a theory that combines General Relativity with a nonlinear electrodynamics \cite{Fan:2016hvf}. Since then, numerous models have been advocated, and the effects of these theories—known as Non Linear Electrodynamics (NLED)—have been investigated in a wide range of contexts, not just those related to cosmology and astrophysics \cite{25,26,27,28,29,30,31,32,33,refe1,Panotopoulos:2020bfl,Panotopoulos:2017hns,Toshmatov:2018tyo,Stuchlik:2019uvf,Rayimbaev:2022hrn,Toshmatov:2014nya,Abdujabbarov:2016hnw,Toshmatov:2017zpr,Stuchlik:2014qja}, but also in non-linear optics \cite{34}, high power laser technologies and plasma physics \cite{35,Laser}, nuclear physics \cite{38,39}, and supeconductors \cite{Panotopoulos:2020bfl}. Many gravitational non-linear electrodynamics (G-NED), extensions of the Reissner-Nordstrom (RN) solutions of the Einstein- Maxwell field equations have gained a lot of attention (see \cite{Rincon:2021gwd,Gonzalez:2021vwp,Rincon:2018dsq,Rincon:2017goj,garcia} and references therein). Additionally, Stuchlík and Schee have demonstrated that models that produce the weak-field limit of Maxwell are considered relevant, as opposed to those that do not provide the correct enlargement of black hole shadows in the absence of charges \cite{Stuchlik:2019uvf}. In particular, the existence of axially symmetric non-linear charged black holes (at least transiently) has been studied \cite{Lambiase2017}, indicating neutrinos as  good probes thanks to their bountiful production in any astrophysical context. As a consequence, it would be interesting, in principle, to investigate the nature of electromagnetism (linear or not), due to different signatures in certain neutrino phenomena, such as neutrino oscillations, spin-flip and r-processes. The effect of non-linear phenomena on the BH shadow, BH thermodynamics, deflection angle of light and also wormholes have been investigated too \cite{Okyay,Ovgun:2019wej,Kumaran_2022,Pantig:2022gih,Javed:2022kzf,Kuang:2018goo,Javed:2019kon,Javed:2020lsg,ElMoumni:2020wrf,Uniyal:2022vdu,Jusufi:2018jof,Halilsoy:2013iza,Allahyari:2019jqz,Vagnozzi:2022moj}. In the context of primordial physics, instead, NLED, when
coupled to a gravitational field,  can give the necessary
negative pressure and enhance cosmic inflation \cite{GARC_A_SALCEDO_2000} and some models also  prevent cosmic singularity at the big bang \cite{Ovgun:2022,Ovgun:2016oit,Ovgun:2017iwg,Otalora:2018bso,OvgunExpo_2021} and ensure matter-antimatter asymmetry \cite{matter_antimatter}. The reason to consider NLED in the primordial Universe comes from the assumption that electromagnetic and gravitational
fields were very strong during the evolution of the early
universe, thereby leading to quantum correction and
giving birth to NLED \cite{Kunze:2013kza,Durrer_2013}.     
Recently, the non-linear electrodynamics has been also
invoked as an available framework for generating the primordial magnetic fields (PMFs) in the Universe \cite{kunz2008,Campanelli2008}. The latter,
indeed, is a still open problem of the modern cosmology,
and although many mechanisms have been proposed, this
issue is far to be solved. Seed of magnetic fields may
arise in different contexts, e.g. string cosmology \cite{10},
inflationary models of the Universe \cite{11,12}, non-minimal electromagnetic-gravitational coupling \cite{14,15},
gauge invariance breakdown \cite{12,16}, density perturbations \cite{17}, gravitational waves in the early Universe
\cite{18}, Lorentz violation \cite{19}, cosmological defects \cite{20}, electroweak anomaly \cite{21}, temporary electric charge non-conservation \cite{22}, trace anomaly \cite{23}, parity violation of the
weak interactions \cite{24}. The current state of art points to an unexplained physical mechanism that creates large-scale magnetic fields and seems to be present in all astrophysical contexts.
They might be remnants of the early Universe that were amplified later in a pregalactic period, according to one idea. To create such large-scale fields, super-horizon correlations can only still be created during inflationary epochs. However, it is still unclear how the electromagnetic  conformal symmetry is broken. Different theoretical techniques have been taken into consideration for this, most notably non-minimal coupling with gravity, which by its very nature broke conformal symmetry (\cite{CarleoPMFs} and reference therein). In a minimal scenario, electromagnetic conformal invariance can also be overcome. In this instance, the major goal is to modify the electromagnetic Lagrangian to a non-linear function of $F\doteq (1/4)F_{\mu \nu}F^{\mu \nu}$, as done in \cite{kunz2008,Campanelli2008,Lambiase2009}.

Since all NLED models significantly depend on scale factors (dimensionless or not), which may cause overlaps with other physics observables, it is obvious that determining the presence of non-linear phenomena is not free of uncertainty. Energy extraction from black holes, which is connected to various significant astrophysical events, including black hole jets and therefore Gamma-ray bursts (GRBs), is one area where NLED effects have not yet been properly studied \cite{komissorov2005}. The Blandford-Znajeck (BZ) process \cite{BZ-1977ds,BZ,BZ1,BZ2,BZ3,BZ4} and the (very recent) magnetic reconnection mechanism \cite{Comisso,Carleo:reconnect}  are the two different energy extraction techniques used today, along with a revised version of the original Penrose process \cite{Wald:1974kya} called magnetic Penrose process \cite{MPP1,MPP2,https://doi.org/10.48550/arxiv.1210.1041}. Among them, the BZ mechanism is still the most widely accepted theory to explain high energy phenomena \cite{Sharma_2021,Takahashi_2021} (even if there are still open questions in certain models or combinations \cite{King_2021,Komissarov_2021,Komissarov_2005}). 
It involves a magnetic field generated by the accretion disk, whose field lines are accumulated during the accretion process and twisted inside the rotating ergosphere. Charged particles within the cylinder of twisted lines can be accelerated away from the black hole, composing the jets. A characteristic feature of this mechanism is that the energy loss rate decays exponentially. This has been confirmed in a good fraction of observations (X-ray light curves) of GRBs \cite{expGRBs}. Furthermore, black holes with brighter accretion disks have more powerful jets implying  a correlation between the two. Even if accretion onto a black hole is the most efficient process for emitting energy from matter it is not able to reach the energy rate of the GRBs, while other energy extraction ways such as the Hawking radiation give predictions on temperature, time-scale and energy rate  highly in conflict  with the observations \cite{RUFFINI_2002}.  Numerical models of black hole accretion systems have
significantly progressed our understanding of relativistic jets indicating two types of jets, one
 associated with the disc that is mass-loaded by disc material and
the other  associated directly with the black hole \cite{Kinney2006}. In the first case, however, jets with high Lorentz factors are not supported. The BZ process, which produces highly relativistic jets by electromagnetically extracting black hole spin energy, remains the most astrophysically plausible mechanism to do so and is in good agreement with direct observations  \cite{Steiner_2012}. 
In this sense, understanding the general relativistic magnetohydrodynamic (GRMHD) model of the
bulk flow dynamics near the black hole (where  relativistic jets
are formed) is essential to study the central engine.

In this paper, in order to determine if non-linear effects may change the rate of energy extraction and the magnetic field configuration surrounding a (non-charged) black hole encircled by its magnetosphere, we will investigate the Blandford-Znajek mechanism in the context of the NLED framework.

The layout of the paper is as follows: in Sec. \ref{sec2} we derive, for the first time, the general version of energy flux up to second order in the spin parameter. Sec. \ref{sec3} is devoted to computing and solving the magnetohydrodynamic problem in Kerr-Schild coordinates, searching, in particular, for separated (monopole and paraboloid) solutions. In  Sec. \ref{sec4} we give some estimates of the energy extraction w.r.t. standard BZ mechanism. We study primordial magnetic fields from (minimally coupled) NLED for different non-linear models in Sec. \ref{sec5}, while discussion and conclusions are drawn in the Sec. \ref{sec6}.   
 In this work, we adopt  natural units  $G=c=1$ and for simplicity set $M=1$ in order to handle adimensional quantities ($r$,$a$,...). The negative metric signature  $(+, -, -, -)$ is also adopted.

\section{Non-linear Magnetohydrodynamics}
\label{sec2}
\setcounter{equation}{0}
In this section, following \cite{BZ-1977ds} and \cite{Gammie_2004}, we derive the energy extraction rate for a spinning, non-charged black hole in presence of stationary, axisymmetric, force-free,
magnetized plasma and an externally sourced magnetic field. In the Kerr-Schild coordinate \footnote{Unlike the classic Kerr coordinates, the Kerr-Schild ones ensure finiteness of the electromagnetic field on the horizon. Notice that here we use a different metric signature than \cite{Gammie_2004} and that in \cite{BZ-1977ds} simpler Kerr coordinates are used.}, the axially symmetric spacetime line element is

\begin{eqnarray}
  ds^{2}& =& \Big( 1-\dfrac{2r}{\Sigma}\Big)dt^{2}-\Big(\dfrac{4r}{\Sigma} \Big)dr dt - \Big( 1+\dfrac{2r}{\Sigma}\Big)dr^{2}-\Sigma d\theta^{2} - {\sin^{2}\theta}\Big[\Sigma + a^{2}\Big( 1+\dfrac{2r}{\Sigma}\Big) \Big]d\phi^{2} \nonumber \\ &+&\Big(\dfrac{4ar\sin^{2}\theta}{\Sigma} \Big)d\phi dt + 2a \Big( 1+\dfrac{2r}{\Sigma} \Big)\sin^{2}\theta d\phi dr, \label{metricG}
 \end{eqnarray}
 
where $\Sigma:=r^{2}+a^{2}\cos^{2}\theta$, $\Delta=r^{2}-2r+a^{2}$. The metric determinant is $g:=|det(g_{\mu\nu})|=-\Sigma^{2}\sin^{2}\theta$. We consider now a general electromagnetic Lagrangian governing the surrounding plasma and call it ${L}_{NLED}$; it is generally a function of the two invariants $X:= (1/4) F_{\mu \nu}F^{\mu \nu}$ and $G:=(1/4) F_{\mu \nu}F^{* \mu \nu}$, where, called $A_{\mu}=(\Phi,-\mathbf{A})$ the four-potential,  $F_{\mu \nu} = \partial_{\mu}A_{\nu} - \partial_{\nu}A_{\mu}$ is the electromagnetic field strength tensor  and $F^{*\mu\nu}=\frac{1}{2}F_{\alpha\beta}\epsilon^{\alpha\beta\mu\nu}$ is its dual ($\epsilon$ is the anti-symmetric Levi-Civita tensor ). Clearly, Maxwell theory is recovered when ${L}_{NLED}=-X$. The energy-momentum tensor, in absence of magnetic charges, is 
\begin{equation}
T_{\mu \nu}^{EM} := \dfrac{2}{\sqrt{-g}}\dfrac{\delta L_{NLED}}{\delta g^{\mu \nu}} = - L(X) g_{\mu \nu}+ L_{X}F_{\mu\rho}F_{\nu\sigma}g^{\rho\sigma},    
\end{equation}
where  with $L_{X}$ we indicate the derivative of $L$ w.r.t. $X$. In principle, the total energy-momentum  tensor should also take matter contribution into account, i.e. $T_{\mu \nu}^{tot}:= T_{\mu \nu}^{EM}+T_{\mu \nu}^{MAT}$, but in the free-force approximation the latter disappears \cite{Gammie_2004}. This leads to
\begin{equation}\label{eq2.2}
    \nabla^{\nu} T_{\mu \nu}^{tot} \approx  \nabla^{\nu} T_{\mu \nu}^{EM} = 0,
\end{equation}
together with the generalized Maxwell equations
\begin{equation}\label{eq2.3}
    \dfrac{1}{\sqrt{-g}}\partial_{\mu}\Big[ \sqrt{-g} L_{X} F^{\mu \nu}   \Big] = - J^{\nu},
\end{equation}
\begin{equation}\label{eq2.4}
    \partial_{\mu} F^{*\mu \nu} = 0,
\end{equation}
with  $J^{\nu}=(\rho,\mathbf{J})$ the four-current density.   Since the plasma is assumed ideal, the electric field in the particle frame, $\mathbf{E'}$, is zero. However, the presence of an external magnetic field leads to a non-zero electric field $\mathbf{E}$, but the ideal MHD approximation implies that $\mathbf{E}\cdot \mathbf{B} =0$, i.e. $G=0$, from which \cite{Gammie_2004}
\begin{equation}\label{eq2.5}
    \dfrac{\partial_{\theta}A_{t}}{\partial_{\theta}A_{\phi}} = \dfrac{\partial_{r}A_{t}}{\partial_{r}A_{\phi}} =: w(r,\theta),
\end{equation}
where we introduced the function $w(r,\theta)$.  With this notation, the electromagnetic tensor is 
\begin{equation}\label{eq2.6}
F_{\mu \nu}=\sqrt{-g}\left(\begin{array}{cccc}
0 & wB_{\theta} & -wB_{r} & 0 \\
-wB_{\theta} & 0 &  -B_{\phi} & B_{\theta} \\
wB_{r} & B_{\phi} & 0 &  -B_{r} \\
0 & -B_{\theta} & B_{r} & 0 
\end{array}\right)
\end{equation}
which automatically satisfies (\ref{eq2.3}). The radial energy and angular momentum flux, as measured by a stationary long-distance observer, are given by
\begin{equation}
    F^{(r)}_{E}:= T^{r}_{t}, \; \; \; \; \; \; F^{(r)}_{L}:= -T^{r}_{\phi} \; .
\end{equation}
Therefore
\begin{equation*}
  F^{(r)}_{E}= - L_ {X} \Big( F_{t\theta}F_{\theta \phi}g^{r\phi}+F_{t\theta}F_{\theta r}g^{rr} - F^{2}_{t \theta}g^{rt}\Big)g^{\theta \theta},
\end{equation*}
and hence 
\begin{equation}\label{eq2.8}
F^{(r)}_{E}= L_ {X} \Big[ 2 B_{r}^{2} w r \Big(w-\dfrac{a}{2r} \Big) + w B_{r}B_{\phi} \Delta \Big]\sin^{2}\theta,
\end{equation}
while the angular momentum flux is $F_{L}^{(r)}=F_{E}^{(r)} / w$. On the horizon, $r_{+}:= 1+\sqrt{1-a^{2}}$, Eq. (\ref{eq2.8}) reads as 
\begin{equation}
    F_{E}(\theta) := -2 L_{X}^{(r_{+})}B_{r}^{2}wr_{+} (\Omega_{H}-w)\sin^{2}\theta,
\end{equation}
where $L_{X}^{(r_{+})}:=L_{X}(r_{+},\theta)$ and  $\Omega_{H}:=a/(2r_{+})$ is the angular velocity of the horizon.
Apart from the factor $L_{X}$, these relations are equal to the linear (Maxwell) case. However, although the change is minimal, the physical consequences could be decisive. Indeed, $ F_{E}(\theta)>0$ not only if $0<w<\Omega_{H}$, but also if $L_{X} < 0$ at the horizon. Moreover, since $L_{X}$ is a function of $X$, and \footnote{$X=\dfrac{1}{2} \left( |\mathbf{B}|^2-|\mathbf{E}|^2 \right)$}
\begin{equation}
    X= \dfrac{1}{2}\Big[B_{r}^{2}(1-w^{2}) + B_{\theta}^{2}(1-w^{2})+B_{\phi}^{2} \Big],
\end{equation}
the energy flux will depend not only on the radial magnetic field $B_{r}$, but in general also on the other two components, namely $B_{\theta}$ and $B_{\phi}$. The power extracted (energy rate) is 
\begin{equation}\label{eq2.11}
    P^{NLED}:=\iint  d\theta d\phi \sqrt{-g}F_{E}(\theta) = 4 \pi \int_{0}^{\pi/2} d\theta \sqrt{-g} F_{E}(\theta).
\end{equation}
In order to evaluate $P^{NLED}$, we need to solve MHD equations and find the expressions for $B_{r}$, $B_{\theta}$ and $B_{\phi}$. This is not an easy task, being quite laborious already in the standard Maxwell theory. As a first approach, we can certainly proceed with a perturbative series expansion in powers of $a$, as originally done in \cite{BZ-1977ds}. Since typically one assumes $w=\Omega / 2$, then $F_{E}\propto a^{2}$ so a Schwarzschild solution (i.e. $a=0$) is fine to obtain an expression for $P^{NLED}$ good up to second order in the spin parameter. It is clear that such a relation would be accurate only in the regime $a \ll 1 $.    \\
Since we want to completely solve the magnetohydrodynamic equations, instead of Eq. (\ref{eq2.2}), we use the (equivalent) set  of equations $F_{\mu\nu}J^{\nu}=0$, coming from free-force approximation. Only two equations are independent, and they give 
\begin{equation}\label{eq2.12}
    J_{r} = -\mu (r,\theta) B_{r} , \; \; \; \; J_{\theta} = -\mu (r,\theta) B_{\theta} ,   \; \; \; \;  J_{\phi} = -\mu (r,\theta) B_{\phi} + J_{t} w
\end{equation}
where we defined $\mu := - J_{\theta}/B_{\theta} = - J_{r}/B_{r}$. The above equations are formally equivalent to those of \cite{BZ-1977ds} and seem not to  depend a priori  on the specific NLED model. However, when coupled to Maxwell equations, difference with the linear theory appears clear. Indeed, in order to find the explicit expression for $\mu$ and $J_{t}$, from Eqs. (\ref{eq2.3}), we get the following set of equations:
\begin{equation}\label{eq2.13}
\begin{array}{cccc}
\partial_{r}\Big[\sin^{2}\theta L_{X} B_{\theta} \Big( \Delta w Y + 4r^{2}w-2ra\Big) \Big] +\partial_{\theta} \Big[ \sin^{2}\theta L_{X} \Big( 2r B_{\phi} - w B_{r} Y \Big) \Big]=-J_{t} \Sigma \sin \theta  \\
\partial_{\theta}\Big[ \sin^{2}\theta L_{X} (2rwB_{r}+\Delta B_{\phi}-aB_{r}) \Big] = -J_{r} \Sigma \sin \theta \\
\partial_{r}\Big[ \sin^{2}\theta L_{X} (2rwB_{r}+\Delta B_{\phi}-aB_{r}) \Big] = J_{\theta} \Sigma \sin \theta \\
\partial_{r}\Big[ \sin^{2} L_{X}\theta \Big( 2rawB_{\theta}-a^{2}B_{\theta}+ \dfrac{\Delta B_{\theta}}{\sin^{2}\theta} \Big)  \Big] + \partial_{\theta}\Big[ \sin^{2}\theta L_{X} \Big( aB_{\phi}-\dfrac{B_{r}}{\sin^{2}\theta} \Big)  \Big] = -J_{\phi} \Sigma \sin \theta \; .
\end{array}
\end{equation}
Together with Eqs. (\ref{eq2.12}) and in a very similar way to \cite{BZ-1977ds}, they lead to \footnote{Notice that our definition for $B_{\phi}$ differs from that of \cite{BZ-1977ds} by a factor $\sqrt{-g}$  (as assumed in \cite{Gammie_2004}) and we use Kerr-Schild coordinates. }
\begin{equation}\label{eq2.14}
    \mu = - \dfrac{d}{d A_{\phi}} \Big[  \sin^{2}\theta L_{X} \Big( \Delta B_{\phi} + 2rwB_{r} -aB_{r} \Big)   \Big] 
\end{equation}
where the explicit dependence on $L_{X}$ is shown. We will call $B_{T}$ the expression in square brackets by analogy with \cite{BZ-1977ds}, even if, in our notation and coordinates, it will be not properly the toroidal field.  \\
By putting $J_{t}$ from  Eq. (\ref{eq2.13}) into Eq. (\ref{eq2.12}) and by using Eq. (\ref{eq2.14}), the important differential equation for $A_{\phi}$ is found:
\begin{equation}\label{eq2.15}
\begin{array}{cccccc}
    B_{\phi}\dfrac{dB_{T}}{dA_{\phi}} = \dfrac{w}{\Sigma \sin\theta}\Bigg[ \partial_{r}\Big(\sin^{2}\theta L_{X}B_{\theta}(\Delta w Y + 4r^{2}w-2ra)\Big) + \partial_{\theta}\Big(\sin^{2}\theta L_{X} ( 2r B_{\phi} - w B_{r} Y )  \Big) \Bigg] \\
    - \dfrac{1}{\Sigma \sin\theta}\partial_{r}\Big[ \sin^{2} L_{X}\theta \Big( 2rawB_{\theta}-a^{2}B_{\theta}+ \dfrac{\Delta B_{\theta}}{\sin^{2}\theta} \Big) \Big] -\dfrac{1}{\Sigma \sin\theta}\partial_{r}\Big[ \sin^{2}\theta L_{X} \Big( aB_{\phi}-\dfrac{B_{r}}{\sin^{2}\theta} \Big) \Big] .
    \end{array}
\end{equation}

Notice that  $w$, $B_{r}$ and $B_{\theta}$ are functions (only) of $A_{\phi}$ by definition of $F_{\mu\nu}$, hence Eq. (\ref{eq2.14}) implies $B_{\phi}$ is only a function of $A_{\phi}$.  In summary, our first unknowns $(B_{r},B_{\theta},B_{\phi},w)$, after using the ideal approximation (\ref{eq2.5}), Maxwell equations (\ref{eq2.3}) and the free-force approximation, have been reduced to one, namely $A_{\phi}$. Eq. (\ref{eq2.15}), for $L_{X}=-1$, is also known as 'stream equation', and its solution $A_{\phi}$ is called  'stream function' \cite{Ghosh_2000}.

\section{Separated Solutions}
\label{sec3}
\setcounter{equation}{0}

\begin{figure*}[t!!!]
\centering
\vspace{0.cm}
\includegraphics[width=7.5cm]{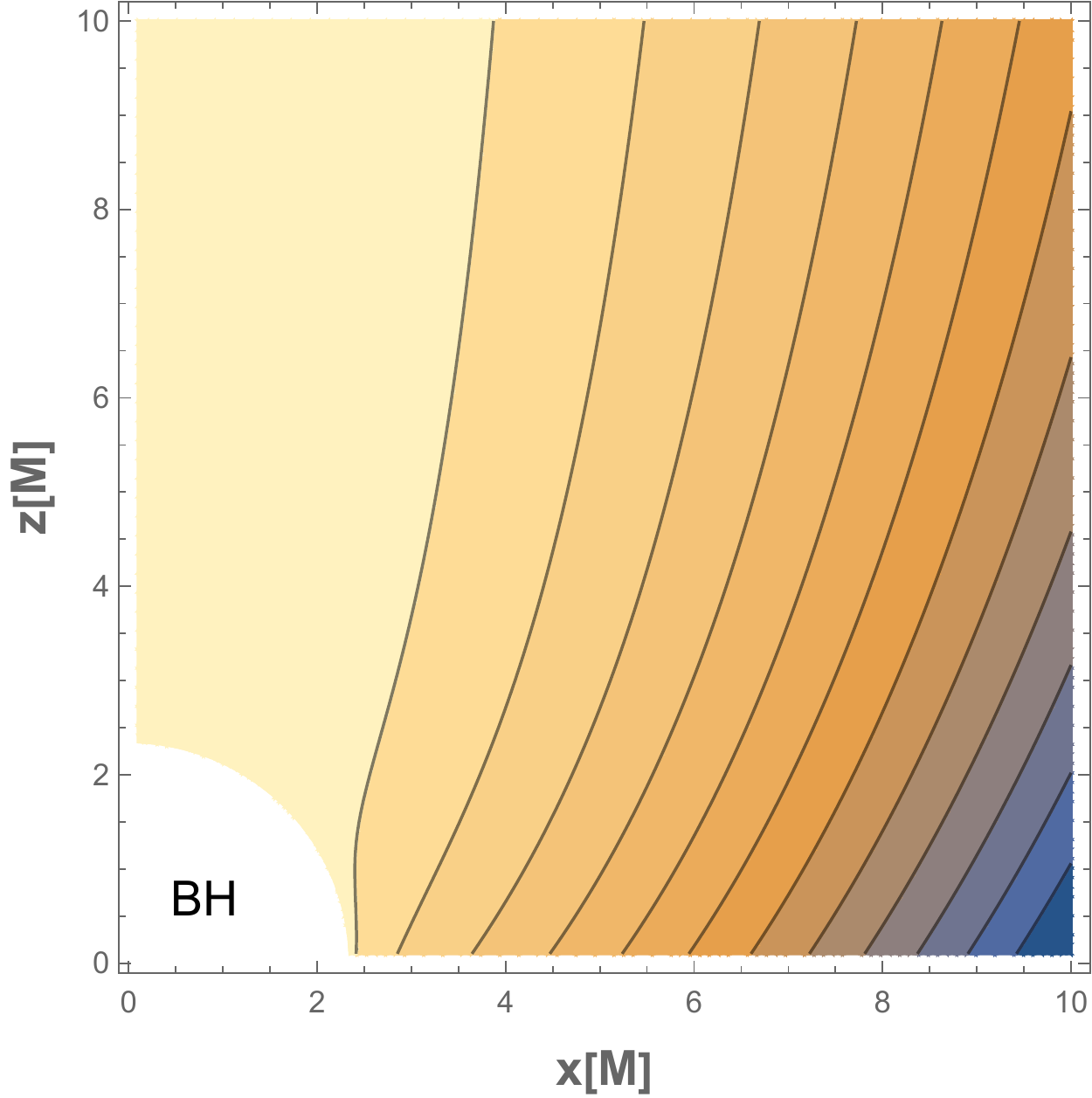}
\includegraphics[width=7.5cm]{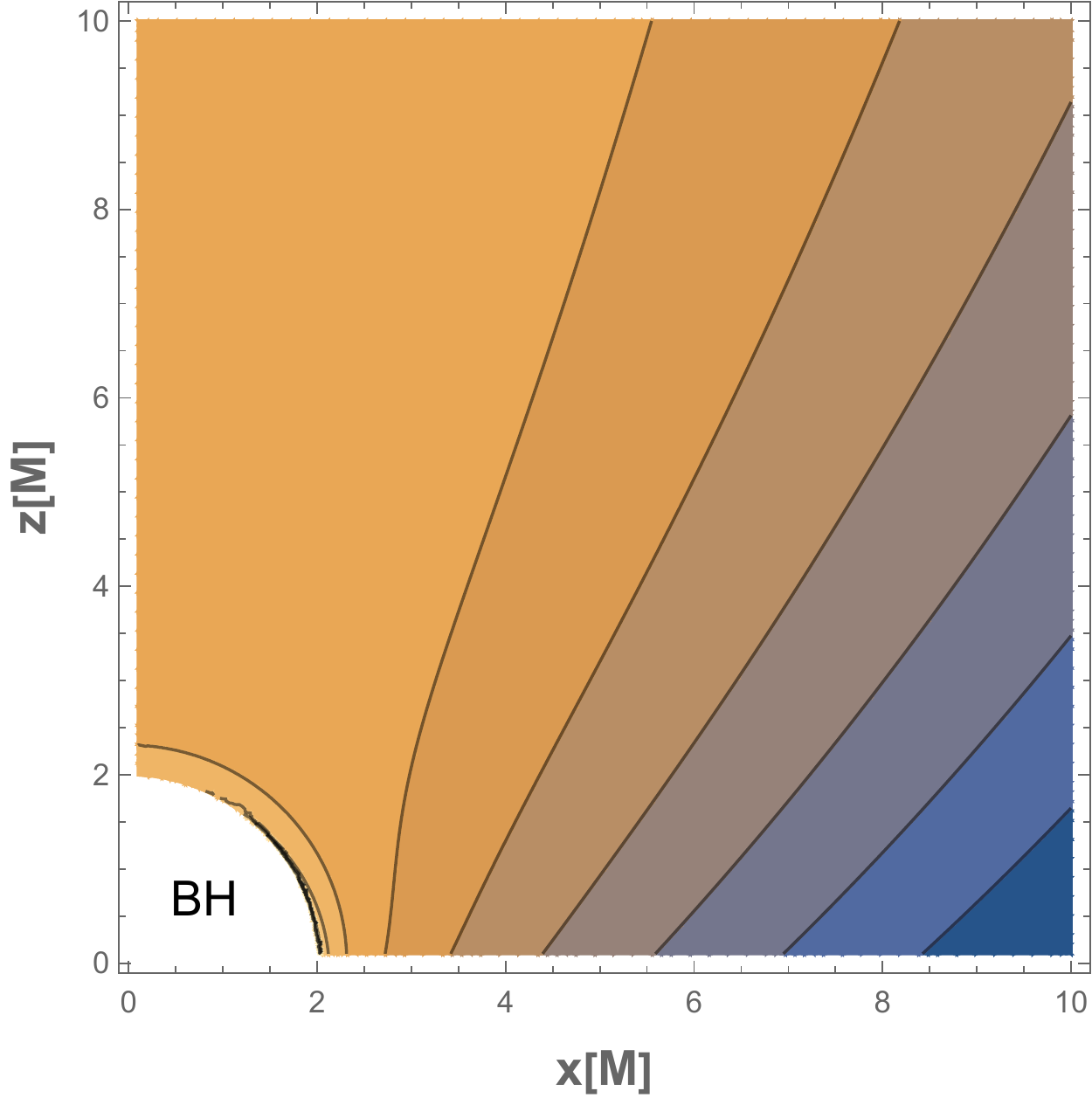}
\caption{(LEFT) Contour-plot representing paraboloidal solution (or stream function)  (\ref{eq3.14}), i.e.  in a power-law NLED model $L_{NLED}=-\gamma X^{2}$, where $X= (1/4)F_{\mu\nu}F^{\mu\nu}$, as function of Cartesian coordinate $x=r\sin\theta$ and $z=r\cos\theta$. The lines shown corresponds to poloidal magnetic field lines around a static ($a=0$, $M=1$) black hole. Colors are purely indicative, since the exact values depend on the integration constants, here assumed to be ideally 1. Any accretion disk (not shown) would 'lie' along the $x$ axis. (RIGHT) Same as before, but in the conventional linear theory (Maxwell). Notice the more pronounced 'verticality' of the non-linear case. Having set $M = 1$, all distances are actually dimensionless.   }
\label{fig:1}
\end{figure*}

In this section, we solve Eq. (\ref{eq2.15}) in the static limit ($a=0$). This will be sufficient to have an expression for the extracted power up to second order in $a$. \\
Following \cite{BZ-1977ds}, we assume that for  $a \ll 1 $

\begin{equation}
    A_{\phi}  = A_{\phi}^{(0)}+a^{2}A_{\phi}^{(2)}+\mathcal{O}(a^{4})  
\end{equation}
\begin{equation}
 B_{\phi}  = a B_{\phi}^{(1)} + \mathcal{O}(a^{3})   
\end{equation}
\begin{equation}
    w  = a w_{\phi}^{(1)} + \mathcal{O}(a^{3})
\end{equation}
while $B_{\phi}=w=0$ when $a=0$. The functions $B_{\phi}^{(1)}$, $w_{\phi}^{(1)}$ and $A_{\phi}^{(2)}$ are unknowns, while $A_{\phi}^{(0)}$ is just the solution for Schwarzschild case. The other components of $\mathbf{B}$ are
\begin{equation}
    B_{r} = - \dfrac{1}{\sqrt{-g}} \Big(\partial_{\theta}A_{\phi}^{(0)} + a^{2}\partial_{\theta}A_{\phi}^{(2)} \Big) \; \; , 
\end{equation}
\begin{equation}\label{Btheta}
  B_{\theta} =  \dfrac{1}{\sqrt{-g}} \Big(\partial_{r}A_{\phi}^{(0)} + a^{2}\partial_{r}A_{\phi}^{(2)} \Big)   
\end{equation}
It is clear that, in the static limit, the only unknown function is $A^{(0)}_{\phi}$. 
Indeed, at zero order in $a$ ($ \sim \mathcal{O}(1)$), Eq. (\ref{eq2.15}) becomes 
\begin{equation}\label{eq3.6}
L A^{(0)}_{\phi} = 0 
\end{equation}
where
\begin{equation}\label{eq3.7}
    L := \dfrac{1}{\sin\theta}\dfrac{\partial}{\partial r} L_{X}^{(0)} \Big( 1 - \dfrac{2}{r} \Big)\dfrac{\partial}{\partial{r}} + \dfrac{1}{r^{2}}\dfrac{\partial}{\partial_{\theta}}\dfrac{L_{X}^{(0)}}{\sin\theta}\dfrac{\partial}{\partial{\theta}}
\end{equation}
where $ L_{X}^{(0)}$ is $L_{X}$ in the Schwarzschild limit \footnote{A solution for $A^{(2)}_{\phi}$ requires a second order equation. See appendix.}. For a power-law model $L_{NLED} = -CX-\gamma X^{\delta}$ \cite{Lambiase2009}, for example, it would be \footnote{Generally, $L_{X}$ is an even function of $a$, i.e $L_{X} = L_{X}^{0} + a^{2}L_{X}^{2} + \mathcal{O}(a^{4}) $. It is essential that $L_{X}^{0} \not= 0$ in order to have a solution.}
\begin{equation}\label{eq3.8}
 L_{X}^{(0)} = -C -\gamma \delta \Big( \dfrac{-1}{2r^{2}\sin\theta} \Big)^{\delta -1}   \Big( \partial_{\theta} A^{(0)}_{\phi}\Big)^{2(\delta-1)}.
\end{equation}
Let us now consider separated solutions for $A_{\phi}$ and also assume a similar form for $L_{X}$, i.e.
\begin{equation}
\begin{array}{ccc}
  A^{(0)}_{\phi} = R(r) \cdot U(\theta)  \\
    L_{X}^{(0)} = f(r) \cdot g(\theta).
    \end{array}
\end{equation}
With this ansatz, Eq. (\ref{eq3.6}) reads as
\begin{equation}
         \dfrac{\partial}{\partial \theta}\left[ \dfrac{g(\theta)}{\sin \theta} \dfrac{\partial U(\theta)}{\partial \theta}\right] = - K \dfrac{g(\theta) U(\theta)}{\sin\theta} ,
\end{equation}
\begin{equation}
         \dfrac{\partial}{\partial r}\left[ f(r)\left( 1- \dfrac{2}{r} \right) \dfrac{\partial R(r)}{\partial r}\right] =  K \dfrac{R(r) f(r)}{r^{2}}
\end{equation}
where $K$ is a separation constant. We will choose $K=0$  so as to obtain the simplest solution (the lowest order\footnote{One in principle can generalize to higher orders as done, for example, in \cite{Ghosh_2000}. } one). \\
From here on, the specific NLED model must be chosen. Assuming a power-law model \footnote{The so-called Kruglov model \cite{KRUGLOV_B}, for example, is not separable, while the Born-Infeld one reduces to a power-law.} and hence Eq. (\ref{eq3.8}), we have to set $C=0$, unless one assumes $L_{X}^{(0)}$ is a function of just one variable, i.e. $f(r)=1$ or $g(\theta)=1$, but this would exclude most of NLED models. As a check, when $\delta=1$, we obtain the known solution as given in \cite{Ghosh_2000,BZ}. For $\delta=2$, the latitudinal part does not change, i.e. 
\begin{equation}\label{eq3.12}
    U(\theta) = \alpha \cos\theta + \beta
\end{equation}
while the radial part strongly changes 
\begin{equation}
    R(r) = c \left( 6r^5 +15r^4 +40r^3 +120r^2+480r+960\ln (r-2)-2192+d \right)^{1/3}
\end{equation}
where $\alpha$, $\beta$, $c$ and $d$ are constants. 
Following \cite{Ghosh_2000}, we note that it is impossible to have a monopole solution\footnote{The logarithmic  singularity, also present in the linear limit, simply means that solutions are valid in regions of space which exclude event horizon.} by default, as there are no combinations of constants to eliminate the radial dependence in $A_{\phi}^{(0)}$ without canceling all $R(r)$; it follows from Eq. 
 (\ref{Btheta}) that $B_{\theta}\not=0$   . A separable paraboloidal solution ($\alpha+\beta =0=d$) is instead possible:
\begin{equation}\label{eq3.14}
    A^{(0)}_{\phi} \sim \big(\cos\theta -1 \big)\big( 6r^5 +15r^4 +40r^3 +120r^2+480r+960\ln (r-2)-2192 \big)^{1/3}.
\end{equation}
For $\delta=3$ and higher values, the angular part will be equal to Eq. (\ref{eq3.12}), while the radial one will be consistent only if $r<2$, i.e. beyond the event horizon, so we discard them. Same epilogue if one chooses negative powers ($\delta<0$): no monopole   solution would exist  and paraboloidal one would  be valid only for $r<r_{+}$.  This could be an interesting point: monopole solutions are actually not physical, while paraboidal magnetic configurations can explain the collimation of the jets \cite{komiss2001,Nathanail_2014}. It must be emphasized that the geometry of the magnetic lines  depends on the distance and thickness of the accretion disk, the only structure capable of generating a magnetic field. Therefore,  exact solutions would require  boundary conditions (see \cite{Ghosh_2000}  and references therein) and therefore specific astrophysical scenarios. Moreover, also numerical simulations could come to our aid as done in \cite{Kinney2006,BZ2,Komissarov_2005,SimulationBZ}. An interesting point of difference of (\ref{eq3.14}) w.r.t. the analogous Maxwell solution is the forward displacement of the flow inversion point ($r\simeq 2.35$ vs $r\simeq 2.31$), i.e. the point in which $A^{(0)}_{\phi}$ change sign (and hence $R(r)=0$). However,  as shown in Fig.  (\ref{fig:1}), the main difference with linear theory is the asymptotic behaviour ($r\gg1$) of the solution,  being $A^{(0)}_{\phi} \sim r^{s}(1-\cos\theta)$  with $s>1$ in the non-linear case ($s=1$ in linear theory). This stronger 'verticality' could favor these kind of solutions in the formation of jets. \\

\section{Some estimates}
\label{sec4}
\begin{figure}[t]
\includegraphics[width=14cm]{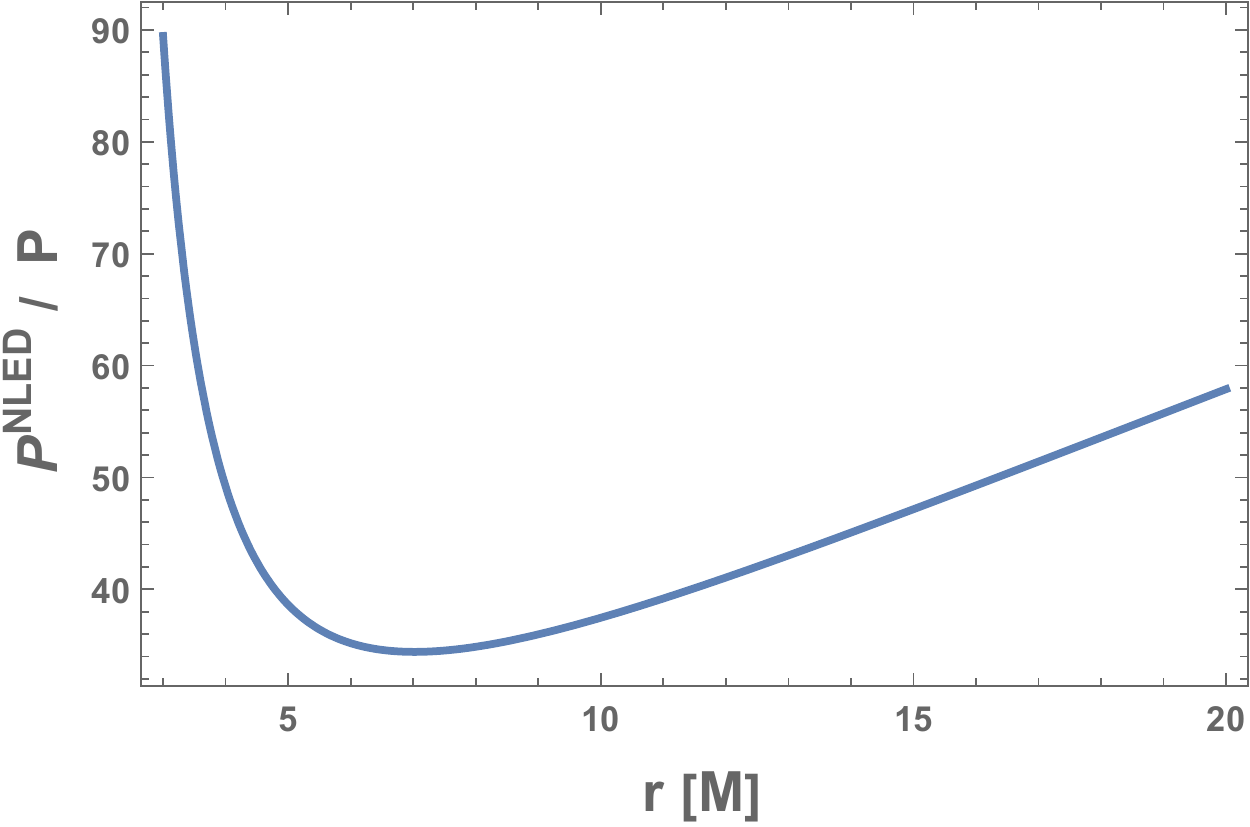}
\caption{\label{fig:2} Estimate of the rate between the extracted power (through BZ mechanism), as function of radial distance, in a fully  non-linear theory (power-law with $\delta=2$) and the equivalent quantity in linear (Maxwell) theory. A paraboloid  solution was taken into consideration in both cases and an assumption of predominant radial field $B_{r}\gg B_{\phi}$ has been made. Having set $M = 1$, all distances are actually dimensionless. }
\end{figure}

\begin{figure}[t]
\includegraphics[width=14cm]{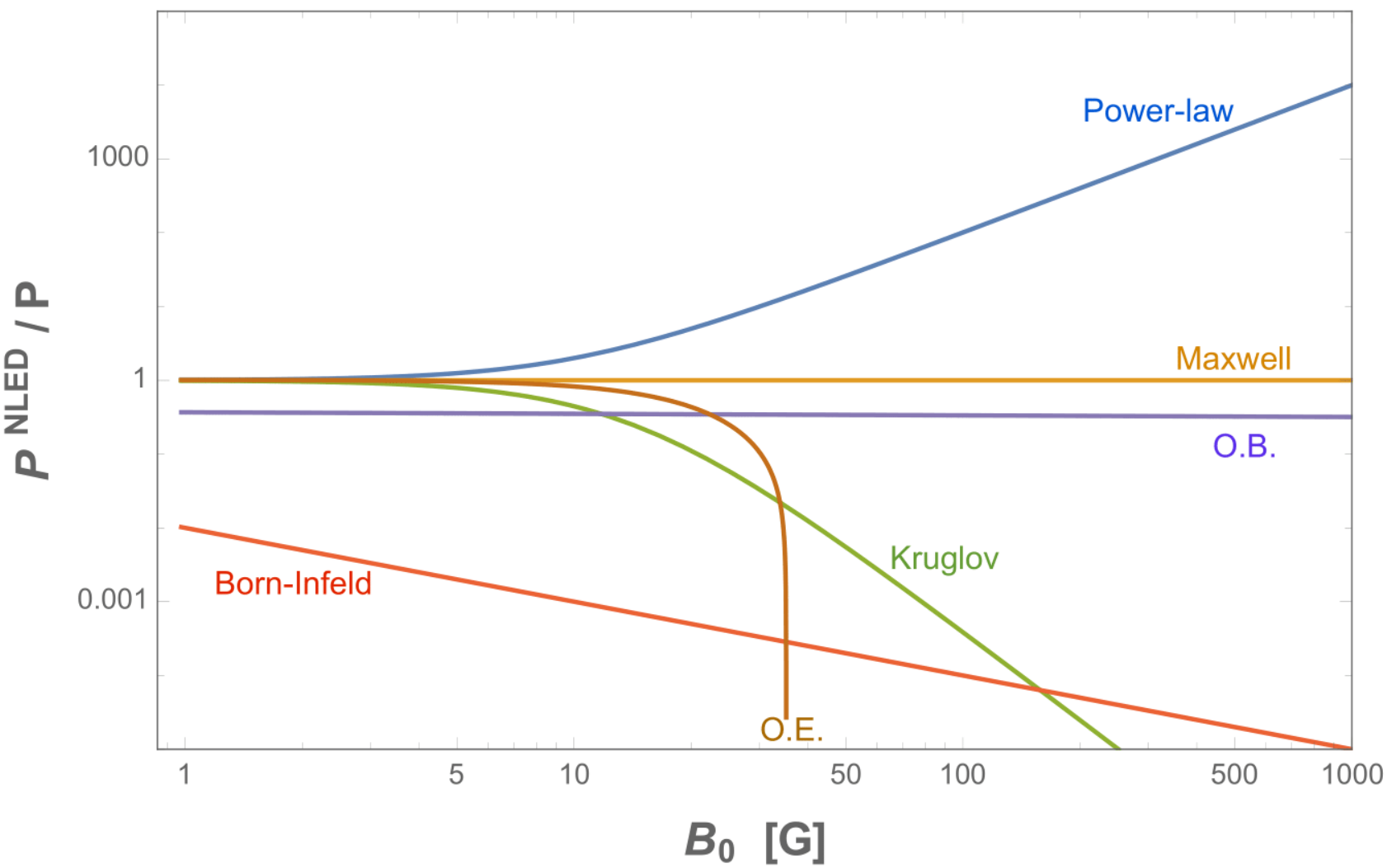}
\caption{\label{fig:3} Estimation of the ratio between the power extracted from a black hole (through BZ mechanism) in presence of non-linear electrodynamics ($P^{NLED}$) and the same quantity in Maxwell theory ($P$), as a function of the magnetic field strength (see Eq. (\ref{generic})). Several NLED models have been taken into account: power-law ($C=1$,$\gamma=0.01$, $\delta=2$), Kruglov  ($\beta=0.01$), Ovgun-Benaoum 
 ($\alpha=\beta=0.01$) and Ovgun Exponential ($\alpha=0.001$, $\beta=0.99$). The ratio is accurate up to second order in the spin parameter for a BH with $M=1$. Only one approximation has been used, namely $B_{\theta}\approx 0$. As it is clear, the behaviour strongly depends on the specific NLED model; however, a power law model with a positive exponent could in principle extract more energy  (while at low magnetic field it behaves like linear electromagnetism), although some regimes could be excluded in order not to exceed the Eddington limit.    }
\end{figure}

\setcounter{equation}{0}
In this section, starting from the result of the previous section, we find an estimate of the extracted power  comparing it with the linear theory (Maxwell) case. Here, we propose two different ways. \\
Given the presence of the singularity at $r=2$ we have to discard this point. In order to use Eq. (\ref{eq2.11}), which is evaluated on the horizon,  we assume the condition $B_{r} \gg B_{\phi}$, which is often used in simulations \footnote{A purely radial magnetic field (monopole), although not realistic, is still considered today being  the  simplest  configuration to implement \cite{komiss2001}, both numerically and analytically. }. From Eq. (\ref{eq2.11}), we find for the power extracted in the Maxwell case\footnote{The (separable) paraboloidal  Schwarzschild  solution in linear theory goes like $A^{(0)}_{\phi} \sim (\cos \theta -1)(r+2 \ln (r-2))$ as reported in \cite{Ghosh_2000}. } $P$, at the second order
\begin{equation}
    P\simeq \dfrac{4\pi}{3 r}\Big[ r + 2 \ln (r-2)\Big]^{2}\Omega_{H}^{2}
\end{equation}
where $r>2$. On the other hand,  in power-law model (\ref{eq3.14}), similar computations lead to 
\begin{equation}
    P^{NLED} \simeq \dfrac{4\pi}{3 r^{5}}\Big( 6r^5 +15r^4 +40r^3 +120r^2+480r+960\ln (r-2)-2192 \Big)^{4/3} \Omega_{H}^{2}
\end{equation}
where we used 
\begin{equation}
    L_{X}^{(r)} = -\dfrac{1}{r^{4}} \Big( 6r^5 +15r^4 +40r^3 +120r^2+480r+960\ln (r-2)-2192 \Big)^{2/3}
\end{equation}
instead of $L_{X}^{(r_{+})}$. Apart from the radial field approximation $B_{r} \gg B_{\phi}$, the rate is quite accurate\footnote{Expressions for $P$ and $P^{NLED}$ are at fault only for a constant depending on the field configuration (monopole, paraboloidal, etc.). We assume that they are of the same order in both cases, as it is plausible.}; it has been plotted as function of $r$ in Fig.  (\ref{fig:2}). From the latter, it is clear that in principle such a NLED model could really extract more energy than in the conventional case. However, it would not have a Maxwellian limit because we had to impose $C=0$ to achieve the analytical solution  (\ref{eq3.14}). \\
The above estimate necessarily requires the stream function, i.e. a solution of the (very involved) stream equation. Moreover, it required to force $C=0$ for the power-law model. We can overcome these issues  in the following way. As before,  let us assume a  radial field  in the form $B_{r}=B_{0}\sin\zeta$, where $B_{0}\sim \sqrt{\sigma_{0}}$ is the magnetic strength as given by plasma magnetization $\sigma_{0}$ ($\zeta$ is the angle between $\mathbf{B}$ and $\hat{\phi}$ at the equator). Unlike before, let us  evaluate  Eq. (\ref{eq2.11}) on the horizon $r=r_{+}$. Just by assuming $B_{\theta}$ negligible, it is straightforward to obtain an expression for $P^{NLED}$ without solving the stream equation and accurate up to second order in the spin parameter. This means that such an estimate would be suitable also for non-separable NLED model, like the Kruglov one $L_{NLED}= -X\cdot(1+\beta X)^{-1}$ \cite{KRUGLOV_B}.  Since in this framework $B_{\phi} = B_{0}\cos\zeta$, the rate w.r.t. Maxwell case simply is 
\begin{equation}
    \dfrac{P^{NLED}}{P} = \dfrac{1}{\left[\frac{\beta}{2}B_{0}^{2}+1\right]^{2}} \; .
\end{equation}
A similar computation was done for $L_{NLED}=-X-\gamma X^{2}$ and a comparison between these two different NLED models has been reported in Fig. (\ref{fig:3}).  It is evident the advantage of power-law model with positive exponent \footnote{A power-law electromagnetic model seems capable of  extracting much more energy than models employing Kerr metric deformations. For example, in the case of a Johannsen metric, the extracted energy  is no more than $\sim$10  times larger \citep{cosimo_2016}. In our framework, the ratio $P^{NLED}/P$ can exceed $10^4$ (see Fig. (\ref{fig:3})).}. In general, we have  
\begin{equation}\label{generic}
    \dfrac{P^{NLED}}{P} = - L_{X} (X_{0})  
\end{equation}
where we defined $X_{0}:= B_{0}^{2}/2$. The strong dependence on the specific NLED model is clearly explicit: the ratio of energy  power in presence of a non-linear electrodynamic model $P^{NLED}$ to linear (Maxwell) case $P$ is simply given by the opposite of the derivative of the lagrangian w.r.t. $X$ evaluated at $X_{0}:= B_{0}^{2}/2 $. 
Notice that the two methods hold true in different regimes. While in  the first way an assumption of type  $B_{r} \gg B_{\phi}$  must be made, in the second estimate one needs $B_{\theta}\ll 1
$.  One can use one or the other depending on the specific context. From an astrophysical point of view, both  observations and simulations suggest that the magnetic field around massive black holes has a poloidal configuration, i.e. the field lines lie in planes containing the axis of rotation  \cite{B_Centro_ViaLattea_2013,Simulaz_poloidal_Liska_2020,EHT_part8_Mag_Fields}.  Therefore, the assumption $B_{r} \gg B_{\phi}$  is  physically achievable. About the second assumption, since a purely radial solution (monopole) is not likely, we  generally expect  $B_{\theta}\not=0$. However, except in extreme paraboidal cases, the polar component of the magnetic field is negligible  for distances well beyond the event horizon (see Fig.(1) in \cite{BZ}).

\section{Primordial Magnetic Field from NLED}
\label{sec5}
\setcounter{equation}{0}
According to General Relativity (GR) primordial fields decayed adiabatically due to conservation of the flux.  i.e. $ a^2 B \sim const $, and hence $B \sim 1/a^2$. Consequently, the magnetic energy density  $\rho_{B} = |\mathbf{B}|^{2}/(8 \pi )$ should have scaled as $  1/a^4$, where $a$ is the scale factor of the (flat) Friedman-Robertson-Walker (FRW) metric. Since the scale factor diverges  during inflation, this type of decay implies  very faint magnetic fields at the end of the inflation period. This scaling is valid for every cosmic energy density present in the Universe, and then also  for the cosmic microwave background (CMB), whose  energy density (assumed almost constant during inflationary era) is given by $\rho_{\gamma} = \pi^{2} T^{4}/ 25$, or, in function of $a$, by $\rho_{\gamma} \sim 1/a^4$ (the extra factor $1/a$ w.r.t to matter, which decays as $\sim 1/a^{3}$, comes from energy redshift).  Therefore, the ratio $r \doteq \rho_{B}/ \rho_{\gamma}$   remained constant until today, with a current value of $r \approx 1$ and this constrain  is a good tool to study  primordial fields. The origin of large-scale magnetic fields has been studied not only in the context of  GR but also in extended or alternative theories of gravity \cite{Kothari:2018aem,Gasperini:1995dh,Atmjeet_2014,Kothari:2018aem}. The main idea behind such works is to assume the non-conservation of the flux, breaking the conformal invariance of the electromagnetic sector and hence making possible a different trend from the adiabatic one for $\mathbf{B}$.  \\
In this section, following \cite{kunz2008,Lambiase2009,CarleoPMFs} and in the context of GR, we  try to find constrains on some NLED models, exploiting existence and survival of PMFs. We start from the action

\begin{equation}\label{actionPrimord}
 S = \int d^{4}x \sqrt{-g} \dfrac{R}{2 \kappa^{2}} + {L}_{NLED}
\end{equation}
where $\kappa^2=8 \pi$  and ${L}_{NLED}(X,G)$  encodes a general electromagnetic theory (see Sec. \ref{sec2}). It is clear that Maxwell theory is obtained when ${L}_{NLED}=-X$. Varying the action w.r.t. the electromagnetic field $A_{\mu}$,   the  field equations are

\begin{equation}\label{eq5.2}
    \partial_{\mu} \Big[ \sqrt{-g} \Big(L_{X}F^{\mu \nu} + L_{G}F^{*\mu \nu} \Big) \Big] = 0
\end{equation}
\begin{equation}
    \partial_{\mu}  F^{*\mu \nu}  = 0
\end{equation}
which are the source free o zero density version of Eqs.~(\ref{eq2.3})-(\ref{eq2.4}). We consider here a conformally  flat FRW  metric
\begin{equation}\label{eq14}
    ds^{2} = a^{2}(\eta)\Big( d\eta^{2} - d\mathbf{x}^2  \Big) = dt^{2} - a(t) d\mathbf{x}^2
\end{equation}
where $\eta = \int_{0}^{t} a^{-1}(t) dt $ is the conformal time and $a(t)$ is a dimensionless scale factor. In this metric and with our signature, $F_{\mu \nu}$ can be written as 
\begin{equation}\label{eq33}
F_{\mu \nu}=a^{2}(\eta)\left(\begin{array}{cccc}
0 & E_{x} & E_{y} & E_{z} \\
-E_{x} & 0 &  -B_{z} & B_{y} \\
-E_{y} & B_{z} & 0 &  -B_{x} \\
-E_{z} & -B_{y} & B_{x} & 0 
\end{array}\right)
\end{equation}
in order to separate highlight the electric and magnetic fields as measured by a comoving (inertial) observer. With this ansatz and assuming the non-existence of magnetic charge, Eq.~(\ref{eq5.2}) becomes 
\begin{equation}
    A_{j}'' + \dfrac{L_{X}'}{L_{X}} A_{j}'- \dfrac{\partial_{i}L_{X}}{L_{X}}(\partial_{i}A_{j}-\partial_{j}A_{i})-\Delta A_{j}=0
\end{equation}
where $j=1,2,3$, $\Delta=:=\delta^{ki}\partial_{k}\partial_{i}$ and a prime denotes derivative w.r.t. conformal time. The above equation can be also written in terms of the electric $E_{j}$ and magnetic $B_{j}$ fields as
\begin{equation}\label{eq5.7}
    \partial_{0}\Big( a^{2}L_{X}\mathbf{E} \Big) - a^{2} \nabla \Big( L_{X} \mathbf{B}\Big) = 0.
\end{equation}
The zero component of Eq.~(\ref{eq5.2}) reads as 
\begin{equation}
    \nabla \Big( L_{X} \mathbf{E} \Big) = 0
\end{equation}
while the Bianchi identity gives 
\begin{equation}\label{eq5.9}
 \partial_{0}\Big( a^{2}\mathbf{B} \Big) + a^{2} \nabla \times \mathbf{E} = 0
\end{equation}
as well as the usual constrain $\nabla \cdot \mathbf{B}=0$. Combining Eq.~(\ref{eq5.7}) and Eq.~(\ref{eq5.9}) one obtains
\begin{equation}\label{eq5.10}
    L_{X}F'' + \Big(\partial_{0}L_{X} \Big)F' = 0
\end{equation}
where we defined $F:=a^{2} B $ with $B:=|\mathbf{B}|$ and assumed the long-wavelength approximation \cite{Lambiase2009}  (i.e. disregarding spatial derivatives).  Notice that in this approximation $F=F(\eta)$. By choosing a power-law model $L_{NLED}=-CX-\gamma X^{\delta}$, we recover the same results of \cite{Lambiase2009}. Here, we focus on other non-linear Lagrangian. As a first model, we consider \cite{Ovgun:2022}
\begin{equation}\label{eq5.11}
    L_{NLED}(X) = - \dfrac{X}{(\beta X^{\alpha} + 1)^{1/\alpha}}
\end{equation}
where $\alpha$ and $\beta$ are two real parameters with $\beta$ controlling the non-linearity contributions.  A suitable condition  for obtaining an analytical solution is the strong regime $(2a^{4}/F^{2})^{\alpha}\ll \beta$, i.e. $B\gg B_{0}$ with $B_{0}:=\dfrac{\sqrt{2}}{\beta^{1/(2\alpha)}}$, it follows from Eq.~(\ref{eq5.10})
\begin{equation}
    \dfrac{d^{2}F}{da^{2}}+\Bigg[ \dfrac{s-1}{a s} - 4 \dfrac{(\alpha+1)}{a}\Bigg]\dfrac{dF}{da}+8\dfrac{(\alpha+1)}{a^{2}}F^{2} = 0
\end{equation}
where $s={-1,2,1}$ depending on which primordial phase we are considering, i.e. inflation, reheating, radiation \footnote{We are assuming a scale factor of the type $a(\eta) = c_{s} \eta^{s}$, where $c_{s}$ is the Hubble constant for the specific primordial era. Notice that the values of $s$ are those of General Relativity; changing the gravitational sector in the action (\ref{actionPrimord}) leads to different $s$ values. See for example \cite{CarleoPMFs}. This is the only point in which, in a minimal approach, gravity comes into play.}. Assuming a power-law solution for $F$, $F\sim a^{p}$, the inflationary exponent is
\begin{equation}
    p_{\pm} = \dfrac{1}{2}\Big[ 3+4 \alpha \pm \sqrt{\Sigma}\Big]
\end{equation}
where we defined $\Sigma := 16 \alpha^2-8\alpha-23$. This solution clearly constrains the parameter $\alpha$ to be either $\alpha>\frac{1}{4}(1+2\sqrt{6})$ or $\alpha<\frac{1}{4}(1-2\sqrt{6})$. When $\alpha=\frac{1}{4}(1\pm2\sqrt{6})$ a pseudo-power-law solution is  possible, namely 
\begin{equation}
    F(a)=c_{1}a^{p}-c_{2}a^{p}\ln{a}(3+4\alpha)
\end{equation}
with $p = 3/2+2\alpha$. \\
In the reheating epoch, the power-law solution $F\sim a^{q}$ reads as 
\begin{equation}
    q_{\pm} = \dfrac{1}{2}\Big[ \frac{1}{2}+4(1+\alpha) \pm \sqrt{\Sigma}\Big]
\end{equation}
with $\Sigma := 16 \alpha^2+4\alpha-47/4$
and is valid when $\alpha<\frac{1}{8}(-1-4\sqrt{3})$ or $\alpha>\frac{1}{8}(-1+4\sqrt{6})$, while a pseudo-power-law solution for the remaining cases has $q = 1/4 + 2(1+\alpha)$. With this solutions, it is possible to express the strong regime assumption in terms of conformal time, i.e. $\eta \gg \eta^{*}$ with $\eta^{*}$
\begin{equation}
    \eta^{*}:= \Big[ \dfrac{\sqrt{2}}{c_{s}^{\lambda-2}}\dfrac{1}{\beta^{1/2\alpha} F}\Big]^{\dfrac{1}{s (\lambda-2)}}
\end{equation}
where $\lambda=\{p,p_{\pm},q,q_{\pm}\}$ and $c_s$ is the Hubble constant.
Following \cite{Lambiase2009}, we can try now to constrain the only present parameter $\alpha$ exploiting the astrophysical observation $r\simeq 1$. Specifically, we found that this is possible only with the combination $(p_{-},q_{+})$, as shown, for different primordial conditions, in Table~\ref{tab1}.  Using these estimations, one can obtain a complete solution for the remaining radiation epoch. For example, taking $\alpha\simeq 15$, the power-law solution for the radiation era goes like $F \sim a^{u}$ with $u \simeq \{2 , 63\}$. 

\begin{table}[t]
\centering
\caption{Estimation of the NLED parameter $\alpha$ in the model (\ref{eq5.11}) from PMFs using the observational constrain $r\simeq1$. Different value of the grand unified theories (GUT) scale, $M_{GUT}$, reheating temperature, $T_{RH}$, and temperature at which plasma effects become dominant, $T^{*}$.}
\label{tab1}
\begin{tabular}{lccc}
\textrm{$M_{GUT}$}&
\textrm{$T_{RH}$}&
\textrm{$T_{*}$}&
\textrm{$\alpha$} \\ \hline\hline 
$10^{16}$ & $10^{9}$ & $10^{12}$ & 28.4\\ \hline
$10^{17}$ & $10^{15}$ & $10^{15}$ & 15.1\\ \hline
$10^{17}$ & $10^{17}$ & $10^{16}$ & 15.4\\ \hline
\end{tabular}
\end{table}

 Similarly to what happens for the Born-Infeld model \footnote{A Born-Infeld model \cite{Born:1934gh,Born1933ModifiedFE}, does not allow either pseudo-power-law or power-law solutions. Expanding the Lagrangian in powers of $X$, however, leads to a power-law model with $\delta=2$ studied in \cite{Lambiase2009}, and hence to the same solutions, which, however, seems to exclude power-law Lagrangian $L_{NLED}=-CX-\gamma X^{\delta}$ with $\delta>2$. }, also the exponential model \cite{Ovgun:2017iwg}
 \begin{equation}\label{eq5.17}
     L_{NLED} = - \dfrac{X e^{-\alpha X}}{\alpha X + \beta}
 \end{equation}
does not allow power-law solutions for $F$. Indeed, in this case, Eq. (\ref{eq5.10}) becomes ($\alpha X \gg \beta$)
\begin{equation}
     \dfrac{d^{2}F}{da^{2}}+\Bigg[ \dfrac{s-1}{a s} - 2 \dfrac{\alpha F^{2}}{a^{5}}\Bigg]\dfrac{dF}{da}+4\dfrac{\alpha F^{3}}{a^{6}} = 0   .
\end{equation}
Alternatively, after expanding, the above model reduces to a power-law one with $\delta = 2$   ($C=1/\beta$ and $\gamma=-\alpha/\beta$) as long as $\beta \leq 1$ and $B\ll \sqrt{2/\alpha}$. As for the Born-Infeld model and based on the results of \cite{Lambiase2009}, such a model  is not compatible with the observation $r\sim 1$. \\
Concluding, while for the model (\ref{eq5.11}) an analytical solution is available and an estimate of the parameter $\alpha$ is possible, the exponential one (\ref{eq5.17}), at least  in the regimes studied to have an analytical solution, is excluded from the observations, in a context of primordial magnetic fields. 

\section{Conclusions}
\label{sec6}
\setcounter{equation}{0}
In the literature, there are no known exact solutions for the magnetosphere of a spinning BH. In this paper, we analytically solved the maghehydrodynamic problem in the context of non-linear electrodynamics (NLED) models, trying to keep an approach as much model-independent as possible. We explored the Blandford-Znajek mechanism  in the light of
this framework in order to establish if non-linear effects can modify the energy extraction
rate and the magnetic field configuration around a (non-charged) black hole surrounded by
its magnetosphere. This attempt goes in the opposite direction to what was done in \cite{Konoplya_2021}, where the electromagnetic sector was unaffected, but a deviation from the familiar Kerr metric was switched on. Unlike \cite{Konoplya_2021}, in our work the black hole spin frequency $\Omega_{H}$  does not undergo changes, but, on the other hand, the formula for  the extracted power deviates (not only numerically but also formally) from conventional one, i.e. with Kerr geometry and Maxwell theory. In particular, we found that the energy flux will depend not only on the radial magnetic field $B_{r}$, but in general also on
the other two components, namely $B_{\theta}$ and $B_{\phi}$ and that, after perturbative expansion in powers of $a$ (as done in \cite{BZ-1977ds}), no monopole solutions exist for power-law models $L_{NLED}=-CX - \gamma X^{\delta}$ and that only the case $\delta=2$ is significant. Paraboloidal solutions, instead, seems to be possible, and they strongly change if compared to linear theory case, especially in the radial part, as shown in Fig. (\ref{fig:1}). An interesting point of difference is the forward displacement of the flow inversion point ($r\simeq 2.35$ vs $r\simeq 2.31$), i.e. the point in which $A^{(0)}_{\phi}$ change sign (and hence $R(r)=0$). However,  as shown in Fig.  (\ref{fig:1}), the main difference with linear theory is the asymptotic behaviour ($r\gg1$) of the solution,  being $A^{(0)}_{\phi} \sim r^{s}(1-\cos\theta)$  with $s>1$ in the non-linear case ($s=1$ in linear theory). The more pronounced 'verticality'   could favor these kinds
of solutions in the formation of jets, even if we do not explored the astrophysical consequences.  \\
We also tried to derive several estimates for the extracted power. We used two different approaches valid in different magnetic regimes ($B_{r} \gg B_{\phi}$  or $B_{\theta}\ll 1$): one requiring the solution of the stream equation and one assuming the magnetic field strength as independent variable (see Figs. (\ref{fig:2}) and (\ref{fig:3}) respectively). In both cases, it appears evident that NLED power-law model with positive power can in principle extract much more energy w.r.t. classical Maxwell theory, and that, on the other hand, models like the Kruglov's, for example, perform no better than the standard EM theory already does, making them rather unlikely. Even if separated solutions are not the only option,  separable paraboloidal solutions seem to be in good agreement with numerical simulations even for rotating black holes, as rotation does not dramatically effects the magnestosphere configuration in this case \cite{Ghosh_2000}. We notice that solutions have been found in the static limit ($a=0$) and that the expression for the extracted power are up to second order in $a$. Moreover, it would be interesting to investigate on higher order solutions ($K\not=0$) as done in Sec. 3 of \cite{Ghosh_2000}, as well as on non-separable ones.  Notice also that  exact solutions would require boundary conditions (see \cite{Ghosh_2000}  and references therein),  therefore only numerical simulations (as done in \cite{Kinney2006,BZ2,Komissarov_2005,SimulationBZ}) could be astrophysically  meaningful. All these observations could be starting points for other works.  \\
Finally,  following \cite{kunz2008,Lambiase2009,CarleoPMFs}, we  tried to find constrains on some NLED models, exploiting existence and survival of PMFs. We focused on two recent NLED models, namely (\ref{eq5.11}) and (\ref{eq5.17}). After solving Maxwell equations, we found the constrain $\alpha\simeq 15$ for the first model, while the second model seems to be incompatible with the observation $r\sim 1$ (see Sec. \ref{sec1}), at least in the regime we studied. In conclusion, our (analytical) results emphasize that the existence and the behavior of non-linear electromagnetic phenomena strongly depend on the model and the physical context, and that  power-law models $L_{NLED}=-CX-\gamma X^{\delta}$ with $\delta \leq 2$  should be further studied. 

\appendix

\section{Second order terms}

Up to second order, the expression for $X$ is 

\begin{equation}
\begin{array}{rl}
    X= \dfrac{1}{2\Sigma^{2}\sin^{2}\theta} \Bigg[ \Big( (\partial_{\theta}A^{(0)}_{\phi})^{2} + a^{4}(\partial_{\theta}A^{(2)}_{\phi})^{2} +2a^{2}\partial_{\theta}A^{(0)}_{\phi}\partial_{\theta}A^{(2)}_{\phi}  \Big) \Big( 1 - a^{2}w^{(1)} \Big) \\ +  a^{4}(\partial_{r}A^{(2)}_{\phi})^{2} \Big( 1 - a^{2}w^{(1)} \Big) + a^{2}{{B_{\phi}^{(1)}}^{2}}\Bigg]
    \end{array}
\end{equation}
while Eq. (\ref{eq2.15}) becomes 
\begin{equation}
    L A^{(2)}_{\phi} = - r^{2}\sin\theta S(r,\theta)
\end{equation}
where L is given in Eq. (\ref{eq3.7}) and 

\begin{equation}
\begin{array}{rl}
    S(r,\theta) := B^{(1)}_{\phi}\dfrac{d\Tilde{B_{T}}}{dA_{\phi}^{(0)}} - \dfrac{w^{(1)}}{r^{2}\sin\theta} \Bigg[ \partial_{\theta} \Big( \dfrac{L_{X}^{0}\sin\theta}{r^{2}}\partial_{r}A^{0}_{\phi}(4r^{2}w^{(1)}-2r+w^{(1)}(r^{4}-4r^{2})) \Big) \\
    + \partial_{\theta}\Big( L_{X}^{(0)}\sin^{2}\theta 2 r B_{\phi}^{(1)} + \dfrac{L_{X}^{0}\sin\theta}{r^{2}}w^{(1)}(r^{2}+2r) \Big) \Bigg] \\ + \dfrac{1}{r^{2}\sin^{2}\theta} \partial_{r}\Big( \dfrac{L_{X}^{0}\sin\theta}{r^{2}}\partial_{r}A^{(0)}_{\phi}( 2r w^{(1)}-1 ) \Big) + \dfrac{1}{r^{2}\sin\theta}\partial_{\theta}\Big(L_{X}^{(0)}\sin^{2}\theta B_{\phi}^{(1)} \Big)
    \end{array}
\end{equation}

with
\begin{equation}
    \Tilde{B_{T}} := L_x^{(0)} \sin ^2 \theta\left(r^2-2 r\right)  B_{\phi}^{(1)}-\dfrac{L_x^{(0)} \sin \theta}{r}  2   w^{(1)}  \partial_\theta  A_{\phi}^{(0)}+\dfrac{L_x^{(0)} \sin \theta}{r^2 }  \partial_\theta A_{\phi}^{(0)}.
\end{equation}


\acknowledgments
We are very grateful to the anonymous reviewers for valuable suggestions that significantly improved the quality of our paper. 
G.L. and A.C. acknowledge the support by the  Istituto Nazionale di Fisica Nucleare (INFN) {\it Iniziativa Specifica} QGSKY. G. L. and A. {\"O}. would like to acknowledge networking support by the COST Action CA18108 - Quantum gravity phenomenology in the multi-messenger approach (QG-MM).




\bibliography{biblio2.bib}

\begin{thebibliography}{131}%
\makeatletter
\providecommand \@ifxundefined [1]{%
 \@ifx{#1\undefined}
}%
\providecommand \@ifnum [1]{%
 \ifnum #1\expandafter \@firstoftwo
 \else \expandafter \@secondoftwo
 \fi
}%
\providecommand \@ifx [1]{%
 \ifx #1\expandafter \@firstoftwo
 \else \expandafter \@secondoftwo
 \fi
}%
\providecommand \natexlab [1]{#1}%
\providecommand \enquote  [1]{``#1''}%
\providecommand \bibnamefont  [1]{#1}%
\providecommand \bibfnamefont [1]{#1}%
\providecommand \citenamefont [1]{#1}%
\providecommand \href@noop [0]{\@secondoftwo}%
\providecommand \href [0]{\begingroup \@sanitize@url \@href}%
\providecommand \@href[1]{\@@startlink{#1}\@@href}%
\providecommand \@@href[1]{\endgroup#1\@@endlink}%
\providecommand \@sanitize@url [0]{\catcode `\\12\catcode `\$12\catcode
  `\&12\catcode `\#12\catcode `\^12\catcode `\_12\catcode `\%12\relax}%
\providecommand \@@startlink[1]{}%
\providecommand \@@endlink[0]{}%
\providecommand \url  [0]{\begingroup\@sanitize@url \@url }%
\providecommand \@url [1]{\endgroup\@href {#1}{\urlprefix }}%
\providecommand \urlprefix  [0]{URL }%
\providecommand \Eprint [0]{\href }%
\providecommand \doibase [0]{http://dx.doi.org/}%
\providecommand \selectlanguage [0]{\@gobble}%
\providecommand \bibinfo  [0]{\@secondoftwo}%
\providecommand \bibfield  [0]{\@secondoftwo}%
\providecommand \translation [1]{[#1]}%
\providecommand \BibitemOpen [0]{}%
\providecommand \bibitemStop [0]{}%
\providecommand \bibitemNoStop [0]{.\EOS\space}%
\providecommand \EOS [0]{\spacefactor3000\relax}%
\providecommand \BibitemShut  [1]{\csname bibitem#1\endcsname}%
\let\auto@bib@innerbib\@empty
\bibitem [{\citenamefont {Born}(1933)}]{Born1933ModifiedFE}%
  \BibitemOpen
  \bibfield  {author} {\bibinfo {author} {\bibfnamefont {Max}\ \bibnamefont
  {Born}},\ }\bibfield  {title} {\enquote {\bibinfo {title} {Modified field
  equations with a finite radius of the electron},}\ }\href@noop {} {\bibfield
  {journal} {\bibinfo  {journal} {Nature}\ }\textbf {\bibinfo {volume} {132}},\
  \bibinfo {pages} {282--282} (\bibinfo {year} {1933})}\BibitemShut {NoStop}%
\bibitem [{\citenamefont {Born}\ and\ \citenamefont
  {Infeld}(1934)}]{Born:1934gh}%
  \BibitemOpen
  \bibfield  {author} {\bibinfo {author} {\bibfnamefont {M.}~\bibnamefont
  {Born}}\ and\ \bibinfo {author} {\bibfnamefont {L.}~\bibnamefont {Infeld}},\
  }\bibfield  {title} {\enquote {\bibinfo {title} {{Foundations of the new
  field theory}},}\ }\href {\doibase 10.1098/rspa.1934.0059} {\bibfield
  {journal} {\bibinfo  {journal} {Proc. Roy. Soc. Lond. A}\ }\textbf {\bibinfo
  {volume} {144}},\ \bibinfo {pages} {425--451} (\bibinfo {year}
  {1934})}\BibitemShut {NoStop}%
\bibitem [{\citenamefont {Gibbons}(2001)}]{Gibbons:2001gy}%
  \BibitemOpen
  \bibfield  {author} {\bibinfo {author} {\bibfnamefont {G~W}\ \bibnamefont
  {Gibbons}},\ }\bibfield  {title} {\enquote {\bibinfo {title} {{Aspects of
  Born-Infeld theory and string / M theory}},}\ }\href {\doibase
  10.1063/1.1419338} {\bibfield  {journal} {\bibinfo  {journal} {AIP Conf.
  Proc.}\ }\textbf {\bibinfo {volume} {589}},\ \bibinfo {pages} {324--350}
  (\bibinfo {year} {2001})},\ \Eprint {http://arxiv.org/abs/hep-th/0106059}
  {arXiv:hep-th/0106059} \BibitemShut {NoStop}%
\bibitem [{\citenamefont {Fradkin}\ and\ \citenamefont
  {Tseytlin}(1985)}]{Fradkin:1985qd}%
  \BibitemOpen
  \bibfield  {author} {\bibinfo {author} {\bibfnamefont {E.~S.}\ \bibnamefont
  {Fradkin}}\ and\ \bibinfo {author} {\bibfnamefont {Arkady~A.}\ \bibnamefont
  {Tseytlin}},\ }\bibfield  {title} {\enquote {\bibinfo {title} {{Nonlinear
  Electrodynamics from Quantized Strings}},}\ }\href {\doibase
  10.1016/0370-2693(85)90205-9} {\bibfield  {journal} {\bibinfo  {journal}
  {Phys. Lett. B}\ }\textbf {\bibinfo {volume} {163}},\ \bibinfo {pages}
  {123--130} (\bibinfo {year} {1985})}\BibitemShut {NoStop}%
\bibitem [{\citenamefont {Davila}\ \emph {et~al.}(2014)\citenamefont {Davila},
  \citenamefont {Schubert},\ and\ \citenamefont {Trejo}}]{Davila:2013wba}%
  \BibitemOpen
  \bibfield  {author} {\bibinfo {author} {\bibfnamefont {Jose~Manuel}\
  \bibnamefont {Davila}}, \bibinfo {author} {\bibfnamefont {Christian}\
  \bibnamefont {Schubert}}, \ and\ \bibinfo {author} {\bibfnamefont
  {Maria~Anabel}\ \bibnamefont {Trejo}},\ }\bibfield  {title} {\enquote
  {\bibinfo {title} {{Photonic processes in Born-Infeld theory}},}\ }\href
  {\doibase 10.1142/S0217751X14501747} {\bibfield  {journal} {\bibinfo
  {journal} {Int. J. Mod. Phys. A}\ }\textbf {\bibinfo {volume} {29}},\
  \bibinfo {pages} {1450174} (\bibinfo {year} {2014})},\ \Eprint
  {http://arxiv.org/abs/1310.8410} {arXiv:1310.8410 [hep-ph]} \BibitemShut
  {NoStop}%
\bibitem [{\citenamefont {Ellis}\ \emph {et~al.}(2017)\citenamefont {Ellis},
  \citenamefont {Mavromatos},\ and\ \citenamefont {You}}]{Ellis:2017edi}%
  \BibitemOpen
  \bibfield  {author} {\bibinfo {author} {\bibfnamefont {John}\ \bibnamefont
  {Ellis}}, \bibinfo {author} {\bibfnamefont {Nick~E.}\ \bibnamefont
  {Mavromatos}}, \ and\ \bibinfo {author} {\bibfnamefont {Tevong}\ \bibnamefont
  {You}},\ }\bibfield  {title} {\enquote {\bibinfo {title} {{Light-by-Light
  Scattering Constraint on Born-Infeld Theory}},}\ }\href {\doibase
  10.1103/PhysRevLett.118.261802} {\bibfield  {journal} {\bibinfo  {journal}
  {Phys. Rev. Lett.}\ }\textbf {\bibinfo {volume} {118}},\ \bibinfo {pages}
  {261802} (\bibinfo {year} {2017})},\ \Eprint
  {http://arxiv.org/abs/1703.08450} {arXiv:1703.08450 [hep-ph]} \BibitemShut
  {NoStop}%
\bibitem [{\citenamefont {Niau~Akmansoy}\ and\ \citenamefont
  {Medeiros}(2018)}]{NiauAkmansoy:2017kbw}%
  \BibitemOpen
  \bibfield  {author} {\bibinfo {author} {\bibfnamefont {P.}~\bibnamefont
  {Niau~Akmansoy}}\ and\ \bibinfo {author} {\bibfnamefont {L.~G.}\ \bibnamefont
  {Medeiros}},\ }\bibfield  {title} {\enquote {\bibinfo {title} {{Constraining
  Born\textendash{}Infeld-like nonlinear electrodynamics using
  hydrogen\textquoteright{}s ionization energy}},}\ }\href {\doibase
  10.1140/epjc/s10052-018-5643-1} {\bibfield  {journal} {\bibinfo  {journal}
  {Eur. Phys. J. C}\ }\textbf {\bibinfo {volume} {78}},\ \bibinfo {pages} {143}
  (\bibinfo {year} {2018})},\ \Eprint {http://arxiv.org/abs/1712.05486}
  {arXiv:1712.05486 [hep-ph]} \BibitemShut {NoStop}%
\bibitem [{\citenamefont {Niau~Akmansoy}\ and\ \citenamefont
  {Medeiros}(2019)}]{NiauAkmansoy:2018ilv}%
  \BibitemOpen
  \bibfield  {author} {\bibinfo {author} {\bibfnamefont {P.}~\bibnamefont
  {Niau~Akmansoy}}\ and\ \bibinfo {author} {\bibfnamefont {L.~G.}\ \bibnamefont
  {Medeiros}},\ }\bibfield  {title} {\enquote {\bibinfo {title} {{Constraining
  nonlinear corrections to Maxwell electrodynamics using $\gamma\gamma$
  scattering}},}\ }\href {\doibase 10.1103/PhysRevD.99.115005} {\bibfield
  {journal} {\bibinfo  {journal} {Phys. Rev. D}\ }\textbf {\bibinfo {volume}
  {99}},\ \bibinfo {pages} {115005} (\bibinfo {year} {2019})},\ \Eprint
  {http://arxiv.org/abs/1809.01296} {arXiv:1809.01296 [hep-ph]} \BibitemShut
  {NoStop}%
\bibitem [{\citenamefont {Neves}\ \emph {et~al.}(2021)\citenamefont {Neves},
  \citenamefont {de~Oliveira}, \citenamefont {Ospedal},\ and\ \citenamefont
  {Helay\"el-Neto}}]{Neves:2021tbt}%
  \BibitemOpen
  \bibfield  {author} {\bibinfo {author} {\bibfnamefont {M.~J.}\ \bibnamefont
  {Neves}}, \bibinfo {author} {\bibfnamefont {Jorge~B.}\ \bibnamefont
  {de~Oliveira}}, \bibinfo {author} {\bibfnamefont {L.~P.~R.}\ \bibnamefont
  {Ospedal}}, \ and\ \bibinfo {author} {\bibfnamefont {J.~A.}\ \bibnamefont
  {Helay\"el-Neto}},\ }\bibfield  {title} {\enquote {\bibinfo {title}
  {{Dispersion relations in nonlinear electrodynamics and the kinematics of the
  Compton effect in a magnetic background}},}\ }\href {\doibase
  10.1103/PhysRevD.104.015006} {\bibfield  {journal} {\bibinfo  {journal}
  {Phys. Rev. D}\ }\textbf {\bibinfo {volume} {104}},\ \bibinfo {pages}
  {015006} (\bibinfo {year} {2021})},\ \Eprint
  {http://arxiv.org/abs/2101.03642} {arXiv:2101.03642 [hep-th]} \BibitemShut
  {NoStop}%
\bibitem [{\citenamefont {Neves}\ \emph {et~al.}(2022)\citenamefont {Neves},
  \citenamefont {Ospedal}, \citenamefont {Helay\"el-Neto},\ and\ \citenamefont
  {Gaete}}]{Neves:2021jdy}%
  \BibitemOpen
  \bibfield  {author} {\bibinfo {author} {\bibfnamefont {M.~J.}\ \bibnamefont
  {Neves}}, \bibinfo {author} {\bibfnamefont {L.~P.~R.}\ \bibnamefont
  {Ospedal}}, \bibinfo {author} {\bibfnamefont {J.~A.}\ \bibnamefont
  {Helay\"el-Neto}}, \ and\ \bibinfo {author} {\bibfnamefont {Patricio}\
  \bibnamefont {Gaete}},\ }\bibfield  {title} {\enquote {\bibinfo {title}
  {{Considerations on anomalous photon and Z-boson self-couplings from the
  Born\textendash{}Infeld weak hypercharge action}},}\ }\href {\doibase
  10.1140/epjc/s10052-022-10296-y} {\bibfield  {journal} {\bibinfo  {journal}
  {Eur. Phys. J. C}\ }\textbf {\bibinfo {volume} {82}},\ \bibinfo {pages} {327}
  (\bibinfo {year} {2022})},\ \Eprint {http://arxiv.org/abs/2109.11004}
  {arXiv:2109.11004 [hep-th]} \BibitemShut {NoStop}%
\bibitem [{\citenamefont {De~Fabritiis}\ \emph {et~al.}(2022)\citenamefont
  {De~Fabritiis}, \citenamefont {Malta},\ and\ \citenamefont
  {Helay\"el-Neto}}]{DeFabritiis:2021qib}%
  \BibitemOpen
  \bibfield  {author} {\bibinfo {author} {\bibfnamefont {P.}~\bibnamefont
  {De~Fabritiis}}, \bibinfo {author} {\bibfnamefont {P.~C.}\ \bibnamefont
  {Malta}}, \ and\ \bibinfo {author} {\bibfnamefont {J.~A.}\ \bibnamefont
  {Helay\"el-Neto}},\ }\bibfield  {title} {\enquote {\bibinfo {title}
  {{Phenomenology of a Born-Infeld extension of the $U(1)_Y$ sector at lepton
  colliders}},}\ }\href {\doibase 10.1103/PhysRevD.105.016007} {\bibfield
  {journal} {\bibinfo  {journal} {Phys. Rev. D}\ }\textbf {\bibinfo {volume}
  {105}},\ \bibinfo {pages} {016007} (\bibinfo {year} {2022})},\ \Eprint
  {http://arxiv.org/abs/2109.12245} {arXiv:2109.12245 [hep-ph]} \BibitemShut
  {NoStop}%
\bibitem [{\citenamefont {Heisenberg}\ and\ \citenamefont
  {Euler}(1936)}]{Heisenberg:1936nmg}%
  \BibitemOpen
  \bibfield  {author} {\bibinfo {author} {\bibfnamefont {W.}~\bibnamefont
  {Heisenberg}}\ and\ \bibinfo {author} {\bibfnamefont {H.}~\bibnamefont
  {Euler}},\ }\bibfield  {title} {\enquote {\bibinfo {title} {{Consequences of
  Dirac's theory of positrons}},}\ }\href {\doibase 10.1007/BF01343663}
  {\bibfield  {journal} {\bibinfo  {journal} {Z. Phys.}\ }\textbf {\bibinfo
  {volume} {98}},\ \bibinfo {pages} {714--732} (\bibinfo {year} {1936})},\
  \Eprint {http://arxiv.org/abs/physics/0605038} {arXiv:physics/0605038}
  \BibitemShut {NoStop}%
\bibitem [{\citenamefont {Kruglov}(2007)}]{Kruglov:2007bh}%
  \BibitemOpen
  \bibfield  {author} {\bibinfo {author} {\bibfnamefont {S.~I.}\ \bibnamefont
  {Kruglov}},\ }\bibfield  {title} {\enquote {\bibinfo {title} {{Vacuum
  birefringence from the effective Lagrangian of the electromagnetic field}},}\
  }\href {\doibase 10.1103/PhysRevD.75.117301} {\bibfield  {journal} {\bibinfo
  {journal} {Phys. Rev. D}\ }\textbf {\bibinfo {volume} {75}},\ \bibinfo
  {pages} {117301} (\bibinfo {year} {2007})}\BibitemShut {NoStop}%
\bibitem [{\citenamefont {Kruglov}(2015{\natexlab{a}})}]{KRUGLOV_B}%
  \BibitemOpen
  \bibfield  {author} {\bibinfo {author} {\bibfnamefont {S.I.}\ \bibnamefont
  {Kruglov}},\ }\bibfield  {title} {\enquote {\bibinfo {title} {A model of
  nonlinear electrodynamics},}\ }\href {\doibase
  https://doi.org/10.1016/j.aop.2014.12.001} {\bibfield  {journal} {\bibinfo
  {journal} {Annals of Physics}\ }\textbf {\bibinfo {volume} {353}},\ \bibinfo
  {pages} {299--306} (\bibinfo {year} {2015}{\natexlab{a}})}\BibitemShut
  {NoStop}%
\bibitem [{\citenamefont {Kruglov}(2015{\natexlab{b}})}]{Kruglov:2014iwa}%
  \BibitemOpen
  \bibfield  {author} {\bibinfo {author} {\bibfnamefont {S.~I.}\ \bibnamefont
  {Kruglov}},\ }\bibfield  {title} {\enquote {\bibinfo {title} {{Nonlinear
  arcsin-electrodynamics}},}\ }\href {\doibase 10.1002/andp.201500142}
  {\bibfield  {journal} {\bibinfo  {journal} {Annalen Phys.}\ }\textbf
  {\bibinfo {volume} {527}},\ \bibinfo {pages} {397--401} (\bibinfo {year}
  {2015}{\natexlab{b}})},\ \Eprint {http://arxiv.org/abs/1410.7633}
  {arXiv:1410.7633 [physics.gen-ph]} \BibitemShut {NoStop}%
\bibitem [{\citenamefont {Kruglov}(2015{\natexlab{c}})}]{Kruglov:2014iqa}%
  \BibitemOpen
  \bibfield  {author} {\bibinfo {author} {\bibfnamefont {S.~I.}\ \bibnamefont
  {Kruglov}},\ }\bibfield  {title} {\enquote {\bibinfo {title} {{On Generalized
  Logarithmic Electrodynamics}},}\ }\href {\doibase
  10.1140/epjc/s10052-015-3314-z} {\bibfield  {journal} {\bibinfo  {journal}
  {Eur. Phys. J. C}\ }\textbf {\bibinfo {volume} {75}},\ \bibinfo {pages} {88}
  (\bibinfo {year} {2015}{\natexlab{c}})},\ \Eprint
  {http://arxiv.org/abs/1411.7741} {arXiv:1411.7741 [hep-th]} \BibitemShut
  {NoStop}%
\bibitem [{\citenamefont {Kruglov}(2015{\natexlab{d}})}]{Kruglov:2015yua}%
  \BibitemOpen
  \bibfield  {author} {\bibinfo {author} {\bibfnamefont {S.~I.}\ \bibnamefont
  {Kruglov}},\ }\bibfield  {title} {\enquote {\bibinfo {title} {{Nonlinear
  electrodynamics and black holes}},}\ }\href {\doibase
  10.1142/S0219887815500735} {\bibfield  {journal} {\bibinfo  {journal} {Int.
  J. Geom. Meth. Mod. Phys.}\ }\textbf {\bibinfo {volume} {12}},\ \bibinfo
  {pages} {1550073} (\bibinfo {year} {2015}{\natexlab{d}})},\ \Eprint
  {http://arxiv.org/abs/1504.03941} {arXiv:1504.03941 [physics.gen-ph]}
  \BibitemShut {NoStop}%
\bibitem [{\citenamefont {Kruglov}(2015{\natexlab{e}})}]{Kruglov:2015fbl}%
  \BibitemOpen
  \bibfield  {author} {\bibinfo {author} {\bibfnamefont {S.~I.}\ \bibnamefont
  {Kruglov}},\ }\bibfield  {title} {\enquote {\bibinfo {title} {{Universe
  acceleration and nonlinear electrodynamics}},}\ }\href {\doibase
  10.1103/PhysRevD.92.123523} {\bibfield  {journal} {\bibinfo  {journal} {Phys.
  Rev. D}\ }\textbf {\bibinfo {volume} {92}},\ \bibinfo {pages} {123523}
  (\bibinfo {year} {2015}{\natexlab{e}})},\ \Eprint
  {http://arxiv.org/abs/1601.06309} {arXiv:1601.06309 [gr-qc]} \BibitemShut
  {NoStop}%
\bibitem [{\citenamefont {Kruglov}(2016{\natexlab{a}})}]{Kruglov:2016cdm}%
  \BibitemOpen
  \bibfield  {author} {\bibinfo {author} {\bibfnamefont {S.~I.}\ \bibnamefont
  {Kruglov}},\ }\bibfield  {title} {\enquote {\bibinfo {title} {{Acceleration
  of Universe by Nonlinear Electromagnetic Fields}},}\ }\href {\doibase
  10.1142/S0218271816400022} {\bibfield  {journal} {\bibinfo  {journal} {Int.
  J. Mod. Phys. D}\ }\textbf {\bibinfo {volume} {25}},\ \bibinfo {pages}
  {1640002} (\bibinfo {year} {2016}{\natexlab{a}})},\ \Eprint
  {http://arxiv.org/abs/1603.07326} {arXiv:1603.07326 [gr-qc]} \BibitemShut
  {NoStop}%
\bibitem [{\citenamefont {Kruglov}(2016{\natexlab{b}})}]{Kruglov:2016ymq}%
  \BibitemOpen
  \bibfield  {author} {\bibinfo {author} {\bibfnamefont {S.~I.}\ \bibnamefont
  {Kruglov}},\ }\bibfield  {title} {\enquote {\bibinfo {title} {{Asymptotic
  Reissner-Nordstr\"om solution within nonlinear electrodynamics}},}\ }\href
  {\doibase 10.1103/PhysRevD.94.044026} {\bibfield  {journal} {\bibinfo
  {journal} {Phys. Rev. D}\ }\textbf {\bibinfo {volume} {94}},\ \bibinfo
  {pages} {044026} (\bibinfo {year} {2016}{\natexlab{b}})},\ \Eprint
  {http://arxiv.org/abs/1608.04275} {arXiv:1608.04275 [gr-qc]} \BibitemShut
  {NoStop}%
\bibitem [{\citenamefont {Kruglov}(2016{\natexlab{c}})}]{Kruglov:2016ezw}%
  \BibitemOpen
  \bibfield  {author} {\bibinfo {author} {\bibfnamefont {S.~I.}\ \bibnamefont
  {Kruglov}},\ }\bibfield  {title} {\enquote {\bibinfo {title} {{Nonlinear
  arcsin-electrodynamics and asymptotic Reissner-Nordstr\"om black holes}},}\
  }\href {\doibase 10.1002/andp.201600027} {\bibfield  {journal} {\bibinfo
  {journal} {Annalen Phys.}\ }\textbf {\bibinfo {volume} {528}},\ \bibinfo
  {pages} {588--596} (\bibinfo {year} {2016}{\natexlab{c}})},\ \Eprint
  {http://arxiv.org/abs/1607.07726} {arXiv:1607.07726 [gr-qc]} \BibitemShut
  {NoStop}%
\bibitem [{\citenamefont {Kruglov}(2017{\natexlab{a}})}]{Kruglov:2017mpj}%
  \BibitemOpen
  \bibfield  {author} {\bibinfo {author} {\bibfnamefont {S.~I.}\ \bibnamefont
  {Kruglov}},\ }\bibfield  {title} {\enquote {\bibinfo {title}
  {{Born\textendash{}Infeld-type electrodynamics and magnetic black holes}},}\
  }\href {\doibase 10.1016/j.aop.2017.06.008} {\bibfield  {journal} {\bibinfo
  {journal} {Annals Phys.}\ }\textbf {\bibinfo {volume} {383}},\ \bibinfo
  {pages} {550--559} (\bibinfo {year} {2017}{\natexlab{a}})},\ \Eprint
  {http://arxiv.org/abs/1707.04495} {arXiv:1707.04495 [gr-qc]} \BibitemShut
  {NoStop}%
\bibitem [{\citenamefont {Kruglov}(2017{\natexlab{b}})}]{Kruglov:2017fck}%
  \BibitemOpen
  \bibfield  {author} {\bibinfo {author} {\bibfnamefont {S.~I.}\ \bibnamefont
  {Kruglov}},\ }\bibfield  {title} {\enquote {\bibinfo {title} {{Black hole as
  a magnetic monopole within exponential nonlinear electrodynamics}},}\ }\href
  {\doibase 10.1016/j.aop.2016.12.036} {\bibfield  {journal} {\bibinfo
  {journal} {Annals Phys.}\ }\textbf {\bibinfo {volume} {378}},\ \bibinfo
  {pages} {59--70} (\bibinfo {year} {2017}{\natexlab{b}})},\ \Eprint
  {http://arxiv.org/abs/1703.02029} {arXiv:1703.02029 [gr-qc]} \BibitemShut
  {NoStop}%
\bibitem [{\citenamefont {Kruglov}(2017{\natexlab{c}})}]{Kruglov:2017xmb}%
  \BibitemOpen
  \bibfield  {author} {\bibinfo {author} {\bibfnamefont {S.~I.}\ \bibnamefont
  {Kruglov}},\ }\bibfield  {title} {\enquote {\bibinfo {title} {{Nonlinear
  Electrodynamics and Magnetic Black Holes}},}\ }\href {\doibase
  10.1002/andp.201700073} {\bibfield  {journal} {\bibinfo  {journal} {Annalen
  Phys.}\ }\textbf {\bibinfo {volume} {529}},\ \bibinfo {pages} {1700073}
  (\bibinfo {year} {2017}{\natexlab{c}})},\ \Eprint
  {http://arxiv.org/abs/1708.07006} {arXiv:1708.07006 [gr-qc]} \BibitemShut
  {NoStop}%
\bibitem [{\citenamefont {Fan}\ and\ \citenamefont {Wang}(2016)}]{Fan:2016hvf}%
  \BibitemOpen
  \bibfield  {author} {\bibinfo {author} {\bibfnamefont {Zhong-Ying}\
  \bibnamefont {Fan}}\ and\ \bibinfo {author} {\bibfnamefont {Xiaobao}\
  \bibnamefont {Wang}},\ }\bibfield  {title} {\enquote {\bibinfo {title}
  {{Construction of Regular Black Holes in General Relativity}},}\ }\href
  {\doibase 10.1103/PhysRevD.94.124027} {\bibfield  {journal} {\bibinfo
  {journal} {Phys. Rev. D}\ }\textbf {\bibinfo {volume} {94}},\ \bibinfo
  {pages} {124027} (\bibinfo {year} {2016})},\ \Eprint
  {http://arxiv.org/abs/1610.02636} {arXiv:1610.02636 [gr-qc]} \BibitemShut
  {NoStop}%
\bibitem [{\citenamefont {Mosquera~Cuesta}\ and\ \citenamefont
  {Lambiase}(2009{\natexlab{a}})}]{25}%
  \BibitemOpen
  \bibfield  {author} {\bibinfo {author} {\bibfnamefont {Herman~J.}\
  \bibnamefont {Mosquera~Cuesta}}\ and\ \bibinfo {author} {\bibfnamefont
  {Gaetano}\ \bibnamefont {Lambiase}},\ }\bibfield  {title} {\enquote {\bibinfo
  {title} {{Primordial magnetic fields and gravitational baryogenesis in
  nonlinear electrodynamics}},}\ }\href {\doibase 10.1103/PhysRevD.80.023013}
  {\bibfield  {journal} {\bibinfo  {journal} {Phys. Rev. D}\ }\textbf {\bibinfo
  {volume} {80}},\ \bibinfo {pages} {023013} (\bibinfo {year}
  {2009}{\natexlab{a}})},\ \Eprint {http://arxiv.org/abs/0907.3678}
  {arXiv:0907.3678 [astro-ph.CO]} \BibitemShut {NoStop}%
\bibitem [{\citenamefont {Mosquera~Cuesta}\ and\ \citenamefont
  {Lambiase}(2011)}]{26}%
  \BibitemOpen
  \bibfield  {author} {\bibinfo {author} {\bibfnamefont {Herman~J.}\
  \bibnamefont {Mosquera~Cuesta}}\ and\ \bibinfo {author} {\bibfnamefont
  {G.}~\bibnamefont {Lambiase}},\ }\bibfield  {title} {\enquote {\bibinfo
  {title} {{Nonlinear electrodynamics and CMB polarization}},}\ }\href
  {\doibase 10.1088/1475-7516/2011/03/033} {\bibfield  {journal} {\bibinfo
  {journal} {JCAP}\ }\textbf {\bibinfo {volume} {03}},\ \bibinfo {pages} {033}
  (\bibinfo {year} {2011})},\ \Eprint {http://arxiv.org/abs/1102.3092}
  {arXiv:1102.3092 [astro-ph.CO]} \BibitemShut {NoStop}%
\bibitem [{\citenamefont {Mosquera~Cuesta}\ and\ \citenamefont
  {Lambiase}(2009{\natexlab{b}})}]{27}%
  \BibitemOpen
  \bibfield  {author} {\bibinfo {author} {\bibfnamefont {Herman~J.}\
  \bibnamefont {Mosquera~Cuesta}}\ and\ \bibinfo {author} {\bibfnamefont
  {Gaetano}\ \bibnamefont {Lambiase}},\ }\bibfield  {title} {\enquote {\bibinfo
  {title} {{Neutrino mass spectrum from neutrino spin-flip-driven gravitational
  waves}},}\ }\href {\doibase 10.1142/S0218271809014571} {\bibfield  {journal}
  {\bibinfo  {journal} {Int. J. Mod. Phys. D}\ }\textbf {\bibinfo {volume}
  {18}},\ \bibinfo {pages} {435--443} (\bibinfo {year}
  {2009}{\natexlab{b}})}\BibitemShut {NoStop}%
\bibitem [{\citenamefont {Corda}\ and\ \citenamefont
  {Mosquera~Cuesta}(2011)}]{28}%
  \BibitemOpen
  \bibfield  {author} {\bibinfo {author} {\bibfnamefont {Christian}\
  \bibnamefont {Corda}}\ and\ \bibinfo {author} {\bibfnamefont {Herman~J.}\
  \bibnamefont {Mosquera~Cuesta}},\ }\bibfield  {title} {\enquote {\bibinfo
  {title} {{Inflation from R\textasciicircum{}2 gravity: a new approach using
  nonlinear electrodynamics}},}\ }\href {\doibase
  10.1016/j.astropartphys.2010.12.002} {\bibfield  {journal} {\bibinfo
  {journal} {Astropart. Phys.}\ }\textbf {\bibinfo {volume} {34}},\ \bibinfo
  {pages} {587--590} (\bibinfo {year} {2011})},\ \Eprint
  {http://arxiv.org/abs/1011.4801} {arXiv:1011.4801 [physics.gen-ph]}
  \BibitemShut {NoStop}%
\bibitem [{\citenamefont {Mosquera~Cuesta}\ and\ \citenamefont
  {Salim}(2004{\natexlab{a}})}]{29}%
  \BibitemOpen
  \bibfield  {author} {\bibinfo {author} {\bibfnamefont {Herman~J.}\
  \bibnamefont {Mosquera~Cuesta}}\ and\ \bibinfo {author} {\bibfnamefont
  {Jose~M.}\ \bibnamefont {Salim}},\ }\bibfield  {title} {\enquote {\bibinfo
  {title} {{Nonlinear electrodynamics and the gravitational redshift of
  pulsars}},}\ }\href {\doibase 10.1111/j.1365-2966.2004.08375.x} {\bibfield
  {journal} {\bibinfo  {journal} {Mon. Not. Roy. Astron. Soc.}\ }\textbf
  {\bibinfo {volume} {354}},\ \bibinfo {pages} {L55--L59} (\bibinfo {year}
  {2004}{\natexlab{a}})},\ \Eprint {http://arxiv.org/abs/astro-ph/0403045}
  {arXiv:astro-ph/0403045} \BibitemShut {NoStop}%
\bibitem [{\citenamefont {Mosquera~Cuesta}\ and\ \citenamefont
  {Salim}(2004{\natexlab{b}})}]{30}%
  \BibitemOpen
  \bibfield  {author} {\bibinfo {author} {\bibfnamefont {Herman~J.}\
  \bibnamefont {Mosquera~Cuesta}}\ and\ \bibinfo {author} {\bibfnamefont
  {Jose~M.}\ \bibnamefont {Salim}},\ }\bibfield  {title} {\enquote {\bibinfo
  {title} {{Nonlinear electrodynamics and the surface redshift of pulsars}},}\
  }\href {\doibase 10.1086/378686} {\bibfield  {journal} {\bibinfo  {journal}
  {Astrophys. J.}\ }\textbf {\bibinfo {volume} {608}},\ \bibinfo {pages}
  {925--929} (\bibinfo {year} {2004}{\natexlab{b}})},\ \Eprint
  {http://arxiv.org/abs/astro-ph/0307513} {arXiv:astro-ph/0307513} \BibitemShut
  {NoStop}%
\bibitem [{\citenamefont {Mosquera~Cuesta}\ \emph {et~al.}(2006)\citenamefont
  {Mosquera~Cuesta}, \citenamefont {de~Freitas~Pacheco},\ and\ \citenamefont
  {Salim}}]{31}%
  \BibitemOpen
  \bibfield  {author} {\bibinfo {author} {\bibfnamefont {Herman~J.}\
  \bibnamefont {Mosquera~Cuesta}}, \bibinfo {author} {\bibfnamefont {Jose~A.}\
  \bibnamefont {de~Freitas~Pacheco}}, \ and\ \bibinfo {author} {\bibfnamefont
  {Jose~M.}\ \bibnamefont {Salim}},\ }\bibfield  {title} {\enquote {\bibinfo
  {title} {{Einstein's gravitational lensing and nonlinear electrodynamics}},}\
  }\href {\doibase 10.1142/S0217751X06025055} {\bibfield  {journal} {\bibinfo
  {journal} {Int. J. Mod. Phys. A}\ }\textbf {\bibinfo {volume} {21}},\
  \bibinfo {pages} {43--55} (\bibinfo {year} {2006})},\ \Eprint
  {http://arxiv.org/abs/astro-ph/0408152} {arXiv:astro-ph/0408152} \BibitemShut
  {NoStop}%
\bibitem [{\citenamefont {Mbelek}\ \emph {et~al.}(2007)\citenamefont {Mbelek},
  \citenamefont {Mosquera~Cuesta}, \citenamefont {Novello},\ and\ \citenamefont
  {Salim}}]{32}%
  \BibitemOpen
  \bibfield  {author} {\bibinfo {author} {\bibfnamefont {Jean~Paul}\
  \bibnamefont {Mbelek}}, \bibinfo {author} {\bibfnamefont {Herman~J.}\
  \bibnamefont {Mosquera~Cuesta}}, \bibinfo {author} {\bibfnamefont
  {M.}~\bibnamefont {Novello}}, \ and\ \bibinfo {author} {\bibfnamefont
  {Jose~M.}\ \bibnamefont {Salim}},\ }\bibfield  {title} {\enquote {\bibinfo
  {title} {{Nonlinear electrodynamics and the Pioneer 10/11 spacecraft
  anomaly}},}\ }\href {\doibase 10.1209/0295-5075/77/19001} {\bibfield
  {journal} {\bibinfo  {journal} {EPL}\ }\textbf {\bibinfo {volume} {77}},\
  \bibinfo {pages} {19001} (\bibinfo {year} {2007})},\ \Eprint
  {http://arxiv.org/abs/astro-ph/0608538} {arXiv:astro-ph/0608538} \BibitemShut
  {NoStop}%
\bibitem [{\citenamefont {Mbelek}\ and\ \citenamefont
  {Mosquera~Cuesta}(2008)}]{33}%
  \BibitemOpen
  \bibfield  {author} {\bibinfo {author} {\bibfnamefont {Jean~Paul}\
  \bibnamefont {Mbelek}}\ and\ \bibinfo {author} {\bibfnamefont {Herman~J.}\
  \bibnamefont {Mosquera~Cuesta}},\ }\bibfield  {title} {\enquote {\bibinfo
  {title} {{Nonlinear electrodynamics and the variation of the fine structure
  constant}},}\ }\href {\doibase 10.1111/j.1365-2966.2008.13503.x} {\bibfield
  {journal} {\bibinfo  {journal} {Mon. Not. Roy. Astron. Soc.}\ }\textbf
  {\bibinfo {volume} {389}},\ \bibinfo {pages} {199} (\bibinfo {year}
  {2008})},\ \Eprint {http://arxiv.org/abs/0707.3288} {arXiv:0707.3288
  [astro-ph]} \BibitemShut {NoStop}%
\bibitem [{\citenamefont {Breton}\ and\ \citenamefont
  {Garcia-Salcedo}(2007)}]{refe1}%
  \BibitemOpen
  \bibfield  {author} {\bibinfo {author} {\bibfnamefont {Nora}\ \bibnamefont
  {Breton}}\ and\ \bibinfo {author} {\bibfnamefont {Ricardo}\ \bibnamefont
  {Garcia-Salcedo}},\ }\href {\doibase 10.48550/ARXIV.HEP-TH/0702008} {\enquote
  {\bibinfo {title} {Nonlinear electrodynamics and black holes},}\ } (\bibinfo
  {year} {2007})\BibitemShut {NoStop}%
\bibitem [{\citenamefont {Panotopoulos}(2021)}]{Panotopoulos:2020bfl}%
  \BibitemOpen
  \bibfield  {author} {\bibinfo {author} {\bibfnamefont {Grigoris}\
  \bibnamefont {Panotopoulos}},\ }\bibfield  {title} {\enquote {\bibinfo
  {title} {{Building (1+1) holographic superconductors in the presence of
  non-linear Electrodynamics}},}\ }\href {\doibase 10.1016/j.cjph.2020.11.023}
  {\bibfield  {journal} {\bibinfo  {journal} {Chin. J. Phys.}\ }\textbf
  {\bibinfo {volume} {69}},\ \bibinfo {pages} {295--302} (\bibinfo {year}
  {2021})},\ \Eprint {http://arxiv.org/abs/2012.09978} {arXiv:2012.09978
  [hep-th]} \BibitemShut {NoStop}%
\bibitem [{\citenamefont {Panotopoulos}\ and\ \citenamefont
  {Rinc\'on}(2017)}]{Panotopoulos:2017hns}%
  \BibitemOpen
  \bibfield  {author} {\bibinfo {author} {\bibfnamefont {Grigoris}\
  \bibnamefont {Panotopoulos}}\ and\ \bibinfo {author} {\bibfnamefont
  {\'Angel}\ \bibnamefont {Rinc\'on}},\ }\bibfield  {title} {\enquote {\bibinfo
  {title} {{Quasinormal modes of black holes in Einstein-power-Maxwell
  theory}},}\ }\href {\doibase 10.1142/S0218271818500347} {\bibfield  {journal}
  {\bibinfo  {journal} {Int. J. Mod. Phys. D}\ }\textbf {\bibinfo {volume}
  {27}},\ \bibinfo {pages} {1850034} (\bibinfo {year} {2017})},\ \Eprint
  {http://arxiv.org/abs/1711.04146} {arXiv:1711.04146 [hep-th]} \BibitemShut
  {NoStop}%
\bibitem [{\citenamefont {Toshmatov}\ \emph {et~al.}(2018)\citenamefont
  {Toshmatov}, \citenamefont {Stuchl\'\i{}k}, \citenamefont {Schee},\ and\
  \citenamefont {Ahmedov}}]{Toshmatov:2018tyo}%
  \BibitemOpen
  \bibfield  {author} {\bibinfo {author} {\bibfnamefont {Bobir}\ \bibnamefont
  {Toshmatov}}, \bibinfo {author} {\bibfnamefont {Zden\v{e}k}\ \bibnamefont
  {Stuchl\'\i{}k}}, \bibinfo {author} {\bibfnamefont {Jan}\ \bibnamefont
  {Schee}}, \ and\ \bibinfo {author} {\bibfnamefont {Bobomurat}\ \bibnamefont
  {Ahmedov}},\ }\bibfield  {title} {\enquote {\bibinfo {title}
  {{Electromagnetic perturbations of black holes in general relativity coupled
  to nonlinear electrodynamics}},}\ }\href {\doibase
  10.1103/PhysRevD.97.084058} {\bibfield  {journal} {\bibinfo  {journal} {Phys.
  Rev. D}\ }\textbf {\bibinfo {volume} {97}},\ \bibinfo {pages} {084058}
  (\bibinfo {year} {2018})},\ \Eprint {http://arxiv.org/abs/1805.00240}
  {arXiv:1805.00240 [gr-qc]} \BibitemShut {NoStop}%
\bibitem [{\citenamefont {Stuchl\'\i{}k}\ and\ \citenamefont
  {Schee}(2019)}]{Stuchlik:2019uvf}%
  \BibitemOpen
  \bibfield  {author} {\bibinfo {author} {\bibfnamefont {Zdenek}\ \bibnamefont
  {Stuchl\'\i{}k}}\ and\ \bibinfo {author} {\bibfnamefont {Jan}\ \bibnamefont
  {Schee}},\ }\bibfield  {title} {\enquote {\bibinfo {title} {{Shadow of the
  regular Bardeen black holes and comparison of the motion of photons and
  neutrinos}},}\ }\href {\doibase 10.1140/epjc/s10052-019-6543-8} {\bibfield
  {journal} {\bibinfo  {journal} {Eur. Phys. J. C}\ }\textbf {\bibinfo {volume}
  {79}},\ \bibinfo {pages} {44} (\bibinfo {year} {2019})}\BibitemShut {NoStop}%
\bibitem [{\citenamefont {Rayimbaev}\ \emph {et~al.}(2022)\citenamefont
  {Rayimbaev}, \citenamefont {Bardiev}, \citenamefont {Abdulxamidov},
  \citenamefont {Abdujabbarov},\ and\ \citenamefont
  {Ahmedov}}]{Rayimbaev:2022hrn}%
  \BibitemOpen
  \bibfield  {author} {\bibinfo {author} {\bibfnamefont {Javlon}\ \bibnamefont
  {Rayimbaev}}, \bibinfo {author} {\bibfnamefont {Dilshodbek}\ \bibnamefont
  {Bardiev}}, \bibinfo {author} {\bibfnamefont {Farrux}\ \bibnamefont
  {Abdulxamidov}}, \bibinfo {author} {\bibfnamefont {Ahmadjon}\ \bibnamefont
  {Abdujabbarov}}, \ and\ \bibinfo {author} {\bibfnamefont {Bobomurat}\
  \bibnamefont {Ahmedov}},\ }\bibfield  {title} {\enquote {\bibinfo {title}
  {{Magnetized and Magnetically Charged Particles Motion around Regular Bardeen
  Black Hole in 4D Einstein Gauss\textendash{}Bonnet Gravity}},}\ }\href
  {\doibase 10.3390/universe8100549} {\bibfield  {journal} {\bibinfo  {journal}
  {Universe}\ }\textbf {\bibinfo {volume} {8}},\ \bibinfo {pages} {549}
  (\bibinfo {year} {2022})}\BibitemShut {NoStop}%
\bibitem [{\citenamefont {Toshmatov}\ \emph {et~al.}(2014)\citenamefont
  {Toshmatov}, \citenamefont {Ahmedov}, \citenamefont {Abdujabbarov},\ and\
  \citenamefont {Stuchlik}}]{Toshmatov:2014nya}%
  \BibitemOpen
  \bibfield  {author} {\bibinfo {author} {\bibfnamefont {Bobir}\ \bibnamefont
  {Toshmatov}}, \bibinfo {author} {\bibfnamefont {Bobomurat}\ \bibnamefont
  {Ahmedov}}, \bibinfo {author} {\bibfnamefont {Ahmadjon}\ \bibnamefont
  {Abdujabbarov}}, \ and\ \bibinfo {author} {\bibfnamefont {Zdenek}\
  \bibnamefont {Stuchlik}},\ }\bibfield  {title} {\enquote {\bibinfo {title}
  {{Rotating Regular Black Hole Solution}},}\ }\href {\doibase
  10.1103/PhysRevD.89.104017} {\bibfield  {journal} {\bibinfo  {journal} {Phys.
  Rev. D}\ }\textbf {\bibinfo {volume} {89}},\ \bibinfo {pages} {104017}
  (\bibinfo {year} {2014})},\ \Eprint {http://arxiv.org/abs/1404.6443}
  {arXiv:1404.6443 [gr-qc]} \BibitemShut {NoStop}%
\bibitem [{\citenamefont {Abdujabbarov}\ \emph {et~al.}(2016)\citenamefont
  {Abdujabbarov}, \citenamefont {Amir}, \citenamefont {Ahmedov},\ and\
  \citenamefont {Ghosh}}]{Abdujabbarov:2016hnw}%
  \BibitemOpen
  \bibfield  {author} {\bibinfo {author} {\bibfnamefont {Ahmadjon}\
  \bibnamefont {Abdujabbarov}}, \bibinfo {author} {\bibfnamefont {Muhammed}\
  \bibnamefont {Amir}}, \bibinfo {author} {\bibfnamefont {Bobomurat}\
  \bibnamefont {Ahmedov}}, \ and\ \bibinfo {author} {\bibfnamefont
  {Sushant~G.}\ \bibnamefont {Ghosh}},\ }\bibfield  {title} {\enquote {\bibinfo
  {title} {{Shadow of rotating regular black holes}},}\ }\href {\doibase
  10.1103/PhysRevD.93.104004} {\bibfield  {journal} {\bibinfo  {journal} {Phys.
  Rev. D}\ }\textbf {\bibinfo {volume} {93}},\ \bibinfo {pages} {104004}
  (\bibinfo {year} {2016})},\ \Eprint {http://arxiv.org/abs/1604.03809}
  {arXiv:1604.03809 [gr-qc]} \BibitemShut {NoStop}%
\bibitem [{\citenamefont {Toshmatov}\ \emph {et~al.}(2017)\citenamefont
  {Toshmatov}, \citenamefont {Stuchl\'\i{}k},\ and\ \citenamefont
  {Ahmedov}}]{Toshmatov:2017zpr}%
  \BibitemOpen
  \bibfield  {author} {\bibinfo {author} {\bibfnamefont {Bobir}\ \bibnamefont
  {Toshmatov}}, \bibinfo {author} {\bibfnamefont {Zden\v{e}k}\ \bibnamefont
  {Stuchl\'\i{}k}}, \ and\ \bibinfo {author} {\bibfnamefont {Bobomurat}\
  \bibnamefont {Ahmedov}},\ }\bibfield  {title} {\enquote {\bibinfo {title}
  {{Generic rotating regular black holes in general relativity coupled to
  nonlinear electrodynamics}},}\ }\href {\doibase 10.1103/PhysRevD.95.084037}
  {\bibfield  {journal} {\bibinfo  {journal} {Phys. Rev. D}\ }\textbf {\bibinfo
  {volume} {95}},\ \bibinfo {pages} {084037} (\bibinfo {year} {2017})},\
  \Eprint {http://arxiv.org/abs/1704.07300} {arXiv:1704.07300 [gr-qc]}
  \BibitemShut {NoStop}%
\bibitem [{\citenamefont {Stuchl\'\i{}k}\ and\ \citenamefont
  {Schee}(2014)}]{Stuchlik:2014qja}%
  \BibitemOpen
  \bibfield  {author} {\bibinfo {author} {\bibfnamefont {Zden\v{e}k}\
  \bibnamefont {Stuchl\'\i{}k}}\ and\ \bibinfo {author} {\bibfnamefont {Jan}\
  \bibnamefont {Schee}},\ }\bibfield  {title} {\enquote {\bibinfo {title}
  {{Circular geodesic of Bardeen and Ayon\textendash{}Beato\textendash{}Garcia
  regular black-hole and no-horizon spacetimes}},}\ }\href {\doibase
  10.1142/S0218271815500200} {\bibfield  {journal} {\bibinfo  {journal} {Int.
  J. Mod. Phys. D}\ }\textbf {\bibinfo {volume} {24}},\ \bibinfo {pages}
  {1550020} (\bibinfo {year} {2014})},\ \Eprint
  {http://arxiv.org/abs/1501.00015} {arXiv:1501.00015 [astro-ph.HE]}
  \BibitemShut {NoStop}%
\bibitem [{\citenamefont {Delphenich}(2006)}]{34}%
  \BibitemOpen
  \bibfield  {author} {\bibinfo {author} {\bibfnamefont {D.~H.}\ \bibnamefont
  {Delphenich}},\ }\bibfield  {title} {\enquote {\bibinfo {title} {{Nonlinear
  optical analogies in quantum electrodynamics}},}\ }\href@noop {} {\
  (\bibinfo {year} {2006})},\ \Eprint {http://arxiv.org/abs/hep-th/0610088}
  {arXiv:hep-th/0610088} \BibitemShut {NoStop}%
\bibitem [{\citenamefont {Lundin}\ \emph {et~al.}(2006)\citenamefont {Lundin},
  \citenamefont {Brodin},\ and\ \citenamefont {Marklund}}]{35}%
  \BibitemOpen
  \bibfield  {author} {\bibinfo {author} {\bibfnamefont {J.}~\bibnamefont
  {Lundin}}, \bibinfo {author} {\bibfnamefont {G.}~\bibnamefont {Brodin}}, \
  and\ \bibinfo {author} {\bibfnamefont {M.}~\bibnamefont {Marklund}},\
  }\bibfield  {title} {\enquote {\bibinfo {title} {{Short wavelength QED
  correction to cold plasma-wave propagation}},}\ }\href {\doibase
  10.1063/1.2356315} {\bibfield  {journal} {\bibinfo  {journal} {Phys.
  Plasmas}\ }\textbf {\bibinfo {volume} {13}},\ \bibinfo {pages} {102102}
  (\bibinfo {year} {2006})},\ \Eprint {http://arxiv.org/abs/physics/0606098}
  {arXiv:physics/0606098} \BibitemShut {NoStop}%
\bibitem [{\citenamefont {Lundstrom}\ \emph {et~al.}(2006)\citenamefont
  {Lundstrom}, \citenamefont {Brodin}, \citenamefont {Lundin}, \citenamefont
  {Marklund}, \citenamefont {Bingham}, \citenamefont {Collier}, \citenamefont
  {Mendonca},\ and\ \citenamefont {Norreys}}]{Laser}%
  \BibitemOpen
  \bibfield  {author} {\bibinfo {author} {\bibfnamefont {E.}~\bibnamefont
  {Lundstrom}}, \bibinfo {author} {\bibfnamefont {G.}~\bibnamefont {Brodin}},
  \bibinfo {author} {\bibfnamefont {J.}~\bibnamefont {Lundin}}, \bibinfo
  {author} {\bibfnamefont {M.}~\bibnamefont {Marklund}}, \bibinfo {author}
  {\bibfnamefont {R.}~\bibnamefont {Bingham}}, \bibinfo {author} {\bibfnamefont
  {J.}~\bibnamefont {Collier}}, \bibinfo {author} {\bibfnamefont {J.~T.}\
  \bibnamefont {Mendonca}}, \ and\ \bibinfo {author} {\bibfnamefont
  {P.}~\bibnamefont {Norreys}},\ }\bibfield  {title} {\enquote {\bibinfo
  {title} {{Using high-power lasers for detection of elastic photon-photon
  scattering}},}\ }\href {\doibase 10.1103/PhysRevLett.96.083602} {\bibfield
  {journal} {\bibinfo  {journal} {Phys. Rev. Lett.}\ }\textbf {\bibinfo
  {volume} {96}},\ \bibinfo {pages} {083602} (\bibinfo {year} {2006})},\
  \Eprint {http://arxiv.org/abs/hep-ph/0510076} {arXiv:hep-ph/0510076}
  \BibitemShut {NoStop}%
\bibitem [{\citenamefont {Ohnishi}\ and\ \citenamefont {Yamamoto}(2014)}]{38}%
  \BibitemOpen
  \bibfield  {author} {\bibinfo {author} {\bibfnamefont {Akira}\ \bibnamefont
  {Ohnishi}}\ and\ \bibinfo {author} {\bibfnamefont {Naoki}\ \bibnamefont
  {Yamamoto}},\ }\bibfield  {title} {\enquote {\bibinfo {title} {{Magnetars and
  the Chiral Plasma Instabilities}},}\ }\href@noop {} {\  (\bibinfo {year}
  {2014})},\ \Eprint {http://arxiv.org/abs/1402.4760} {arXiv:1402.4760
  [astro-ph.HE]} \BibitemShut {NoStop}%
\bibitem [{\citenamefont {Akamatsu}\ and\ \citenamefont {Yamamoto}(2013)}]{39}%
  \BibitemOpen
  \bibfield  {author} {\bibinfo {author} {\bibfnamefont {Yukinao}\ \bibnamefont
  {Akamatsu}}\ and\ \bibinfo {author} {\bibfnamefont {Naoki}\ \bibnamefont
  {Yamamoto}},\ }\bibfield  {title} {\enquote {\bibinfo {title} {{Chiral Plasma
  Instabilities}},}\ }\href {\doibase 10.1103/PhysRevLett.111.052002}
  {\bibfield  {journal} {\bibinfo  {journal} {Phys. Rev. Lett.}\ }\textbf
  {\bibinfo {volume} {111}},\ \bibinfo {pages} {052002} (\bibinfo {year}
  {2013})},\ \Eprint {http://arxiv.org/abs/1302.2125} {arXiv:1302.2125
  [nucl-th]} \BibitemShut {NoStop}%
\bibitem [{\citenamefont {Rincon}\ \emph {et~al.}(2022)\citenamefont {Rincon},
  \citenamefont {Gonzalez}, \citenamefont {Panotopoulos}, \citenamefont
  {Saavedra},\ and\ \citenamefont {Vasquez}}]{Rincon:2021gwd}%
  \BibitemOpen
  \bibfield  {author} {\bibinfo {author} {\bibfnamefont {Angel}\ \bibnamefont
  {Rincon}}, \bibinfo {author} {\bibfnamefont {P.~A.}\ \bibnamefont
  {Gonzalez}}, \bibinfo {author} {\bibfnamefont {Grigoris}\ \bibnamefont
  {Panotopoulos}}, \bibinfo {author} {\bibfnamefont {Joel}\ \bibnamefont
  {Saavedra}}, \ and\ \bibinfo {author} {\bibfnamefont {Yerko}\ \bibnamefont
  {Vasquez}},\ }\bibfield  {title} {\enquote {\bibinfo {title} {{Quasinormal
  modes for a non-minimally coupled scalar field in a five-dimensional
  Einstein\textendash{}Power\textendash{}Maxwell background}},}\ }\href
  {\doibase 10.1140/epjp/s13360-022-03438-4} {\bibfield  {journal} {\bibinfo
  {journal} {Eur. Phys. J. Plus}\ }\textbf {\bibinfo {volume} {137}},\ \bibinfo
  {pages} {1278} (\bibinfo {year} {2022})},\ \Eprint
  {http://arxiv.org/abs/2112.04793} {arXiv:2112.04793 [gr-qc]} \BibitemShut
  {NoStop}%
\bibitem [{\citenamefont {Gonz\'alez}\ \emph {et~al.}(2021)\citenamefont
  {Gonz\'alez}, \citenamefont {Rinc\'on}, \citenamefont {Saavedra},\ and\
  \citenamefont {V\'asquez}}]{Gonzalez:2021vwp}%
  \BibitemOpen
  \bibfield  {author} {\bibinfo {author} {\bibfnamefont {P.~A.}\ \bibnamefont
  {Gonz\'alez}}, \bibinfo {author} {\bibfnamefont {\'Angel}\ \bibnamefont
  {Rinc\'on}}, \bibinfo {author} {\bibfnamefont {Joel}\ \bibnamefont
  {Saavedra}}, \ and\ \bibinfo {author} {\bibfnamefont {Yerko}\ \bibnamefont
  {V\'asquez}},\ }\bibfield  {title} {\enquote {\bibinfo {title} {{Superradiant
  instability and charged scalar quasinormal modes for (2+1)-dimensional
  Coulomb-like AdS black holes from nonlinear electrodynamics}},}\ }\href
  {\doibase 10.1103/PhysRevD.104.084047} {\bibfield  {journal} {\bibinfo
  {journal} {Phys. Rev. D}\ }\textbf {\bibinfo {volume} {104}},\ \bibinfo
  {pages} {084047} (\bibinfo {year} {2021})},\ \Eprint
  {http://arxiv.org/abs/2107.08611} {arXiv:2107.08611 [gr-qc]} \BibitemShut
  {NoStop}%
\bibitem [{\citenamefont {Rinc\'on}\ \emph {et~al.}(2018)\citenamefont
  {Rinc\'on}, \citenamefont {Contreras}, \citenamefont {Bargue\~no},
  \citenamefont {Koch},\ and\ \citenamefont {Panotopoulos}}]{Rincon:2018dsq}%
  \BibitemOpen
  \bibfield  {author} {\bibinfo {author} {\bibfnamefont {\'Angel}\ \bibnamefont
  {Rinc\'on}}, \bibinfo {author} {\bibfnamefont {Ernesto}\ \bibnamefont
  {Contreras}}, \bibinfo {author} {\bibfnamefont {Pedro}\ \bibnamefont
  {Bargue\~no}}, \bibinfo {author} {\bibfnamefont {Benjamin}\ \bibnamefont
  {Koch}}, \ and\ \bibinfo {author} {\bibfnamefont {Grigorios}\ \bibnamefont
  {Panotopoulos}},\ }\bibfield  {title} {\enquote {\bibinfo {title}
  {{Scale-dependent ( $2+1$ )-dimensional electrically charged black holes in
  Einstein-power-Maxwell theory}},}\ }\href {\doibase
  10.1140/epjc/s10052-018-6106-4} {\bibfield  {journal} {\bibinfo  {journal}
  {Eur. Phys. J. C}\ }\textbf {\bibinfo {volume} {78}},\ \bibinfo {pages} {641}
  (\bibinfo {year} {2018})},\ \Eprint {http://arxiv.org/abs/1807.08047}
  {arXiv:1807.08047 [hep-th]} \BibitemShut {NoStop}%
\bibitem [{\citenamefont {Rinc\'on}\ \emph {et~al.}(2017)\citenamefont
  {Rinc\'on}, \citenamefont {Contreras}, \citenamefont {Bargue\~no},
  \citenamefont {Koch}, \citenamefont {Panotopoulos},\ and\ \citenamefont
  {Hern\'andez-Arboleda}}]{Rincon:2017goj}%
  \BibitemOpen
  \bibfield  {author} {\bibinfo {author} {\bibfnamefont {\'Angel}\ \bibnamefont
  {Rinc\'on}}, \bibinfo {author} {\bibfnamefont {Ernesto}\ \bibnamefont
  {Contreras}}, \bibinfo {author} {\bibfnamefont {Pedro}\ \bibnamefont
  {Bargue\~no}}, \bibinfo {author} {\bibfnamefont {Benjamin}\ \bibnamefont
  {Koch}}, \bibinfo {author} {\bibfnamefont {Grigorios}\ \bibnamefont
  {Panotopoulos}}, \ and\ \bibinfo {author} {\bibfnamefont {Alejandro}\
  \bibnamefont {Hern\'andez-Arboleda}},\ }\bibfield  {title} {\enquote
  {\bibinfo {title} {{Scale dependent three-dimensional charged black holes in
  linear and non-linear electrodynamics}},}\ }\href {\doibase
  10.1140/epjc/s10052-017-5045-9} {\bibfield  {journal} {\bibinfo  {journal}
  {Eur. Phys. J. C}\ }\textbf {\bibinfo {volume} {77}},\ \bibinfo {pages} {494}
  (\bibinfo {year} {2017})},\ \Eprint {http://arxiv.org/abs/1704.04845}
  {arXiv:1704.04845 [hep-th]} \BibitemShut {NoStop}%
\bibitem [{\citenamefont {Diaz-alonso}\ and\ \citenamefont
  {Rubiera-Garcia}(2013)}]{garcia}%
  \BibitemOpen
  \bibfield  {author} {\bibinfo {author} {\bibfnamefont {Joaquin}\ \bibnamefont
  {Diaz-alonso}}\ and\ \bibinfo {author} {\bibfnamefont {Diego}\ \bibnamefont
  {Rubiera-Garcia}},\ }\bibfield  {title} {\enquote {\bibinfo {title} {Charged
  black hole solutions of non-linear electrodynamics and generalized gauge
  field theories},}\ \ }(\bibinfo {year} {2013})\BibitemShut {NoStop}%
\bibitem [{\citenamefont {Mosquera~Cuesta}\ \emph {et~al.}(2017)\citenamefont
  {Mosquera~Cuesta}, \citenamefont {Lambiase},\ and\ \citenamefont
  {Pereira}}]{Lambiase2017}%
  \BibitemOpen
  \bibfield  {author} {\bibinfo {author} {\bibfnamefont {Herman~J.}\
  \bibnamefont {Mosquera~Cuesta}}, \bibinfo {author} {\bibfnamefont {Gaetano}\
  \bibnamefont {Lambiase}}, \ and\ \bibinfo {author} {\bibfnamefont {Jonas~P.}\
  \bibnamefont {Pereira}},\ }\bibfield  {title} {\enquote {\bibinfo {title}
  {{Probing nonlinear electrodynamics in slowly rotating spacetimes through
  neutrino astrophysics}},}\ }\href {\doibase 10.1103/PhysRevD.95.025011}
  {\bibfield  {journal} {\bibinfo  {journal} {Phys. Rev. D}\ }\textbf {\bibinfo
  {volume} {95}},\ \bibinfo {pages} {025011} (\bibinfo {year} {2017})},\
  \Eprint {http://arxiv.org/abs/1701.00431} {arXiv:1701.00431 [gr-qc]}
  \BibitemShut {NoStop}%
\bibitem [{\citenamefont {{Okyay}}\ and\ \citenamefont
  {{{\"O}vg{\"u}n}}(2022)}]{Okyay}%
  \BibitemOpen
  \bibfield  {author} {\bibinfo {author} {\bibfnamefont {M.}~\bibnamefont
  {{Okyay}}}\ and\ \bibinfo {author} {\bibfnamefont {A.}~\bibnamefont
  {{{\"O}vg{\"u}n}}},\ }\bibfield  {title} {\enquote {\bibinfo {title}
  {{Nonlinear electrodynamics effects on the black hole shadow, deflection
  angle, quasinormal modes and greybody factors}},}\ }\href {\doibase
  10.1088/1475-7516/2022/01/009} {\bibfield  {journal} {\bibinfo  {journal}
  {jcap}\ }\textbf {\bibinfo {volume} {2022}},\ \bibinfo {eid} {009} (\bibinfo
  {year} {2022})},\ \Eprint {http://arxiv.org/abs/2108.07766} {arXiv:2108.07766
  [gr-qc]} \BibitemShut {NoStop}%
\bibitem [{\citenamefont {\"Ovg\"un}(2019)}]{Ovgun:2019wej}%
  \BibitemOpen
  \bibfield  {author} {\bibinfo {author} {\bibfnamefont {A.}~\bibnamefont
  {\"Ovg\"un}},\ }\bibfield  {title} {\enquote {\bibinfo {title} {{Weak field
  deflection angle by regular black holes with cosmic strings using the
  Gauss-Bonnet theorem}},}\ }\href {\doibase 10.1103/PhysRevD.99.104075}
  {\bibfield  {journal} {\bibinfo  {journal} {Phys. Rev. D}\ }\textbf {\bibinfo
  {volume} {99}},\ \bibinfo {pages} {104075} (\bibinfo {year} {2019})},\
  \Eprint {http://arxiv.org/abs/1902.04411} {arXiv:1902.04411 [gr-qc]}
  \BibitemShut {NoStop}%
\bibitem [{\citenamefont {Kumaran}\ and\ \citenamefont
  {Övgün}(2022)}]{Kumaran_2022}%
  \BibitemOpen
  \bibfield  {author} {\bibinfo {author} {\bibfnamefont {Yashmitha}\
  \bibnamefont {Kumaran}}\ and\ \bibinfo {author} {\bibfnamefont {Ali}\
  \bibnamefont {Övgün}},\ }\bibfield  {title} {\enquote {\bibinfo {title}
  {Deflection angle and shadow of the reissner{\textendash}nordström black
  hole with higher-order magnetic correction in einstein-nonlinear-maxwell
  fields},}\ }\href {\doibase 10.3390/sym14102054} {\bibfield  {journal}
  {\bibinfo  {journal} {Symmetry}\ }\textbf {\bibinfo {volume} {14}},\ \bibinfo
  {pages} {2054} (\bibinfo {year} {2022})}\BibitemShut {NoStop}%
\bibitem [{\citenamefont {Pantig}\ \emph {et~al.}(2022)\citenamefont {Pantig},
  \citenamefont {Mastrototaro}, \citenamefont {Lambiase},\ and\ \citenamefont
  {\"Ovg\"un}}]{Pantig:2022gih}%
  \BibitemOpen
  \bibfield  {author} {\bibinfo {author} {\bibfnamefont {Reggie~C.}\
  \bibnamefont {Pantig}}, \bibinfo {author} {\bibfnamefont {Leonardo}\
  \bibnamefont {Mastrototaro}}, \bibinfo {author} {\bibfnamefont {Gaetano}\
  \bibnamefont {Lambiase}}, \ and\ \bibinfo {author} {\bibfnamefont {Ali}\
  \bibnamefont {\"Ovg\"un}},\ }\bibfield  {title} {\enquote {\bibinfo {title}
  {{Shadow, lensing, quasinormal modes, greybody bounds and neutrino
  propagation by dyonic ModMax black holes}},}\ }\href {\doibase
  10.1140/epjc/s10052-022-11125-y} {\bibfield  {journal} {\bibinfo  {journal}
  {Eur. Phys. J. C}\ }\textbf {\bibinfo {volume} {82}},\ \bibinfo {pages}
  {1155} (\bibinfo {year} {2022})},\ \Eprint {http://arxiv.org/abs/2208.06664}
  {arXiv:2208.06664 [gr-qc]} \BibitemShut {NoStop}%
\bibitem [{\citenamefont {Javed}\ \emph {et~al.}(2022)\citenamefont {Javed},
  \citenamefont {Aqib},\ and\ \citenamefont {\"Ovg\"un}}]{Javed:2022kzf}%
  \BibitemOpen
  \bibfield  {author} {\bibinfo {author} {\bibfnamefont {Wajiha}\ \bibnamefont
  {Javed}}, \bibinfo {author} {\bibfnamefont {Muhammad}\ \bibnamefont {Aqib}},
  \ and\ \bibinfo {author} {\bibfnamefont {Ali}\ \bibnamefont {\"Ovg\"un}},\
  }\bibfield  {title} {\enquote {\bibinfo {title} {{Effect of the magnetic
  charge on weak deflection angle and greybody bound of the black hole in
  Einstein-Gauss-Bonnet gravity}},}\ }\href {\doibase
  10.1016/j.physletb.2022.137114} {\bibfield  {journal} {\bibinfo  {journal}
  {Phys. Lett. B}\ }\textbf {\bibinfo {volume} {829}},\ \bibinfo {pages}
  {137114} (\bibinfo {year} {2022})},\ \Eprint
  {http://arxiv.org/abs/2204.07864} {arXiv:2204.07864 [gr-qc]} \BibitemShut
  {NoStop}%
\bibitem [{\citenamefont {Kuang}\ \emph {et~al.}(2018)\citenamefont {Kuang},
  \citenamefont {Liu},\ and\ \citenamefont {\"Ovg\"un}}]{Kuang:2018goo}%
  \BibitemOpen
  \bibfield  {author} {\bibinfo {author} {\bibfnamefont {Xiao-Mei}\
  \bibnamefont {Kuang}}, \bibinfo {author} {\bibfnamefont {Bo}~\bibnamefont
  {Liu}}, \ and\ \bibinfo {author} {\bibfnamefont {Ali}\ \bibnamefont
  {\"Ovg\"un}},\ }\bibfield  {title} {\enquote {\bibinfo {title} {{Nonlinear
  electrodynamics AdS black hole and related phenomena in the extended
  thermodynamics}},}\ }\href {\doibase 10.1140/epjc/s10052-018-6320-0}
  {\bibfield  {journal} {\bibinfo  {journal} {Eur. Phys. J. C}\ }\textbf
  {\bibinfo {volume} {78}},\ \bibinfo {pages} {840} (\bibinfo {year} {2018})},\
  \Eprint {http://arxiv.org/abs/1807.10447} {arXiv:1807.10447 [gr-qc]}
  \BibitemShut {NoStop}%
\bibitem [{\citenamefont {Javed}\ \emph {et~al.}(2019)\citenamefont {Javed},
  \citenamefont {Abbas},\ and\ \citenamefont {\"Ovg\"un}}]{Javed:2019kon}%
  \BibitemOpen
  \bibfield  {author} {\bibinfo {author} {\bibfnamefont {Wajiha}\ \bibnamefont
  {Javed}}, \bibinfo {author} {\bibfnamefont {Jameela}\ \bibnamefont {Abbas}},
  \ and\ \bibinfo {author} {\bibfnamefont {Ali}\ \bibnamefont {\"Ovg\"un}},\
  }\bibfield  {title} {\enquote {\bibinfo {title} {{Deflection angle of photon
  from magnetized black hole and effect of nonlinear electrodynamics}},}\
  }\href {\doibase 10.1140/epjc/s10052-019-7208-3} {\bibfield  {journal}
  {\bibinfo  {journal} {Eur. Phys. J. C}\ }\textbf {\bibinfo {volume} {79}},\
  \bibinfo {pages} {694} (\bibinfo {year} {2019})},\ \Eprint
  {http://arxiv.org/abs/1908.09632} {arXiv:1908.09632 [physics.gen-ph]}
  \BibitemShut {NoStop}%
\bibitem [{\citenamefont {Javed}\ \emph {et~al.}(2020)\citenamefont {Javed},
  \citenamefont {Hamza},\ and\ \citenamefont {\"Ovg\"un}}]{Javed:2020lsg}%
  \BibitemOpen
  \bibfield  {author} {\bibinfo {author} {\bibfnamefont {Wajiha}\ \bibnamefont
  {Javed}}, \bibinfo {author} {\bibfnamefont {Ali}\ \bibnamefont {Hamza}}, \
  and\ \bibinfo {author} {\bibfnamefont {Ali}\ \bibnamefont {\"Ovg\"un}},\
  }\bibfield  {title} {\enquote {\bibinfo {title} {{Effect of nonlinear
  electrodynamics on the weak field deflection angle by a black hole}},}\
  }\href {\doibase 10.20944/preprints201911.0142.v1} {\bibfield  {journal}
  {\bibinfo  {journal} {Phys. Rev. D}\ }\textbf {\bibinfo {volume} {101}},\
  \bibinfo {pages} {103521} (\bibinfo {year} {2020})},\ \Eprint
  {http://arxiv.org/abs/2005.09464} {arXiv:2005.09464 [gr-qc]} \BibitemShut
  {NoStop}%
\bibitem [{\citenamefont {El~Moumni}\ \emph {et~al.}(2022)\citenamefont
  {El~Moumni}, \citenamefont {Masmar},\ and\ \citenamefont
  {\"Ovg\"un}}]{ElMoumni:2020wrf}%
  \BibitemOpen
  \bibfield  {author} {\bibinfo {author} {\bibfnamefont {H.}~\bibnamefont
  {El~Moumni}}, \bibinfo {author} {\bibfnamefont {K.}~\bibnamefont {Masmar}}, \
  and\ \bibinfo {author} {\bibfnamefont {Ali}\ \bibnamefont {\"Ovg\"un}},\
  }\bibfield  {title} {\enquote {\bibinfo {title} {{Weak deflection angle of
  light in two classes of black holes in nonlinear electrodynamics via
  Gauss\textendash{}Bonnet theorem}},}\ }\href {\doibase
  10.1142/S0219887822500943} {\bibfield  {journal} {\bibinfo  {journal} {Int.
  J. Geom. Meth. Mod. Phys.}\ }\textbf {\bibinfo {volume} {19}},\ \bibinfo
  {pages} {2250094} (\bibinfo {year} {2022})},\ \Eprint
  {http://arxiv.org/abs/2008.06711} {arXiv:2008.06711 [gr-qc]} \BibitemShut
  {NoStop}%
\bibitem [{\citenamefont {Uniyal}\ \emph {et~al.}(2023)\citenamefont {Uniyal},
  \citenamefont {Pantig},\ and\ \citenamefont {\"Ovg\"un}}]{Uniyal:2022vdu}%
  \BibitemOpen
  \bibfield  {author} {\bibinfo {author} {\bibfnamefont {Akhil}\ \bibnamefont
  {Uniyal}}, \bibinfo {author} {\bibfnamefont {Reggie~C.}\ \bibnamefont
  {Pantig}}, \ and\ \bibinfo {author} {\bibfnamefont {Ali}\ \bibnamefont
  {\"Ovg\"un}},\ }\bibfield  {title} {\enquote {\bibinfo {title} {{Probing a
  non-linear electrodynamics black hole with thin accretion disk, shadow, and
  deflection angle with M87* and Sgr A* from EHT}},}\ }\href {\doibase
  10.1016/j.dark.2023.101178} {\bibfield  {journal} {\bibinfo  {journal} {Phys.
  Dark Univ.}\ }\textbf {\bibinfo {volume} {40}},\ \bibinfo {pages} {101178}
  (\bibinfo {year} {2023})},\ \Eprint {http://arxiv.org/abs/2205.11072}
  {arXiv:2205.11072 [gr-qc]} \BibitemShut {NoStop}%
\bibitem [{\citenamefont {Jusufi}\ \emph {et~al.}(2018)\citenamefont {Jusufi},
  \citenamefont {\"Ovg\"un}, \citenamefont {Saavedra}, \citenamefont
  {V\'asquez},\ and\ \citenamefont {Gonz\'alez}}]{Jusufi:2018jof}%
  \BibitemOpen
  \bibfield  {author} {\bibinfo {author} {\bibfnamefont {Kimet}\ \bibnamefont
  {Jusufi}}, \bibinfo {author} {\bibfnamefont {Ali}\ \bibnamefont {\"Ovg\"un}},
  \bibinfo {author} {\bibfnamefont {Joel}\ \bibnamefont {Saavedra}}, \bibinfo
  {author} {\bibfnamefont {Yerko}\ \bibnamefont {V\'asquez}}, \ and\ \bibinfo
  {author} {\bibfnamefont {P.~A.}\ \bibnamefont {Gonz\'alez}},\ }\bibfield
  {title} {\enquote {\bibinfo {title} {{Deflection of light by rotating regular
  black holes using the Gauss-Bonnet theorem}},}\ }\href {\doibase
  10.1103/PhysRevD.97.124024} {\bibfield  {journal} {\bibinfo  {journal} {Phys.
  Rev. D}\ }\textbf {\bibinfo {volume} {97}},\ \bibinfo {pages} {124024}
  (\bibinfo {year} {2018})},\ \Eprint {http://arxiv.org/abs/1804.00643}
  {arXiv:1804.00643 [gr-qc]} \BibitemShut {NoStop}%
\bibitem [{\citenamefont {Halilsoy}\ \emph {et~al.}(2014)\citenamefont
  {Halilsoy}, \citenamefont {Ovgun},\ and\ \citenamefont
  {Mazharimousavi}}]{Halilsoy:2013iza}%
  \BibitemOpen
  \bibfield  {author} {\bibinfo {author} {\bibfnamefont {M.}~\bibnamefont
  {Halilsoy}}, \bibinfo {author} {\bibfnamefont {A.}~\bibnamefont {Ovgun}}, \
  and\ \bibinfo {author} {\bibfnamefont {S.~Habib}\ \bibnamefont
  {Mazharimousavi}},\ }\bibfield  {title} {\enquote {\bibinfo {title}
  {{Thin-shell wormholes from the regular Hayward black hole}},}\ }\href
  {\doibase 10.1140/epjc/s10052-014-2796-4} {\bibfield  {journal} {\bibinfo
  {journal} {Eur. Phys. J. C}\ }\textbf {\bibinfo {volume} {74}},\ \bibinfo
  {pages} {2796} (\bibinfo {year} {2014})},\ \Eprint
  {http://arxiv.org/abs/1312.6665} {arXiv:1312.6665 [gr-qc]} \BibitemShut
  {NoStop}%
\bibitem [{\citenamefont {Allahyari}\ \emph {et~al.}(2020)\citenamefont
  {Allahyari}, \citenamefont {Khodadi}, \citenamefont {Vagnozzi},\ and\
  \citenamefont {Mota}}]{Allahyari:2019jqz}%
  \BibitemOpen
  \bibfield  {author} {\bibinfo {author} {\bibfnamefont {Alireza}\ \bibnamefont
  {Allahyari}}, \bibinfo {author} {\bibfnamefont {Mohsen}\ \bibnamefont
  {Khodadi}}, \bibinfo {author} {\bibfnamefont {Sunny}\ \bibnamefont
  {Vagnozzi}}, \ and\ \bibinfo {author} {\bibfnamefont {David~F.}\ \bibnamefont
  {Mota}},\ }\bibfield  {title} {\enquote {\bibinfo {title} {{Magnetically
  charged black holes from non-linear electrodynamics and the Event Horizon
  Telescope}},}\ }\href {\doibase 10.1088/1475-7516/2020/02/003} {\bibfield
  {journal} {\bibinfo  {journal} {JCAP}\ }\textbf {\bibinfo {volume} {02}},\
  \bibinfo {pages} {003} (\bibinfo {year} {2020})},\ \Eprint
  {http://arxiv.org/abs/1912.08231} {arXiv:1912.08231 [gr-qc]} \BibitemShut
  {NoStop}%
\bibitem [{\citenamefont {Vagnozzi}\ \emph {et~al.}(2022)\citenamefont
  {Vagnozzi} \emph {et~al.}}]{Vagnozzi:2022moj}%
  \BibitemOpen
  \bibfield  {author} {\bibinfo {author} {\bibfnamefont {Sunny}\ \bibnamefont
  {Vagnozzi}} \emph {et~al.},\ }\bibfield  {title} {\enquote {\bibinfo {title}
  {{Horizon-scale tests of gravity theories and fundamental physics from the
  Event Horizon Telescope image of Sagittarius A$^*$}},}\ }\href@noop {} {\
  (\bibinfo {year} {2022})},\ \Eprint {http://arxiv.org/abs/2205.07787}
  {arXiv:2205.07787 [gr-qc]} \BibitemShut {NoStop}%
\bibitem [{\citenamefont {Garcia-Salcedo}\ and\ \citenamefont
  {Breton}(2000)}]{GARC_A_SALCEDO_2000}%
  \BibitemOpen
  \bibfield  {author} {\bibinfo {author} {\bibfnamefont {Ricardo}\ \bibnamefont
  {Garcia-Salcedo}}\ and\ \bibinfo {author} {\bibfnamefont {Nora}\ \bibnamefont
  {Breton}},\ }\bibfield  {title} {\enquote {\bibinfo {title} {{Born-Infeld
  cosmologies}},}\ }\href {\doibase 10.1016/S0217-751X(00)00216-9} {\bibfield
  {journal} {\bibinfo  {journal} {Int. J. Mod. Phys. A}\ }\textbf {\bibinfo
  {volume} {15}},\ \bibinfo {pages} {4341--4354} (\bibinfo {year} {2000})},\
  \Eprint {http://arxiv.org/abs/gr-qc/0004017} {arXiv:gr-qc/0004017}
  \BibitemShut {NoStop}%
\bibitem [{\citenamefont {Benaoum}\ \emph {et~al.}(2022)\citenamefont
  {Benaoum}, \citenamefont {Leon}, \citenamefont {Ovgun},\ and\ \citenamefont
  {Quevedo}}]{Ovgun:2022}%
  \BibitemOpen
  \bibfield  {author} {\bibinfo {author} {\bibfnamefont {H.~B.}\ \bibnamefont
  {Benaoum}}, \bibinfo {author} {\bibfnamefont {Genly}\ \bibnamefont {Leon}},
  \bibinfo {author} {\bibfnamefont {A.}~\bibnamefont {Ovgun}}, \ and\ \bibinfo
  {author} {\bibfnamefont {H.}~\bibnamefont {Quevedo}},\ }\bibfield  {title}
  {\enquote {\bibinfo {title} {{Inflation Driven by Non-Linear
  Electrodynamics}},}\ }\href@noop {} {\  (\bibinfo {year} {2022})},\ \Eprint
  {http://arxiv.org/abs/2206.13157} {arXiv:2206.13157 [gr-qc]} \BibitemShut
  {NoStop}%
\bibitem [{\citenamefont {\"Ovg\"un}(2017)}]{Ovgun:2016oit}%
  \BibitemOpen
  \bibfield  {author} {\bibinfo {author} {\bibfnamefont {A.}~\bibnamefont
  {\"Ovg\"un}},\ }\bibfield  {title} {\enquote {\bibinfo {title} {{Inflation
  and Acceleration of the Universe by Nonlinear Magnetic Monopole Fields}},}\
  }\href {\doibase 10.1140/epjc/s10052-017-4673-4} {\bibfield  {journal}
  {\bibinfo  {journal} {Eur. Phys. J. C}\ }\textbf {\bibinfo {volume} {77}},\
  \bibinfo {pages} {105} (\bibinfo {year} {2017})},\ \Eprint
  {http://arxiv.org/abs/1604.01837} {arXiv:1604.01837 [gr-qc]} \BibitemShut
  {NoStop}%
\bibitem [{\citenamefont {\"Ovg\"un}\ \emph {et~al.}(2018)\citenamefont
  {\"Ovg\"un}, \citenamefont {Leon}, \citenamefont {Maga\~na},\ and\
  \citenamefont {Jusufi}}]{Ovgun:2017iwg}%
  \BibitemOpen
  \bibfield  {author} {\bibinfo {author} {\bibfnamefont {Ali}\ \bibnamefont
  {\"Ovg\"un}}, \bibinfo {author} {\bibfnamefont {Genly}\ \bibnamefont {Leon}},
  \bibinfo {author} {\bibfnamefont {Juan}\ \bibnamefont {Maga\~na}}, \ and\
  \bibinfo {author} {\bibfnamefont {Kimet}\ \bibnamefont {Jusufi}},\ }\bibfield
   {title} {\enquote {\bibinfo {title} {{Falsifying cosmological models based
  on a non-linear electrodynamics}},}\ }\href {\doibase
  10.1140/epjc/s10052-018-5936-4} {\bibfield  {journal} {\bibinfo  {journal}
  {Eur. Phys. J. C}\ }\textbf {\bibinfo {volume} {78}},\ \bibinfo {pages} {462}
  (\bibinfo {year} {2018})},\ \Eprint {http://arxiv.org/abs/1709.09794}
  {arXiv:1709.09794 [gr-qc]} \BibitemShut {NoStop}%
\bibitem [{\citenamefont {Otalora}\ \emph {et~al.}(2018)\citenamefont
  {Otalora}, \citenamefont {\"Ovg\"un}, \citenamefont {Saavedra},\ and\
  \citenamefont {Videla}}]{Otalora:2018bso}%
  \BibitemOpen
  \bibfield  {author} {\bibinfo {author} {\bibfnamefont {Giovanni}\
  \bibnamefont {Otalora}}, \bibinfo {author} {\bibfnamefont {Ali}\ \bibnamefont
  {\"Ovg\"un}}, \bibinfo {author} {\bibfnamefont {Joel}\ \bibnamefont
  {Saavedra}}, \ and\ \bibinfo {author} {\bibfnamefont {Nelson}\ \bibnamefont
  {Videla}},\ }\bibfield  {title} {\enquote {\bibinfo {title} {{Inflation from
  a nonlinear magnetic monopole field nonminimally coupled to curvature}},}\
  }\href {\doibase 10.1088/1475-7516/2018/06/003} {\bibfield  {journal}
  {\bibinfo  {journal} {JCAP}\ }\textbf {\bibinfo {volume} {06}},\ \bibinfo
  {pages} {003} (\bibinfo {year} {2018})},\ \Eprint
  {http://arxiv.org/abs/1803.11358} {arXiv:1803.11358 [gr-qc]} \BibitemShut
  {NoStop}%
\bibitem [{\citenamefont {Joseph}\ and\ \citenamefont
  {\"Ovg\"un}(2022)}]{OvgunExpo_2021}%
  \BibitemOpen
  \bibfield  {author} {\bibinfo {author} {\bibfnamefont {Gabriel~W.}\
  \bibnamefont {Joseph}}\ and\ \bibinfo {author} {\bibfnamefont {Ali}\
  \bibnamefont {\"Ovg\"un}},\ }\bibfield  {title} {\enquote {\bibinfo {title}
  {{Cosmology with variable G and nonlinear electrodynamics}},}\ }\href
  {\doibase 10.1007/s12648-021-02110-4} {\bibfield  {journal} {\bibinfo
  {journal} {Indian J. Phys.}\ }\textbf {\bibinfo {volume} {96}},\ \bibinfo
  {pages} {1861--1866} (\bibinfo {year} {2022})},\ \Eprint
  {http://arxiv.org/abs/2104.11066} {arXiv:2104.11066 [gr-qc]} \BibitemShut
  {NoStop}%
\bibitem [{\citenamefont {Benaoum}\ and\ \citenamefont
  {Ovgun}(2021)}]{matter_antimatter}%
  \BibitemOpen
  \bibfield  {author} {\bibinfo {author} {\bibfnamefont {H.~B.}\ \bibnamefont
  {Benaoum}}\ and\ \bibinfo {author} {\bibfnamefont {A.}~\bibnamefont
  {Ovgun}},\ }\bibfield  {title} {\enquote {\bibinfo {title}
  {{Matter-antimatter asymmetry induced by non-linear electrodynamics}},}\
  }\href {\doibase 10.1088/1361-6382/abfd90} {\bibfield  {journal} {\bibinfo
  {journal} {Class. Quant. Grav.}\ }\textbf {\bibinfo {volume} {38}},\ \bibinfo
  {pages} {135019} (\bibinfo {year} {2021})},\ \Eprint
  {http://arxiv.org/abs/2105.07695} {arXiv:2105.07695 [gr-qc]} \BibitemShut
  {NoStop}%
\bibitem [{\citenamefont {Kunze}(2013)}]{Kunze:2013kza}%
  \BibitemOpen
  \bibfield  {author} {\bibinfo {author} {\bibfnamefont {Kerstin~E.}\
  \bibnamefont {Kunze}},\ }\bibfield  {title} {\enquote {\bibinfo {title}
  {{Cosmological Magnetic Fields}},}\ }\href {\doibase
  10.1088/0741-3335/55/12/124026} {\bibfield  {journal} {\bibinfo  {journal}
  {Plasma Phys. Control. Fusion}\ }\textbf {\bibinfo {volume} {55}},\ \bibinfo
  {pages} {124026} (\bibinfo {year} {2013})},\ \Eprint
  {http://arxiv.org/abs/1307.2153} {arXiv:1307.2153 [astro-ph.CO]} \BibitemShut
  {NoStop}%
\bibitem [{\citenamefont {Durrer}\ and\ \citenamefont
  {Neronov}(2013)}]{Durrer_2013}%
  \BibitemOpen
  \bibfield  {author} {\bibinfo {author} {\bibfnamefont {Ruth}\ \bibnamefont
  {Durrer}}\ and\ \bibinfo {author} {\bibfnamefont {Andrii}\ \bibnamefont
  {Neronov}},\ }\bibfield  {title} {\enquote {\bibinfo {title} {Cosmological
  magnetic fields: their generation, evolution and observation},}\ }\href
  {\doibase 10.1007/s00159-013-0062-7} {\bibfield  {journal} {\bibinfo
  {journal} {The Astronomy and Astrophysics Review}\ }\textbf {\bibinfo
  {volume} {21}} (\bibinfo {year} {2013}),\
  10.1007/s00159-013-0062-7}\BibitemShut {NoStop}%
\bibitem [{\citenamefont {Kunze}(2008)}]{kunz2008}%
  \BibitemOpen
  \bibfield  {author} {\bibinfo {author} {\bibfnamefont {Kerstin~E.}\
  \bibnamefont {Kunze}},\ }\bibfield  {title} {\enquote {\bibinfo {title}
  {Primordial magnetic fields and nonlinear electrodynamics},}\ }\href
  {\doibase 10.1103/PhysRevD.77.023530} {\bibfield  {journal} {\bibinfo
  {journal} {Phys. Rev. D}\ }\textbf {\bibinfo {volume} {77}},\ \bibinfo
  {pages} {023530} (\bibinfo {year} {2008})}\BibitemShut {NoStop}%
\bibitem [{\citenamefont {Campanelli}\ \emph {et~al.}(2008)\citenamefont
  {Campanelli}, \citenamefont {Cea}, \citenamefont {Fogli},\ and\ \citenamefont
  {Tedesco}}]{Campanelli2008}%
  \BibitemOpen
  \bibfield  {author} {\bibinfo {author} {\bibfnamefont {L.}~\bibnamefont
  {Campanelli}}, \bibinfo {author} {\bibfnamefont {P.}~\bibnamefont {Cea}},
  \bibinfo {author} {\bibfnamefont {G.~L.}\ \bibnamefont {Fogli}}, \ and\
  \bibinfo {author} {\bibfnamefont {L.}~\bibnamefont {Tedesco}},\ }\bibfield
  {title} {\enquote {\bibinfo {title} {Inflation-produced magnetic fields in
  nonlinear electrodynamics},}\ }\href {\doibase 10.1103/PhysRevD.77.043001}
  {\bibfield  {journal} {\bibinfo  {journal} {Phys. Rev. D}\ }\textbf {\bibinfo
  {volume} {77}},\ \bibinfo {pages} {043001} (\bibinfo {year}
  {2008})}\BibitemShut {NoStop}%
\bibitem [{\citenamefont {Gasperini}\ \emph
  {et~al.}(1995{\natexlab{a}})\citenamefont {Gasperini}, \citenamefont
  {Giovannini},\ and\ \citenamefont {Veneziano}}]{10}%
  \BibitemOpen
  \bibfield  {author} {\bibinfo {author} {\bibfnamefont {M.}~\bibnamefont
  {Gasperini}}, \bibinfo {author} {\bibfnamefont {M.}~\bibnamefont
  {Giovannini}}, \ and\ \bibinfo {author} {\bibfnamefont {G.}~\bibnamefont
  {Veneziano}},\ }\bibfield  {title} {\enquote {\bibinfo {title} {Primordial
  magnetic fields from string cosmology},}\ }\href {\doibase
  10.1103/PhysRevLett.75.3796} {\bibfield  {journal} {\bibinfo  {journal}
  {Phys. Rev. Lett.}\ }\textbf {\bibinfo {volume} {75}},\ \bibinfo {pages}
  {3796--3799} (\bibinfo {year} {1995}{\natexlab{a}})}\BibitemShut {NoStop}%
\bibitem [{\citenamefont {Bamba}\ \emph {et~al.}(2008)\citenamefont {Bamba},
  \citenamefont {Geng},\ and\ \citenamefont {Ho}}]{11}%
  \BibitemOpen
  \bibfield  {author} {\bibinfo {author} {\bibfnamefont {K}~\bibnamefont
  {Bamba}}, \bibinfo {author} {\bibfnamefont {C~Q}\ \bibnamefont {Geng}}, \
  and\ \bibinfo {author} {\bibfnamefont {S~H}\ \bibnamefont {Ho}},\ }\bibfield
  {title} {\enquote {\bibinfo {title} {Large-scale magnetic fields from
  inflation due to chern{\textendash}simons-like effective interaction},}\
  }\href {\doibase 10.1088/1475-7516/2008/11/013} {\bibfield  {journal}
  {\bibinfo  {journal} {Journal of Cosmology and Astroparticle Physics}\
  }\textbf {\bibinfo {volume} {2008}},\ \bibinfo {pages} {013} (\bibinfo {year}
  {2008})}\BibitemShut {NoStop}%
\bibitem [{\citenamefont {Turner}\ and\ \citenamefont {Widrow}(1988)}]{12}%
  \BibitemOpen
  \bibfield  {author} {\bibinfo {author} {\bibfnamefont {Michael~S.}\
  \bibnamefont {Turner}}\ and\ \bibinfo {author} {\bibfnamefont {Lawrence~M.}\
  \bibnamefont {Widrow}},\ }\bibfield  {title} {\enquote {\bibinfo {title}
  {Inflation-produced, large-scale magnetic fields},}\ }\href {\doibase
  10.1103/PhysRevD.37.2743} {\bibfield  {journal} {\bibinfo  {journal} {Phys.
  Rev. D}\ }\textbf {\bibinfo {volume} {37}},\ \bibinfo {pages} {2743--2754}
  (\bibinfo {year} {1988})}\BibitemShut {NoStop}%
\bibitem [{\citenamefont {Opher}\ and\ \citenamefont {Wichoski}(1997)}]{14}%
  \BibitemOpen
  \bibfield  {author} {\bibinfo {author} {\bibfnamefont {Reuven}\ \bibnamefont
  {Opher}}\ and\ \bibinfo {author} {\bibfnamefont {Ubirajara~F.}\ \bibnamefont
  {Wichoski}},\ }\bibfield  {title} {\enquote {\bibinfo {title} {Origin of
  magnetic fields in the universe due to nonminimal
  gravitational-electromagnetic coupling},}\ }\href {\doibase
  10.1103/physrevlett.78.787} {\bibfield  {journal} {\bibinfo  {journal}
  {Physical Review Letters}\ }\textbf {\bibinfo {volume} {78}},\ \bibinfo
  {pages} {787--790} (\bibinfo {year} {1997})}\BibitemShut {NoStop}%
\bibitem [{\citenamefont {Bamba}\ \emph {et~al.}(2010)\citenamefont {Bamba},
  \citenamefont {Geng},\ and\ \citenamefont {Lee}}]{15}%
  \BibitemOpen
  \bibfield  {author} {\bibinfo {author} {\bibfnamefont {Kazuharu}\
  \bibnamefont {Bamba}}, \bibinfo {author} {\bibfnamefont {Chao-Qiang}\
  \bibnamefont {Geng}}, \ and\ \bibinfo {author} {\bibfnamefont {Chung-Chi}\
  \bibnamefont {Lee}},\ }\bibfield  {title} {\enquote {\bibinfo {title}
  {Cosmological evolution in exponential gravity},}\ }\href {\doibase
  10.1088/1475-7516/2010/08/021} {\bibfield  {journal} {\bibinfo  {journal}
  {Journal of Cosmology and Astroparticle Physics}\ }\textbf {\bibinfo {volume}
  {2010}},\ \bibinfo {pages} {021--021} (\bibinfo {year} {2010})}\BibitemShut
  {NoStop}%
\bibitem [{\citenamefont {Mazzitelli}\ and\ \citenamefont
  {Spedalieri}(1995)}]{16}%
  \BibitemOpen
  \bibfield  {author} {\bibinfo {author} {\bibfnamefont {Francisco~D.}\
  \bibnamefont {Mazzitelli}}\ and\ \bibinfo {author} {\bibfnamefont
  {Federico~M.}\ \bibnamefont {Spedalieri}},\ }\bibfield  {title} {\enquote
  {\bibinfo {title} {Scalar electrodynamics and primordial magnetic fields},}\
  }\href {\doibase 10.1103/physrevd.52.6694} {\bibfield  {journal} {\bibinfo
  {journal} {Physical Review D}\ }\textbf {\bibinfo {volume} {52}},\ \bibinfo
  {pages} {6694--6699} (\bibinfo {year} {1995})}\BibitemShut {NoStop}%
\bibitem [{\citenamefont {Matarrese}\ \emph {et~al.}(2005)\citenamefont
  {Matarrese}, \citenamefont {Mollerach}, \citenamefont {Notari},\ and\
  \citenamefont {Riotto}}]{17}%
  \BibitemOpen
  \bibfield  {author} {\bibinfo {author} {\bibfnamefont {S.}~\bibnamefont
  {Matarrese}}, \bibinfo {author} {\bibfnamefont {S.}~\bibnamefont
  {Mollerach}}, \bibinfo {author} {\bibfnamefont {A.}~\bibnamefont {Notari}}, \
  and\ \bibinfo {author} {\bibfnamefont {A.}~\bibnamefont {Riotto}},\
  }\bibfield  {title} {\enquote {\bibinfo {title} {Large-scale magnetic fields
  from density perturbations},}\ }\href {\doibase 10.1103/PhysRevD.71.043502}
  {\bibfield  {journal} {\bibinfo  {journal} {Phys. Rev. D}\ }\textbf {\bibinfo
  {volume} {71}},\ \bibinfo {pages} {043502} (\bibinfo {year}
  {2005})}\BibitemShut {NoStop}%
\bibitem [{\citenamefont {Tsagas}\ \emph {et~al.}(2003)\citenamefont {Tsagas},
  \citenamefont {Dunsby},\ and\ \citenamefont {Marklund}}]{18}%
  \BibitemOpen
  \bibfield  {author} {\bibinfo {author} {\bibfnamefont {Christos~G}\
  \bibnamefont {Tsagas}}, \bibinfo {author} {\bibfnamefont {Peter~K.S}\
  \bibnamefont {Dunsby}}, \ and\ \bibinfo {author} {\bibfnamefont {Mattias}\
  \bibnamefont {Marklund}},\ }\bibfield  {title} {\enquote {\bibinfo {title}
  {Gravitational wave amplification of seed magnetic fields},}\ }\href
  {\doibase 10.1016/s0370-2693(03)00415-5} {\bibfield  {journal} {\bibinfo
  {journal} {Physics Letters B}\ }\textbf {\bibinfo {volume} {561}},\ \bibinfo
  {pages} {17--25} (\bibinfo {year} {2003})}\BibitemShut {NoStop}%
\bibitem [{\citenamefont {Bertolami}\ and\ \citenamefont {Mota}(1999)}]{19}%
  \BibitemOpen
  \bibfield  {author} {\bibinfo {author} {\bibfnamefont {O.}~\bibnamefont
  {Bertolami}}\ and\ \bibinfo {author} {\bibfnamefont {D.F.}\ \bibnamefont
  {Mota}},\ }\bibfield  {title} {\enquote {\bibinfo {title} {Primordial
  magnetic fields via spontaneous breaking of lorentz invariance},}\ }\href
  {\doibase 10.1016/s0370-2693(99)00418-9} {\bibfield  {journal} {\bibinfo
  {journal} {Physics Letters B}\ }\textbf {\bibinfo {volume} {455}},\ \bibinfo
  {pages} {96--103} (\bibinfo {year} {1999})}\BibitemShut {NoStop}%
\bibitem [{\citenamefont {Vachaspati}\ and\ \citenamefont
  {Vilenkin}(1991)}]{20}%
  \BibitemOpen
  \bibfield  {author} {\bibinfo {author} {\bibfnamefont {Tanmay}\ \bibnamefont
  {Vachaspati}}\ and\ \bibinfo {author} {\bibfnamefont {Alexander}\
  \bibnamefont {Vilenkin}},\ }\bibfield  {title} {\enquote {\bibinfo {title}
  {Large-scale structure from wiggly cosmic strings},}\ }\href {\doibase
  10.1103/PhysRevLett.67.1057} {\bibfield  {journal} {\bibinfo  {journal}
  {Phys. Rev. Lett.}\ }\textbf {\bibinfo {volume} {67}},\ \bibinfo {pages}
  {1057--1061} (\bibinfo {year} {1991})}\BibitemShut {NoStop}%
\bibitem [{\citenamefont {Joyce}\ and\ \citenamefont
  {Shaposhnikov}(1997)}]{21}%
  \BibitemOpen
  \bibfield  {author} {\bibinfo {author} {\bibfnamefont {M.}~\bibnamefont
  {Joyce}}\ and\ \bibinfo {author} {\bibfnamefont {M.}~\bibnamefont
  {Shaposhnikov}},\ }\bibfield  {title} {\enquote {\bibinfo {title} {Primordial
  magnetic fields, right electrons, and the abelian anomaly},}\ }\href
  {\doibase 10.1103/PhysRevLett.79.1193} {\bibfield  {journal} {\bibinfo
  {journal} {Phys. Rev. Lett.}\ }\textbf {\bibinfo {volume} {79}},\ \bibinfo
  {pages} {1193--1196} (\bibinfo {year} {1997})}\BibitemShut {NoStop}%
\bibitem [{\citenamefont {Dolgov}\ and\ \citenamefont {Silk}(1993)}]{22}%
  \BibitemOpen
  \bibfield  {author} {\bibinfo {author} {\bibfnamefont {Alexandre}\
  \bibnamefont {Dolgov}}\ and\ \bibinfo {author} {\bibfnamefont {Joseph}\
  \bibnamefont {Silk}},\ }\bibfield  {title} {\enquote {\bibinfo {title}
  {Electric charge asymmetry of the universe and magnetic field generation},}\
  }\href {\doibase 10.1103/PhysRevD.47.3144} {\bibfield  {journal} {\bibinfo
  {journal} {Phys. Rev. D}\ }\textbf {\bibinfo {volume} {47}},\ \bibinfo
  {pages} {3144--3150} (\bibinfo {year} {1993})}\BibitemShut {NoStop}%
\bibitem [{\citenamefont {Dolgov}(1993)}]{23}%
  \BibitemOpen
  \bibfield  {author} {\bibinfo {author} {\bibfnamefont {A.~D.}\ \bibnamefont
  {Dolgov}},\ }\bibfield  {title} {\enquote {\bibinfo {title} {Breaking of
  conformal invariance and electromagnetic field generation in the universe},}\
  }\href {\doibase 10.1103/PhysRevD.48.2499} {\bibfield  {journal} {\bibinfo
  {journal} {Phys. Rev. D}\ }\textbf {\bibinfo {volume} {48}},\ \bibinfo
  {pages} {2499--2501} (\bibinfo {year} {1993})}\BibitemShut {NoStop}%
\bibitem [{\citenamefont {Semikoz}\ and\ \citenamefont {Sokoloff}(2004)}]{24}%
  \BibitemOpen
  \bibfield  {author} {\bibinfo {author} {\bibfnamefont {V.~B.}\ \bibnamefont
  {Semikoz}}\ and\ \bibinfo {author} {\bibfnamefont {D.~D.}\ \bibnamefont
  {Sokoloff}},\ }\bibfield  {title} {\enquote {\bibinfo {title} {Large-scale
  magnetic field generation by $\ensuremath{\alpha}$ effect driven by
  collective neutrino-plasma interaction},}\ }\href {\doibase
  10.1103/PhysRevLett.92.131301} {\bibfield  {journal} {\bibinfo  {journal}
  {Phys. Rev. Lett.}\ }\textbf {\bibinfo {volume} {92}},\ \bibinfo {pages}
  {131301} (\bibinfo {year} {2004})}\BibitemShut {NoStop}%
\bibitem [{\citenamefont {Capozziello}\ \emph {et~al.}(2022)\citenamefont
  {Capozziello}, \citenamefont {Carleo},\ and\ \citenamefont
  {Lambiase}}]{CarleoPMFs}%
  \BibitemOpen
  \bibfield  {author} {\bibinfo {author} {\bibfnamefont {S.}~\bibnamefont
  {Capozziello}}, \bibinfo {author} {\bibfnamefont {A.}~\bibnamefont {Carleo}},
  \ and\ \bibinfo {author} {\bibfnamefont {G.}~\bibnamefont {Lambiase}},\
  }\bibfield  {title} {\enquote {\bibinfo {title} {{The amplification of
  cosmological magnetic fields in extended f(T,B) teleparallel gravity}},}\
  }\href {\doibase 10.1088/1475-7516/2022/10/020} {\bibfield  {journal}
  {\bibinfo  {journal} {JCAP}\ }\textbf {\bibinfo {volume} {10}},\ \bibinfo
  {pages} {020} (\bibinfo {year} {2022})},\ \Eprint
  {http://arxiv.org/abs/2208.11186} {arXiv:2208.11186 [gr-qc]} \BibitemShut
  {NoStop}%
\bibitem [{\citenamefont {Cuesta}\ and\ \citenamefont
  {Lambiase}(2009)}]{Lambiase2009}%
  \BibitemOpen
  \bibfield  {author} {\bibinfo {author} {\bibfnamefont {Herman J.~Mosquera}\
  \bibnamefont {Cuesta}}\ and\ \bibinfo {author} {\bibfnamefont {Gaetano}\
  \bibnamefont {Lambiase}},\ }\bibfield  {title} {\enquote {\bibinfo {title}
  {Primordial magnetic fields and gravitational baryogenesis in nonlinear
  electrodynamics},}\ }\href {\doibase 10.1103/physrevd.80.023013} {\bibfield
  {journal} {\bibinfo  {journal} {Physical Review D}\ }\textbf {\bibinfo
  {volume} {80}} (\bibinfo {year} {2009}),\
  10.1103/physrevd.80.023013}\BibitemShut {NoStop}%
\bibitem [{\citenamefont {{Komissarov}}(2005)}]{komissorov2005}%
  \BibitemOpen
  \bibfield  {author} {\bibinfo {author} {\bibfnamefont {S.~S.}\ \bibnamefont
  {{Komissarov}}},\ }\bibfield  {title} {\enquote {\bibinfo {title}
  {{Observations of the Blandford-Znajek process and the magnetohydrodynamic
  Penrose process in computer simulations of black hole magnetospheres}},}\
  }\href {\doibase 10.1111/j.1365-2966.2005.08974.x} {\bibfield  {journal}
  {\bibinfo  {journal} {mnras}\ }\textbf {\bibinfo {volume} {359}},\ \bibinfo
  {pages} {801--808} (\bibinfo {year} {2005})},\ \Eprint
  {http://arxiv.org/abs/astro-ph/0501599} {arXiv:astro-ph/0501599 [astro-ph]}
  \BibitemShut {NoStop}%
\bibitem [{\citenamefont {Blandford}\ and\ \citenamefont
  {Znajek}(1977)}]{BZ-1977ds}%
  \BibitemOpen
  \bibfield  {author} {\bibinfo {author} {\bibfnamefont {R.~D.}\ \bibnamefont
  {Blandford}}\ and\ \bibinfo {author} {\bibfnamefont {R.~L.}\ \bibnamefont
  {Znajek}},\ }\bibfield  {title} {\enquote {\bibinfo {title} {{Electromagnetic
  extractions of energy from Kerr black holes}},}\ }\href {\doibase
  10.1093/mnras/179.3.433} {\bibfield  {journal} {\bibinfo  {journal} {Mon.
  Not. Roy. Astron. Soc.}\ }\textbf {\bibinfo {volume} {179}},\ \bibinfo
  {pages} {433--456} (\bibinfo {year} {1977})}\BibitemShut {NoStop}%
\bibitem [{\citenamefont {Tchekhovskoy}\ \emph {et~al.}(2010)\citenamefont
  {Tchekhovskoy}, \citenamefont {Narayan},\ and\ \citenamefont
  {McKinney}}]{BZ}%
  \BibitemOpen
  \bibfield  {author} {\bibinfo {author} {\bibfnamefont {Alexander}\
  \bibnamefont {Tchekhovskoy}}, \bibinfo {author} {\bibfnamefont {Ramesh}\
  \bibnamefont {Narayan}}, \ and\ \bibinfo {author} {\bibfnamefont
  {Jonathan~C.}\ \bibnamefont {McKinney}},\ }\bibfield  {title} {\enquote
  {\bibinfo {title} {{Black Hole Spin and the Radio Loud/Quiet Dichotomy of
  Active Galactic Nuclei}},}\ }\href {\doibase 10.1088/0004-637X/711/1/50}
  {\bibfield  {journal} {\bibinfo  {journal} {Astrophys. J.}\ }\textbf
  {\bibinfo {volume} {711}},\ \bibinfo {pages} {50--63} (\bibinfo {year}
  {2010})},\ \Eprint {http://arxiv.org/abs/0911.2228} {arXiv:0911.2228
  [astro-ph.HE]} \BibitemShut {NoStop}%
\bibitem [{\citenamefont {Lee}\ \emph {et~al.}(2000)\citenamefont {Lee},
  \citenamefont {Wijers},\ and\ \citenamefont {Brown}}]{BZ1}%
  \BibitemOpen
  \bibfield  {author} {\bibinfo {author} {\bibfnamefont {H.~K.}\ \bibnamefont
  {Lee}}, \bibinfo {author} {\bibfnamefont {R.~A. M.~J.}\ \bibnamefont
  {Wijers}}, \ and\ \bibinfo {author} {\bibfnamefont {G.~E.}\ \bibnamefont
  {Brown}},\ }\bibfield  {title} {\enquote {\bibinfo {title} {{The
  Blandford-Znajek process as a central engine for a gamma-ray burst}},}\
  }\href {\doibase 10.1016/S0370-1573(99)00084-8} {\bibfield  {journal}
  {\bibinfo  {journal} {Phys. Rept.}\ }\textbf {\bibinfo {volume} {325}},\
  \bibinfo {pages} {83--114} (\bibinfo {year} {2000})},\ \Eprint
  {http://arxiv.org/abs/astro-ph/9906213} {arXiv:astro-ph/9906213} \BibitemShut
  {NoStop}%
\bibitem [{\citenamefont {Tchekhovskoy}\ \emph {et~al.}(2008)\citenamefont
  {Tchekhovskoy}, \citenamefont {McKinney},\ and\ \citenamefont
  {Narayan}}]{BZ2}%
  \BibitemOpen
  \bibfield  {author} {\bibinfo {author} {\bibfnamefont {Alexander}\
  \bibnamefont {Tchekhovskoy}}, \bibinfo {author} {\bibfnamefont {Jonathan~C.}\
  \bibnamefont {McKinney}}, \ and\ \bibinfo {author} {\bibfnamefont {Ramesh}\
  \bibnamefont {Narayan}},\ }\bibfield  {title} {\enquote {\bibinfo {title}
  {{Simulations of Ultrarelativistic Magnetodynamic Jets from Gamma-ray Burst
  Engines}},}\ }\href {\doibase 10.1111/j.1365-2966.2008.13425.x} {\bibfield
  {journal} {\bibinfo  {journal} {Mon. Not. Roy. Astron. Soc.}\ }\textbf
  {\bibinfo {volume} {388}},\ \bibinfo {pages} {551} (\bibinfo {year}
  {2008})},\ \Eprint {http://arxiv.org/abs/0803.3807} {arXiv:0803.3807
  [astro-ph]} \BibitemShut {NoStop}%
\bibitem [{\citenamefont {Komissarov}\ and\ \citenamefont
  {Barkov}(2009)}]{BZ3}%
  \BibitemOpen
  \bibfield  {author} {\bibinfo {author} {\bibfnamefont {S.~S.}\ \bibnamefont
  {Komissarov}}\ and\ \bibinfo {author} {\bibfnamefont {M.~V.}\ \bibnamefont
  {Barkov}},\ }\bibfield  {title} {\enquote {\bibinfo {title} {{Activation of
  the Blandford-Znajek mechanism in collapsing stars}},}\ }\href {\doibase
  10.1111/j.1365-2966.2009.14831.x} {\bibfield  {journal} {\bibinfo  {journal}
  {Mon. Not. Roy. Astron. Soc.}\ }\textbf {\bibinfo {volume} {397}},\ \bibinfo
  {pages} {1153} (\bibinfo {year} {2009})},\ \Eprint
  {http://arxiv.org/abs/0902.2881} {arXiv:0902.2881 [astro-ph.HE]} \BibitemShut
  {NoStop}%
\bibitem [{\citenamefont {Ruffini}\ and\ \citenamefont {Wilson}(1975)}]{BZ4}%
  \BibitemOpen
  \bibfield  {author} {\bibinfo {author} {\bibfnamefont {Remo}\ \bibnamefont
  {Ruffini}}\ and\ \bibinfo {author} {\bibfnamefont {James~R.}\ \bibnamefont
  {Wilson}},\ }\bibfield  {title} {\enquote {\bibinfo {title} {{Relativistic
  Magnetohydrodynamical Effects of Plasma Accreting Into a Black Hole}},}\
  }\href {\doibase 10.1103/PhysRevD.12.2959} {\bibfield  {journal} {\bibinfo
  {journal} {Phys. Rev. D}\ }\textbf {\bibinfo {volume} {12}},\ \bibinfo
  {pages} {2959} (\bibinfo {year} {1975})}\BibitemShut {NoStop}%
\bibitem [{\citenamefont {Comisso}\ and\ \citenamefont
  {Asenjo}(2021)}]{Comisso}%
  \BibitemOpen
  \bibfield  {author} {\bibinfo {author} {\bibfnamefont {Luca}\ \bibnamefont
  {Comisso}}\ and\ \bibinfo {author} {\bibfnamefont {Felipe~A.}\ \bibnamefont
  {Asenjo}},\ }\bibfield  {title} {\enquote {\bibinfo {title} {Magnetic
  reconnection as a mechanism for energy extraction from rotating black
  holes},}\ }\href {\doibase 10.1103/physrevd.103.023014} {\bibfield  {journal}
  {\bibinfo  {journal} {Physical Review D}\ }\textbf {\bibinfo {volume} {103}}
  (\bibinfo {year} {2021}),\ 10.1103/physrevd.103.023014}\BibitemShut {NoStop}%
\bibitem [{\citenamefont {Carleo}\ \emph {et~al.}(2022)\citenamefont {Carleo},
  \citenamefont {Lambiase},\ and\ \citenamefont
  {Mastrototaro}}]{Carleo:reconnect}%
  \BibitemOpen
  \bibfield  {author} {\bibinfo {author} {\bibfnamefont {Amodio}\ \bibnamefont
  {Carleo}}, \bibinfo {author} {\bibfnamefont {Gaetano}\ \bibnamefont
  {Lambiase}}, \ and\ \bibinfo {author} {\bibfnamefont {Leonardo}\ \bibnamefont
  {Mastrototaro}},\ }\bibfield  {title} {\enquote {\bibinfo {title} {{Energy
  extraction via magnetic reconnection in Lorentz breaking Kerr\textendash{}Sen
  and Kiselev black holes}},}\ }\href {\doibase
  10.1140/epjc/s10052-022-10751-w} {\bibfield  {journal} {\bibinfo  {journal}
  {Eur. Phys. J. C}\ }\textbf {\bibinfo {volume} {82}},\ \bibinfo {pages} {776}
  (\bibinfo {year} {2022})},\ \Eprint {http://arxiv.org/abs/2206.12988}
  {arXiv:2206.12988 [gr-qc]} \BibitemShut {NoStop}%
\bibitem [{\citenamefont {Wald}(1974)}]{Wald:1974kya}%
  \BibitemOpen
  \bibfield  {author} {\bibinfo {author} {\bibfnamefont {Robert~M.}\
  \bibnamefont {Wald}},\ }\bibfield  {title} {\enquote {\bibinfo {title}
  {{Energy Limits on the Penrose Process}},}\ }\href {\doibase 10.1086/152959}
  {\bibfield  {journal} {\bibinfo  {journal} {Astrophys. J.}\ }\textbf
  {\bibinfo {volume} {191}},\ \bibinfo {pages} {231} (\bibinfo {year}
  {1974})}\BibitemShut {NoStop}%
\bibitem [{\citenamefont {Stuchlík}\ \emph {et~al.}(2021)\citenamefont
  {Stuchlík}, \citenamefont {Kološ},\ and\ \citenamefont {Tursunov}}]{MPP1}%
  \BibitemOpen
  \bibfield  {author} {\bibinfo {author} {\bibfnamefont {Zdeněk}\ \bibnamefont
  {Stuchlík}}, \bibinfo {author} {\bibfnamefont {Martin}\ \bibnamefont
  {Kološ}}, \ and\ \bibinfo {author} {\bibfnamefont {Arman}\ \bibnamefont
  {Tursunov}},\ }\bibfield  {title} {\enquote {\bibinfo {title} {Penrose
  process: Its variants and astrophysical applications},}\ }\href {\doibase
  10.3390/universe7110416} {\bibfield  {journal} {\bibinfo  {journal}
  {Universe}\ }\textbf {\bibinfo {volume} {7}} (\bibinfo {year} {2021}),\
  10.3390/universe7110416}\BibitemShut {NoStop}%
\bibitem [{\citenamefont {Tursunov}\ \emph {et~al.}(2020)\citenamefont
  {Tursunov}, \citenamefont {Stuchl{\'{\i}}k}, \citenamefont {Kolo{\v{s}}},
  \citenamefont {Dadhich},\ and\ \citenamefont {Ahmedov}}]{MPP2}%
  \BibitemOpen
  \bibfield  {author} {\bibinfo {author} {\bibfnamefont {Arman}\ \bibnamefont
  {Tursunov}}, \bibinfo {author} {\bibfnamefont {Zden{\v{e} }k}\ \bibnamefont
  {Stuchl{\'{\i}}k}}, \bibinfo {author} {\bibfnamefont {Martin}\ \bibnamefont
  {Kolo{\v{s}}}}, \bibinfo {author} {\bibfnamefont {Naresh}\ \bibnamefont
  {Dadhich}}, \ and\ \bibinfo {author} {\bibfnamefont {Bobomurat}\ \bibnamefont
  {Ahmedov}},\ }\bibfield  {title} {\enquote {\bibinfo {title} {Supermassive
  black holes as possible sources of ultrahigh-energy cosmic rays},}\ }\href
  {\doibase 10.3847/1538-4357/ab8ae9} {\bibfield  {journal} {\bibinfo
  {journal} {The Astrophysical Journal}\ }\textbf {\bibinfo {volume} {895}},\
  \bibinfo {pages} {14} (\bibinfo {year} {2020})}\BibitemShut {NoStop}%
\bibitem [{\citenamefont
  {Dadhich}(2012)}]{https://doi.org/10.48550/arxiv.1210.1041}%
  \BibitemOpen
  \bibfield  {author} {\bibinfo {author} {\bibfnamefont {Naresh}\ \bibnamefont
  {Dadhich}},\ }\href {\doibase 10.48550/ARXIV.1210.1041} {\enquote {\bibinfo
  {title} {Magnetic penrose process and blanford-zanejk mechanism: A
  clarification},}\ } (\bibinfo {year} {2012})\BibitemShut {NoStop}%
\bibitem [{\citenamefont {Sharma}\ \emph {et~al.}(2021)\citenamefont {Sharma},
  \citenamefont {Iyyani},\ and\ \citenamefont {Bhattacharya}}]{Sharma_2021}%
  \BibitemOpen
  \bibfield  {author} {\bibinfo {author} {\bibfnamefont {Vidushi}\ \bibnamefont
  {Sharma}}, \bibinfo {author} {\bibfnamefont {Shabnam}\ \bibnamefont
  {Iyyani}}, \ and\ \bibinfo {author} {\bibfnamefont {Dipankar}\ \bibnamefont
  {Bhattacharya}},\ }\bibfield  {title} {\enquote {\bibinfo {title}
  {Identifying black hole central engines in gamma-ray bursts},}\ }\href
  {\doibase 10.3847/2041-8213/abd53f} {\bibfield  {journal} {\bibinfo
  {journal} {The Astrophysical Journal}\ }\textbf {\bibinfo {volume} {908}},\
  \bibinfo {pages} {L2} (\bibinfo {year} {2021})}\BibitemShut {NoStop}%
\bibitem [{\citenamefont {Takahashi}\ \emph {et~al.}(2021)\citenamefont
  {Takahashi}, \citenamefont {Kino},\ and\ \citenamefont
  {Pu}}]{Takahashi_2021}%
  \BibitemOpen
  \bibfield  {author} {\bibinfo {author} {\bibfnamefont {Masaaki}\ \bibnamefont
  {Takahashi}}, \bibinfo {author} {\bibfnamefont {Motoki}\ \bibnamefont
  {Kino}}, \ and\ \bibinfo {author} {\bibfnamefont {Hung-Yi}\ \bibnamefont
  {Pu}},\ }\bibfield  {title} {\enquote {\bibinfo {title} {Relativistic jet
  acceleration region in a black hole magnetosphere},}\ }\href {\doibase
  10.1103/physrevd.104.103004} {\bibfield  {journal} {\bibinfo  {journal}
  {Physical Review D}\ }\textbf {\bibinfo {volume} {104}} (\bibinfo {year}
  {2021}),\ 10.1103/physrevd.104.103004}\BibitemShut {NoStop}%
\bibitem [{\citenamefont {King}\ and\ \citenamefont
  {Pringle}(2021)}]{King_2021}%
  \BibitemOpen
  \bibfield  {author} {\bibinfo {author} {\bibfnamefont {A.~R.}\ \bibnamefont
  {King}}\ and\ \bibinfo {author} {\bibfnamefont {J.~E.}\ \bibnamefont
  {Pringle}},\ }\bibfield  {title} {\enquote {\bibinfo {title} {Can the
  blandford{\textendash}znajek mechanism power steady jets?}}\ }\href {\doibase
  10.3847/2041-8213/ac19a1} {\bibfield  {journal} {\bibinfo  {journal} {The
  Astrophysical Journal Letters}\ }\textbf {\bibinfo {volume} {918}},\ \bibinfo
  {pages} {L22} (\bibinfo {year} {2021})}\BibitemShut {NoStop}%
\bibitem [{\citenamefont {Komissarov}(2021)}]{Komissarov_2021}%
  \BibitemOpen
  \bibfield  {author} {\bibinfo {author} {\bibfnamefont {Serguei~S}\
  \bibnamefont {Komissarov}},\ }\bibfield  {title} {\enquote {\bibinfo {title}
  {Electrically charged black holes and the blandford{\textendash}znajek
  mechanism},}\ }\href {\doibase 10.1093/mnras/stab2686} {\bibfield  {journal}
  {\bibinfo  {journal} {Monthly Notices of the Royal Astronomical Society}\
  }\textbf {\bibinfo {volume} {512}},\ \bibinfo {pages} {2798--2805} (\bibinfo
  {year} {2021})}\BibitemShut {NoStop}%
\bibitem [{\citenamefont {Komissarov}(2005)}]{Komissarov_2005}%
  \BibitemOpen
  \bibfield  {author} {\bibinfo {author} {\bibfnamefont {S.~S.}\ \bibnamefont
  {Komissarov}},\ }\bibfield  {title} {\enquote {\bibinfo {title} {Observations
  of the blandford-znajek process and the magnetohydrodynamic penrose process
  in computer simulations of black hole magnetospheres},}\ }\href {\doibase
  10.1111/j.1365-2966.2005.08974.x} {\bibfield  {journal} {\bibinfo  {journal}
  {Monthly Notices of the Royal Astronomical Society}\ }\textbf {\bibinfo
  {volume} {359}},\ \bibinfo {pages} {801--808} (\bibinfo {year}
  {2005})}\BibitemShut {NoStop}%
\bibitem [{\citenamefont {Contopoulos}\ \emph {et~al.}(2017)\citenamefont
  {Contopoulos}, \citenamefont {Nathanail},\ and\ \citenamefont
  {Strantzalis}}]{expGRBs}%
  \BibitemOpen
  \bibfield  {author} {\bibinfo {author} {\bibfnamefont {Ioannis}\ \bibnamefont
  {Contopoulos}}, \bibinfo {author} {\bibfnamefont {Antonios}\ \bibnamefont
  {Nathanail}}, \ and\ \bibinfo {author} {\bibfnamefont {Achillies}\
  \bibnamefont {Strantzalis}},\ }\bibfield  {title} {\enquote {\bibinfo {title}
  {The signature of the blandford-znajek mechanism in grb light curves},}\
  }\href {\doibase 10.3390/galaxies5020021} {\bibfield  {journal} {\bibinfo
  {journal} {Galaxies}\ }\textbf {\bibinfo {volume} {5}} (\bibinfo {year}
  {2017}),\ 10.3390/galaxies5020021}\BibitemShut {NoStop}%
\bibitem [{\citenamefont {RUFFINI}(2002)}]{RUFFINI_2002}%
  \BibitemOpen
  \bibfield  {author} {\bibinfo {author} {\bibfnamefont {REMO~J.}\ \bibnamefont
  {RUFFINI}},\ }\bibfield  {title} {\enquote {\bibinfo {title} {{BLACK} {HOLES}
  {AND} {GAMMA} {RAY} {BURSTS}: {BACKGROUND} {FOR} {THE} {THEORETICAL}
  {MODEL}},}\ }in\ \href {\doibase 10.1142/9789812777386_0024} {\emph {\bibinfo
  {booktitle} {The Ninth Marcel Grossmann Meeting}}}\ (\bibinfo  {publisher}
  {World Scientific Publishing Company},\ \bibinfo {year} {2002})\ pp.\
  \bibinfo {pages} {347--379}\BibitemShut {NoStop}%
\bibitem [{\citenamefont {{McKinney}}(2006)}]{Kinney2006}%
  \BibitemOpen
  \bibfield  {author} {\bibinfo {author} {\bibfnamefont {Jonathan~C.}\
  \bibnamefont {{McKinney}}},\ }\bibfield  {title} {\enquote {\bibinfo {title}
  {{General relativistic magnetohydrodynamic simulations of the jet formation
  and large-scale propagation from black hole accretion systems}},}\ }\href
  {\doibase 10.1111/j.1365-2966.2006.10256.x} {\bibfield  {journal} {\bibinfo
  {journal} {mnras}\ }\textbf {\bibinfo {volume} {368}},\ \bibinfo {pages}
  {1561--1582} (\bibinfo {year} {2006})},\ \Eprint
  {http://arxiv.org/abs/astro-ph/0603045} {arXiv:astro-ph/0603045 [astro-ph]}
  \BibitemShut {NoStop}%
\bibitem [{\citenamefont {Steiner}\ \emph {et~al.}(2012)\citenamefont
  {Steiner}, \citenamefont {McClintock},\ and\ \citenamefont
  {Narayan}}]{Steiner_2012}%
  \BibitemOpen
  \bibfield  {author} {\bibinfo {author} {\bibfnamefont {James~F.}\
  \bibnamefont {Steiner}}, \bibinfo {author} {\bibfnamefont {Jeffrey~E.}\
  \bibnamefont {McClintock}}, \ and\ \bibinfo {author} {\bibfnamefont {Ramesh}\
  \bibnamefont {Narayan}},\ }\bibfield  {title} {\enquote {\bibinfo {title}
  {{Jet} {power} {and} {black} {hole} {spin}: {testing} {an} {emprical}
  {relationship} {and} {using} {it} {to} {predict} {the} {spins} {of} {six}
  {black} {holes}},}\ }\href {\doibase 10.1088/0004-637x/762/2/104} {\bibfield
  {journal} {\bibinfo  {journal} {The Astrophysical Journal}\ }\textbf
  {\bibinfo {volume} {762}},\ \bibinfo {pages} {104} (\bibinfo {year}
  {2012})}\BibitemShut {NoStop}%
\bibitem [{\citenamefont {McKinney}\ and\ \citenamefont
  {Gammie}(2004)}]{Gammie_2004}%
  \BibitemOpen
  \bibfield  {author} {\bibinfo {author} {\bibfnamefont {Jonathan~C.}\
  \bibnamefont {McKinney}}\ and\ \bibinfo {author} {\bibfnamefont {Charles~F.}\
  \bibnamefont {Gammie}},\ }\bibfield  {title} {\enquote {\bibinfo {title} {A
  measurement of the electromagnetic luminosity of a kerr black hole},}\ }\href
  {\doibase 10.1086/422244} {\bibfield  {journal} {\bibinfo  {journal} {The
  Astrophysical Journal}\ }\textbf {\bibinfo {volume} {611}},\ \bibinfo {pages}
  {977--995} (\bibinfo {year} {2004})}\BibitemShut {NoStop}%
\bibitem [{\citenamefont {Ghosh}(2000)}]{Ghosh_2000}%
  \BibitemOpen
  \bibfield  {author} {\bibinfo {author} {\bibfnamefont {P.}~\bibnamefont
  {Ghosh}},\ }\bibfield  {title} {\enquote {\bibinfo {title} {The structure of
  black hole magnetospheres -- i. schwarzschild black holes},}\ }\href
  {\doibase 10.1046/j.1365-8711.2000.03410.x} {\bibfield  {journal} {\bibinfo
  {journal} {Monthly Notices of the Royal Astronomical Society}\ }\textbf
  {\bibinfo {volume} {315}},\ \bibinfo {pages} {89--97} (\bibinfo {year}
  {2000})}\BibitemShut {NoStop}%
\bibitem [{\citenamefont {{Komissarov}}(2001)}]{komiss2001}%
  \BibitemOpen
  \bibfield  {author} {\bibinfo {author} {\bibfnamefont {S.~S.}\ \bibnamefont
  {{Komissarov}}},\ }\bibfield  {title} {\enquote {\bibinfo {title} {{Direct
  numerical simulations of the Blandford-Znajek effect}},}\ }\href {\doibase
  10.1046/j.1365-8711.2001.04863.x} {\bibfield  {journal} {\bibinfo  {journal}
  {MNRAS}\ }\textbf {\bibinfo {volume} {326}},\ \bibinfo {pages} {L41--L44}
  (\bibinfo {year} {2001})}\BibitemShut {NoStop}%
\bibitem [{\citenamefont {Nathanail}\ and\ \citenamefont
  {Contopoulos}(2014)}]{Nathanail_2014}%
  \BibitemOpen
  \bibfield  {author} {\bibinfo {author} {\bibfnamefont {Antonios}\
  \bibnamefont {Nathanail}}\ and\ \bibinfo {author} {\bibfnamefont {Ioannis}\
  \bibnamefont {Contopoulos}},\ }\bibfield  {title} {\enquote {\bibinfo {title}
  {{Black Hole Magnetospheres}},}\ }\href {\doibase
  10.1088/0004-637X/788/2/186} {\bibfield  {journal} {\bibinfo  {journal}
  {Astrophys. J.}\ }\textbf {\bibinfo {volume} {788}},\ \bibinfo {pages} {186}
  (\bibinfo {year} {2014})},\ \Eprint {http://arxiv.org/abs/1404.0549}
  {arXiv:1404.0549 [astro-ph.HE]} \BibitemShut {NoStop}%
\bibitem [{\citenamefont {Penna}\ \emph {et~al.}(2013)\citenamefont {Penna},
  \citenamefont {Narayan},\ and\ \citenamefont {S{\k{a}
  }dowski}}]{SimulationBZ}%
  \BibitemOpen
  \bibfield  {author} {\bibinfo {author} {\bibfnamefont {Robert~F.}\
  \bibnamefont {Penna}}, \bibinfo {author} {\bibfnamefont {Ramesh}\
  \bibnamefont {Narayan}}, \ and\ \bibinfo {author} {\bibfnamefont
  {Aleksander}\ \bibnamefont {S{\k{a} }dowski}},\ }\bibfield  {title} {\enquote
  {\bibinfo {title} {General relativistic magnetohydrodynamic simulations of
  blandford{\textendash}znajek jets and the membrane paradigm},}\ }\href
  {\doibase 10.1093/mnras/stt1860} {\bibfield  {journal} {\bibinfo  {journal}
  {Monthly Notices of the Royal Astronomical Society}\ }\textbf {\bibinfo
  {volume} {436}},\ \bibinfo {pages} {3741--3758} (\bibinfo {year}
  {2013})}\BibitemShut {NoStop}%
\bibitem [{\citenamefont {Pei}\ \emph {et~al.}(2016)\citenamefont {Pei},
  \citenamefont {Nampalliwar}, \citenamefont {Bambi},\ and\ \citenamefont
  {Middleton}}]{cosimo_2016}%
  \BibitemOpen
  \bibfield  {author} {\bibinfo {author} {\bibfnamefont {Guancheng}\
  \bibnamefont {Pei}}, \bibinfo {author} {\bibfnamefont {Sourabh}\ \bibnamefont
  {Nampalliwar}}, \bibinfo {author} {\bibfnamefont {Cosimo}\ \bibnamefont
  {Bambi}}, \ and\ \bibinfo {author} {\bibfnamefont {Matthew~J.}\ \bibnamefont
  {Middleton}},\ }\bibfield  {title} {\enquote {\bibinfo {title}
  {Blandford{\textendash}znajek mechanism in black holes in alternative
  theories of gravity},}\ }\href {\doibase 10.1140/epjc/s10052-016-4387-z}
  {\bibfield  {journal} {\bibinfo  {journal} {The European Physical Journal C}\
  }\textbf {\bibinfo {volume} {76}} (\bibinfo {year} {2016}),\
  10.1140/epjc/s10052-016-4387-z}\BibitemShut {NoStop}%
\bibitem [{\citenamefont {Eatough}\ \emph {et~al.}(2013)\citenamefont
  {Eatough}, \citenamefont {Falcke}, \citenamefont {Karuppusamy}, \citenamefont
  {Lee}, \citenamefont {Champion}, \citenamefont {Keane}, \citenamefont
  {Desvignes}, \citenamefont {Schnitzeler}, \citenamefont {Spitler},
  \citenamefont {Kramer}, \citenamefont {Klein}, \citenamefont {Bassa},
  \citenamefont {Bower}, \citenamefont {Brunthaler}, \citenamefont {Cognard},
  \citenamefont {Deller}, \citenamefont {Demorest}, \citenamefont {Freire},
  \citenamefont {Kraus}, \citenamefont {Lyne}, \citenamefont {Noutsos},
  \citenamefont {Stappers},\ and\ \citenamefont
  {Wex}}]{B_Centro_ViaLattea_2013}%
  \BibitemOpen
  \bibfield  {author} {\bibinfo {author} {\bibfnamefont {R.~P.}\ \bibnamefont
  {Eatough}}, \bibinfo {author} {\bibfnamefont {H.}~\bibnamefont {Falcke}},
  \bibinfo {author} {\bibfnamefont {R.}~\bibnamefont {Karuppusamy}}, \bibinfo
  {author} {\bibfnamefont {K.~J.}\ \bibnamefont {Lee}}, \bibinfo {author}
  {\bibfnamefont {D.~J.}\ \bibnamefont {Champion}}, \bibinfo {author}
  {\bibfnamefont {E.~F.}\ \bibnamefont {Keane}}, \bibinfo {author}
  {\bibfnamefont {G.}~\bibnamefont {Desvignes}}, \bibinfo {author}
  {\bibfnamefont {D.~H. F.~M.}\ \bibnamefont {Schnitzeler}}, \bibinfo {author}
  {\bibfnamefont {L.~G.}\ \bibnamefont {Spitler}}, \bibinfo {author}
  {\bibfnamefont {M.}~\bibnamefont {Kramer}}, \bibinfo {author} {\bibfnamefont
  {B.}~\bibnamefont {Klein}}, \bibinfo {author} {\bibfnamefont
  {C.}~\bibnamefont {Bassa}}, \bibinfo {author} {\bibfnamefont {G.~C.}\
  \bibnamefont {Bower}}, \bibinfo {author} {\bibfnamefont {A.}~\bibnamefont
  {Brunthaler}}, \bibinfo {author} {\bibfnamefont {I.}~\bibnamefont {Cognard}},
  \bibinfo {author} {\bibfnamefont {A.~T.}\ \bibnamefont {Deller}}, \bibinfo
  {author} {\bibfnamefont {P.~B.}\ \bibnamefont {Demorest}}, \bibinfo {author}
  {\bibfnamefont {P.~C.~C.}\ \bibnamefont {Freire}}, \bibinfo {author}
  {\bibfnamefont {A.}~\bibnamefont {Kraus}}, \bibinfo {author} {\bibfnamefont
  {A.~G.}\ \bibnamefont {Lyne}}, \bibinfo {author} {\bibfnamefont
  {A.}~\bibnamefont {Noutsos}}, \bibinfo {author} {\bibfnamefont
  {B.}~\bibnamefont {Stappers}}, \ and\ \bibinfo {author} {\bibfnamefont
  {N.}~\bibnamefont {Wex}},\ }\bibfield  {title} {\enquote {\bibinfo {title} {A
  strong magnetic field around the supermassive black hole at the centre of the
  galaxy},}\ }\href {\doibase 10.1038/nature12499} {\bibfield  {journal}
  {\bibinfo  {journal} {Nature}\ }\textbf {\bibinfo {volume} {501}},\ \bibinfo
  {pages} {391--394} (\bibinfo {year} {2013})}\BibitemShut {NoStop}%
\bibitem [{\citenamefont {Liska}\ \emph {et~al.}(2020)\citenamefont {Liska},
  \citenamefont {Tchekhovskoy},\ and\ \citenamefont
  {Quataert}}]{Simulaz_poloidal_Liska_2020}%
  \BibitemOpen
  \bibfield  {author} {\bibinfo {author} {\bibfnamefont {M}~\bibnamefont
  {Liska}}, \bibinfo {author} {\bibfnamefont {A}~\bibnamefont {Tchekhovskoy}},
  \ and\ \bibinfo {author} {\bibfnamefont {E}~\bibnamefont {Quataert}},\
  }\bibfield  {title} {\enquote {\bibinfo {title} {Large-scale poloidal
  magnetic field dynamo leads to powerful jets in {GRMHD} simulations of black
  hole accretion with toroidal field},}\ }\href {\doibase
  10.1093/mnras/staa955} {\bibfield  {journal} {\bibinfo  {journal} {Monthly
  Notices of the Royal Astronomical Society}\ }\textbf {\bibinfo {volume}
  {494}},\ \bibinfo {pages} {3656--3662} (\bibinfo {year} {2020})}\BibitemShut
  {NoStop}%
\bibitem [{\citenamefont {{Event Horizon Telescope
  Collaboration}}(2021)}]{EHT_part8_Mag_Fields}%
  \BibitemOpen
  \bibfield  {author} {\bibinfo {author} {\bibnamefont {{Event Horizon
  Telescope Collaboration}}},\ }\bibfield  {title} {\enquote {\bibinfo {title}
  {{First M87 Event Horizon Telescope Results. VIII. Magnetic Field Structure
  near The Event Horizon}},}\ }\href {\doibase 10.3847/2041-8213/abe4de}
  {\bibfield  {journal} {\bibinfo  {journal} {APJ}\ }\textbf {\bibinfo {volume}
  {910}},\ \bibinfo {eid} {L13} (\bibinfo {year} {2021})},\ \Eprint
  {http://arxiv.org/abs/2105.01173} {arXiv:2105.01173 [astro-ph.HE]}
  \BibitemShut {NoStop}%
\bibitem [{\citenamefont {Kothari}\ \emph {et~al.}(2020)\citenamefont
  {Kothari}, \citenamefont {Saketh},\ and\ \citenamefont
  {Jain}}]{Kothari:2018aem}%
  \BibitemOpen
  \bibfield  {author} {\bibinfo {author} {\bibfnamefont {Rahul}\ \bibnamefont
  {Kothari}}, \bibinfo {author} {\bibfnamefont {M.~V.~S.}\ \bibnamefont
  {Saketh}}, \ and\ \bibinfo {author} {\bibfnamefont {Pankaj}\ \bibnamefont
  {Jain}},\ }\bibfield  {title} {\enquote {\bibinfo {title} {{Torsion driven
  Inflationary Magnetogenesis}},}\ }\href {\doibase
  10.1103/PhysRevD.102.024008} {\bibfield  {journal} {\bibinfo  {journal}
  {Phys. Rev. D}\ }\textbf {\bibinfo {volume} {102}},\ \bibinfo {pages}
  {024008} (\bibinfo {year} {2020})},\ \Eprint
  {http://arxiv.org/abs/1806.02505} {arXiv:1806.02505 [astro-ph.CO]}
  \BibitemShut {NoStop}%
\bibitem [{\citenamefont {Gasperini}\ \emph
  {et~al.}(1995{\natexlab{b}})\citenamefont {Gasperini}, \citenamefont
  {Giovannini},\ and\ \citenamefont {Veneziano}}]{Gasperini:1995dh}%
  \BibitemOpen
  \bibfield  {author} {\bibinfo {author} {\bibfnamefont {M.}~\bibnamefont
  {Gasperini}}, \bibinfo {author} {\bibfnamefont {Massimo}\ \bibnamefont
  {Giovannini}}, \ and\ \bibinfo {author} {\bibfnamefont {G.}~\bibnamefont
  {Veneziano}},\ }\bibfield  {title} {\enquote {\bibinfo {title} {{Primordial
  magnetic fields from string cosmology}},}\ }\href {\doibase
  10.1103/PhysRevLett.75.3796} {\bibfield  {journal} {\bibinfo  {journal}
  {Phys. Rev. Lett.}\ }\textbf {\bibinfo {volume} {75}},\ \bibinfo {pages}
  {3796--3799} (\bibinfo {year} {1995}{\natexlab{b}})},\ \Eprint
  {http://arxiv.org/abs/hep-th/9504083} {arXiv:hep-th/9504083} \BibitemShut
  {NoStop}%
\bibitem [{\citenamefont {Atmjeet}\ \emph {et~al.}(2014)\citenamefont
  {Atmjeet}, \citenamefont {Pahwa}, \citenamefont {Seshadri},\ and\
  \citenamefont {Subramanian}}]{Atmjeet_2014}%
  \BibitemOpen
  \bibfield  {author} {\bibinfo {author} {\bibfnamefont {Kumar}\ \bibnamefont
  {Atmjeet}}, \bibinfo {author} {\bibfnamefont {Isha}\ \bibnamefont {Pahwa}},
  \bibinfo {author} {\bibfnamefont {T.{\hspace{0.167em} }R.}\ \bibnamefont
  {Seshadri}}, \ and\ \bibinfo {author} {\bibfnamefont {Kandaswamy}\
  \bibnamefont {Subramanian}},\ }\bibfield  {title} {\enquote {\bibinfo {title}
  {Cosmological magnetogenesis from extra-dimensional gauss-bonnet gravity},}\
  }\href {\doibase 10.1103/physrevd.89.063002} {\bibfield  {journal} {\bibinfo
  {journal} {Physical Review D}\ }\textbf {\bibinfo {volume} {89}} (\bibinfo
  {year} {2014}),\ 10.1103/physrevd.89.063002}\BibitemShut {NoStop}%
\bibitem [{\citenamefont {Konoplya}\ \emph {et~al.}(2021)\citenamefont
  {Konoplya}, \citenamefont {Kunz},\ and\ \citenamefont
  {Zhidenko}}]{Konoplya_2021}%
  \BibitemOpen
  \bibfield  {author} {\bibinfo {author} {\bibfnamefont {R.A.}\ \bibnamefont
  {Konoplya}}, \bibinfo {author} {\bibfnamefont {J.}~\bibnamefont {Kunz}}, \
  and\ \bibinfo {author} {\bibfnamefont {A.}~\bibnamefont {Zhidenko}},\
  }\bibfield  {title} {\enquote {\bibinfo {title} {Blandford-znajek mechanism
  in the general stationary axially-symmetric black-hole spacetime},}\ }\href
  {\doibase 10.1088/1475-7516/2021/12/002} {\bibfield  {journal} {\bibinfo
  {journal} {Journal of Cosmology and Astroparticle Physics}\ }\textbf
  {\bibinfo {volume} {2021}},\ \bibinfo {pages} {002} (\bibinfo {year}
  {2021})}\BibitemShut {NoStop}%
\end{thebibliography}%


\end{document}